\pdfoutput=1
\documentclass[twocolumn]{aa} 
\usepackage{natbib}      
\bibpunct{(}{)}{;}{a}{}{,} 
\usepackage{amsmath}
\usepackage{txfonts}
\usepackage{caption}
\usepackage{subcaption}
\usepackage{graphicx}
\usepackage{bm}
\usepackage[utf8]{inputenc}
\usepackage{float}
\usepackage{stfloats}
\usepackage{hyperref}
\hypersetup{
colorlinks=true,
linkcolor=blue,
urlcolor=blue,
citecolor=blue,
}
\begin{document} 
%
\newcommand{\graphics}{plots_finalfinal}
\newcommand{\graphicsmore}{plots_finalfinal}
\newcommand{\graphicsmoree}{plots_finalfinal}
\newcommand{\gf}{plots_finalfinal}
\newcommand{\gfstr}{plots_finalfinal/streams}
\newcommand{\cmg}{\rm\, cm^2 \, g^{-1}} 
\newcommand{\gcm}{\rm\, g \, cm^{-3}} 
   \title{Global 3D radiation-hydrodynamic simulations of gas accretion: The opacity dependent growth of Saturn-mass planets}
	 \titlerunning{The opacity dependent growth of Saturn-mass planets}
   \author{M. Schulik,
          \inst{1}
					A. Johansen,
					\inst{1}
          B. Bitsch, 
					\inst{2}
					E. Lega
					\inst{3}}
	\authorrunning{Schulik et al.}
   \institute{Lund Observatory, Box 43, S\"olvegatan 27, SE-22100 Lund, Sweden, 
              \email{schulik@astro.lu.se}
						\and
						Max-Planck Institut f\"ur Astronomie, K\"onigsstuhl 17, 69117 Heidelberg, Germany
						\and
             Laboratoire Lagrange, UMR7293, Universit\'e de la C\^ote d'Azur, Boulevard de la Observatoire, 06304 Nice Cedex 4, France
             }
   \date{Received ...}
  \abstract
   {The full spatial structure and temporal evolution of the accretion flow into the envelopes of growing gas giants in their nascent discs is only accessible in simulations. 
Such simulations are constrained in their approach of computing the formation of gas giants by dimensionality, resolution, consideration of self-gravity, energy treatment and the adopted opacity law.
Our study explores how a number of these parameters affect that measured accretion rate of a Saturn-mass planet.
We present a global 3D radiative hydrodynamics framework using the FARGOCA-code. 
The planet is represented by a gravitational potential with a smoothing length at the location of the planet. No mass or energy sink is used, instead luminosity and gas accretion rates are self-consistently computed.
We find that the gravitational smoothing length must be resolved by at least 10 grid cells to obtain converged measurements of the gas accretion rates. Secondly, we find gas accretion rates into planetary envelopes compatible with previous studies, and continue to explain those via the structure of our planetary envelopes and their luminosities. Our measured gas accretion rates are formally in the stage of Kelvin-Helmholtz contraction due to the modest entropy loss that can be obtained over the simulation time-scale, but our accretion rates are compatible with those expected during late run-away accretion. Our detailed simulations of the gas flow into the envelope of a Saturn-mass planet provides a framework for understanding the general problem of gas accretion during planet formation and highlights circulation features that develop inside the planetary envelopes. Those circulation features feedback into the envelope energetics and can have further implications for transporting dust into the inner regions of the envelope.}
	%
   \maketitle
%
\section{Introduction}
Gas giants form an important class of planets in the Solar System as well as in extrasolar planetary systems. As the most massive planetary class, their gravity determines the evolution of planetary systems after their formation \citep{davies2014}.
Ever since the first discovery of gas giant exoplanets \citep{mayor1995}, their occurence rates and correlations with host star properties have been mapped observationally. 
It is clear now that gas giants orbit primarily around stars that are rich in heavy elements \citep{gonzalez1997, fischer2005, mayor2011}; this is particularly pronounced for gas giants on eccentric orbits that likely resulted from planet-planet scattering in systems of multiple gas giants \citep{dawson2013, buchhave2018}. The significant increase in the formation rate of gas giants with metallicity supports the core accretion model for the formation of such planets \citep{pollack1996}.
The formation of a planetary core and subsequent gas accretion must happen during the gaseous disc phase of duration between 1-10 Myrs . The then following formation of one or more gas giants can have dramatic influences on further planet formation. The growing gas giants will form gaps, first in the dust and, if the planet becomes sufficiently massive, also in the gas.
These gaps in dust \citep{paardekooper2004, lambrechts2014, bitsch2018, ataiee2018}) and the gas distributions \citep{ward1997, lubow2005, crida2006}, will cut off or at least heavily modify the influx of material onto the inner planets. The edges of planetary gas gaps can destabilize, become turbulent and thus influence further disc evolution via changing the strength of turbulence or regulating their own depth \citep{hallam2017}. The gap structure also feedbacks on the planet. Gaps slow planetary migration through the parent disc \citep{lin1986} and both gap depth and migration speed can affect accretion independently \citep{duermann2017, kanagawa2018}.
Migrating gas giants will also excite the orbits of other planets in the system \citep{raymond2016} and change their migration speed \citep{masset2001}. They can further influence the evolution of systems through trailing swarms of pebbles that accumulate at their gap edge and form seed locations for the growth of new planets \citep{ronnet2018}.
It is thus an important priority of planet formation theory to clarify the gas accretion rates of young gas giants.
The early one-dimensional gas accretion models of \cite{mizuno1980} determined that beyond a certain critical core mass, the gaseous envelope of a young planet cannot remain in hydrostatic equilibrium. Subsequent work has then often focused on finding this critical mass as function of disc parameters with increasing model complexities \citep{bodenheimerpollack1986, wuchterl1990, pollack1996}. Important ingredients in letting planets pass the critical core mass was firstly the build-up of mass via accretion of solids, and secondly the cessation of their accretion, which allows for the circumplanetary gas envelope to cool, contract and go into runaway gas accretion. Those approaches have been refined over the years, e.g. with more detailed opacity laws and structure models \citep{pisoyoudin2014, piso2015}.
However, effects of higher dimensionality have been shown to be important even before crossing critical mass. Horseshoe-flows that penetrate the planetary Hill-sphere in the midplane and interfere with the contracting planetary envelope can only be accurately modeled in at least 2-D \citep{ormel2015}. Vertical outflows as well as cooling of the gas into space even necessitate full 3-D models \citep{tanigawa2012, lambrechts2017}.

As well as dimensionality, also the treatment of the energy equation(s) for gas dynamics plays a crucial role in calculating the release of initial accretion heat, allowing the accretion process to continue.
The treatment of the energy equation has gone through several iterations, ranging from isothermal \citep[e.g.][]{stonenorman1992, dangelo2003, tanigawa2012, fung2015}  or adiabatic, or mixes between them  \citep[e.g.][]{machida2010} over Newtonian-cooling with density-dependent cooling times \citep{gressel2013, zhu2016} to radiative transport simulations \citep[e.g.][]{ayliffebate2009, ayliffebate2012, dangelobodenheimer2013} which approximate the transitions from optically thick to thin regions. 

Since 2-D or 3-D simulations are computationally expensive, there are limitations on the maximum size of the simulation domain around the planet and limitations on how close to the planet the simulation can resolve physics correctly. This has led to two approaches: 

Local shearing boxes, resolve the planetary Hill-sphere well by the use of mesh-refinement techniques \citep[e.g.][]{zhu2016}, but can have problems representing gap-opening planets and keeping the global energy balance. Global simulations have large radial and full azimuthal extent, and are sophisticated enough to resolve the planetary Hill-sphere \citep[e.g.][]{fung2015, lambrechts2017} or even the planetary radius \citep[e.g.][]{dangelo2003, dangelobodenheimer2013, judith2014}, but usually suffer from run-time restrictions that makes it impractical to simulate evolutionary timescales of planets. A notable exception here is the work from \cite{ayliffebate2012} who managed to simulate the runaway collapse of planetary envelopes in 3D radiative SPH-hydrodynamics globally with self-gravity.

Finally, accretion of the planet is also subject to different treatments. It is general practice to either use a global simulation with a certain fraction of mass taken out of the central 10-50\% of the planetary Hill-radius \citep{kley1989, klahrkley2006}, or in simulations with very high spatial resolution to let the cooling process proceed naturally, without taking out mass artificially \citep{dangelobodenheimer2013}.

In this work we aim to measure mass accretion rates onto giant planets, while relaxing physical assumptions as much as possible.
To this end, we present a grid-based, global, 3D radiation hydrodynamical two-temperature setting where neither the accretion rate nor the accretion luminosity is imposed, but self-consistently calculated from the contraction of the gas, viscous heating and heat advection. 
We resolve the planetary Hill-sphere down to 5\% of the Hill-radius, where this again is maximally resolved by up to $\approx$$8$-$20$ grid cells, in order to assess the convergence of the results with increasing resolution of the Hill radius.
We also investigate the influence of the two numerical parameters resolution and gravitational smoothing length, as well as of the key physical parameter regulating the cooling of the planetary envelope which is the opacity law, on the measured accretion rates.

This is intended part as feasibility study, part as physical investigation. 
As the outcome of any accretion signal should be an unambigous increase in mass, we searched the literature for helpful hints concerning the required parameters to maximize this increase. The work of \cite{ayliffebate2009} and \cite{bodenheimer2013}, where our setup mostly resembles the former, showed that a Saturn-mass planet should yield a maximum in accretion rate. This results on one hand from the fact that accretion rates increase with planetary masses, but on the other hand that massive, gap-opening planets deplete their own feeding zones through the action of gravitational torques, decreasing accretion rates as they grow even further. Therefore we focus here on Saturn-mass protoplanets and leave the dependence of the results on the mass of the protoplanet to a future work.

As our planets are massive enough to form a gap in the gas distribution and are beyond their pebble isolation mass \citep{lambrechts2014, bitsch2018}, we do not set any additional heating sources from solids, which has been explored in a similar setting to ours in \cite{lambrechts2017} for low-mass planets. Similar work to theirs has also been done in \cite{dangelobodenheimer2013} for low-mass planets between $5$ and $15$ Earth masses, although those authors did a more careful modelisation of the thermodynamics, particularly the dependence of the adiabatic index $\gamma$ on temperature. They, however, assumed matter-photon energy equilibrium, which implies a one-temperature approach, an approximation which we can relax in order to perform a more realistic exploration of the cooling behaviour of high-mass planets.

This paper is organized as follows. In Sect. \ref{sec:model} we present a brief overview of the physical model used and the free parameters it depends on. Section \ref{sec:methods} outlines the numerical methods used to solve the physical model and explains three individual steps we use to keep computation time reasonable. Section \ref{sec:numericalresolution} discusses a limited set of measured accretion rates and their dependency on the numerical resolution, which is shown to be the key parameter for this problem. This section also defines our concept of a well-resolved simulation.
Section \ref{sec:opacsmoothing} follows up investigating accretion rates as function of smoothing length and opacities for well-resolved simulations.
Section \ref{sec:envelope} discusses the envelope structure of a few selected simulations to highlight the influence of the simulation parameters and connect those back to the accretion rate measurments.
Finally, Sect. \ref{sec:discussion} discusses implications and various findings from our simulation runs. In Appendices \ref{sec:appendix_fld} - \ref{sec:appendix_entropy} we expand on the main text by justifying the flux-limited diffusion approximation and luminosity measurments, as well as deriving the potential temperature, a tool we use in this paper.

\section{Physical model}
\label{sec:model}

The modeling of gas accretion rates requires the treatment of density, momentum and two energy conservation equations. In this section we present those, and discuss a few of their important properties.
	
\subsection{The Navier-Stokes equations}

We use the standard, time-dependent system of compressional mass conservation and Navier-Stokes equations for viscous fluids
\begin{align}
\frac{\partial \rho}{\partial t} + \vec \nabla \cdot \left( \rho \vec u \right) &= 0, 
\label{eq:massconservation}
\\
\frac{\partial \vec u}{\partial t}  + \vec u \cdot \vec \nabla \vec u &= -\frac{1}{\rho} \vec\nabla P + \vec \nabla \cdot \vec \Pi + \vec g
\label{eq:navierstokes},
\end{align}
where $t$ is the time, $\rm \rho$ is the scalar volume mass density of the gas, $\bm{u}$ is the local vectorial velocity field, $P$ is the pressure scalar field, $\bm{\Pi}$ is the viscous stress tensor, defined as
\begin{align}
\vec \Pi = 2 \rho \nu \left[ \vec D - \frac{1}{3} ( \vec \nabla \cdot \vec u )  \vec I \right]
\label{eq:viscousstresstensor}
\end{align}
is a function of density, physical viscosity $\nu$, shear tensor $D_{ij} = \frac{1}{2} \left( \frac{\partial u_i}{\partial x_j} + \frac{\partial u_j}{\partial x_i} \right)$ and velocity divergence times the unit matrix $\bm{I}$.
Finally, $\bm{g}$ is the local vectorial gravity field. The field of gravity $\bm{g}$ contains the contributions from the star, as well as the planetary core. The planetary gravity field is additionally smoothed out in the innermost grid-cells. Here we use the gravitational smoothing prescription from \citet{klahrkley2006}, which sets the planetary gravitational potential to 
\begin{align}
\mathit{\Phi}_{\rm p} = - m_{\rm p} G \left[ \frac{\mathit{r}^3}{\mathit{r}_s^4} - 2 \frac{\mathit{r}^2}{\mathit{r}_s^3} + \frac{2}{\mathit{r}_s} \right] ,
\label{eq:planetarypotential}
\end{align}
where $\vec g = - \bm{\nabla} \mathit{\Phi}(r)$, $r$ is the planetocentric radius, $m_{\rm p}$ the planetary mass, $G$ the gravitational constant and $r_{\rm s}$ is the smoothing length. At this point we also introduce the dimensionless gravitational smoothing length $\tilde{r}_{\rm s} = r_{\rm s}/r_{\rm H}$, where $r_{\rm H}$ is the planetary Hill-radius 
\begin{align}
r_{\rm H} = r_{\rm p} \left( \frac{m_p}{3 m_{\odot}} \right)^{1/3},
\label{eq:hillradius}
\end{align}
with $r_{\rm p}$ the planet-to-star semi-major axis and $m_{\odot}$ the stellar mass.
The gravitational smoothing is 
necessary to reach numerically convergent solutions when increasing the simulation resolution. For works that study migration \citep[i.e.][]{kley2009} the gravitational smoothing additionally takes care of avoiding singular values of the gravity that could occur when the planet migrates very close to a cell midpoint. We neglect momentum transfer forces from the photons onto the gas, as their ratio w.r.t. the planetary gravity is only $\sim$ $10^{-5}$ in the worst case of the highest opacities used in our model.

Equations \ref{eq:massconservation} and \ref{eq:navierstokes} are solved in FARGOCA \citep{lega2014} via a classic operator split method \citep{stonenorman1992}.  
More information on this methodology can be found in \citet{masset2000}.
As frame of reference we chose the planetary co-rotating frame, while the planetary radial position is fixed at $r_{\rm p}=5.2 \rm\, AU$ distance from the star. Inertial forces resulting from this transformation are the Coriolis and centrifugal force and are not written explicitly into eq. \ref{eq:navierstokes}.

\subsection{The energy equations}
\label{sec:equations}


Our simulations solve two energy equations, one for the internal energy of the gas, $e_{\rm int}$, and one for the radiative energy of photons, $e_{\rm rad}$. Technically, this corresponds also to two temperatures, one for the gas and one for the radiation. We nevertheless only name one of those by name, the gas temperature or just the temperature $T$, which connects to the internal energy over the equation of state $P=\left( \gamma-1 \right) e_{\rm int} $, with the constant adiabatic index $\gamma$ set to $1.4$ in our simulations. While $\gamma$ can vary in a fully realistic setting, we restrict our simulations to potential depths and thus temperatures which would vary $\gamma$ only weakly \citep{popovas2016}. Only our deepest potentials reach $T\approx 2000\, \rm K$ and only in the few grid cells closest to the planet. Using a variable $\gamma$ would then change our results only negligibly (O. Gressel (private communication)).  

The two energy equations are \citep{commercon2011}
\begin{align}
\frac{\partial e_{\rm int}}{\partial t} + \left( \vec u \cdot \vec \nabla \right) e_{\rm int} &= -e_{\rm int}\vec\nabla\cdot\vec u -\rho \kappa_{\rm P} \left(B -c e_{\rm rad} \right) + Q_{\rm visc} \label{eq:energyequation1}, \\
\frac{\partial e_{\rm rad}}{\partial t} + \vec \nabla \cdot \vec F &= \rho \kappa_{\rm P} \left(B -c e_{\rm rad} \right), \label{eq:energyequation2}	
\end{align}
where $\kappa_{\rm P}$ is the Planck-mean opacity,
$B$ is the black-body energy radiation density with $B=4\sigma T^4$, $\sigma$ the radiation constant, $c$ the speed of light, $Q_{\rm visc} = \nu \rho \vec \Pi^2$ is the self-contraction of the stress tensor and the radiative flux density is $\bm{F}$.
Mechanisms for local change of internal energy correspond to the processes of advection, compression, coupling to the radiation field and viscous heating, from left to right in eq. \ref{eq:energyequation1}. 

The radiative fluxes $\bm{F}$ are computed via the classical Flux-Limited Diffusion (FLD). FLD is a formalism introduced by \cite{levermore1981} that presses the radiation energy fluxes in the optically thin limit in a form compatible with the optically thick limit, in order to allow for solutions in both cases with only one solver. In appendix \ref{sec:appendix_fld} we show how the fluxes transition from the optically thick to the optically thin limit. In general, the FLD radiative fluxes have the form
\begin{align}
\bm{F} = \frac{\lambda c}{\rho \kappa_{\rm R}} \vec \nabla e_{\rm rad},
\label{eq:fluxesfld}
\end{align}
with $\kappa_{\rm R}$ the Rosseland mean opacity and the interpolation function $\lambda$ being a function $\lambda = f(\rho \kappa_R/L_e)$ of the local optical depth per grid-cell divided by the local radiative energy gradient scale length $L_e$. Here $\lambda$ is calculated for each cell-interface in order to decide between optically thick or thin fluxes. Specific choices for $\kappa_{\rm P}$ and $\kappa_{\rm R}$ are discussed in the next section. The implicit FLD solver that we use is based on the work of \citet{commercon2011}, also described there in great detail, and has already been successfully used e.g. in \citet{bitsch2013a} and \citet{lega2015}. 

A useful reformulation of the internal energy equation into that of the specific entropy $S$ is \citep[see][Chapt. 1.6]{vallis}
\begin{align}
\rho \frac{d S}{dt} = -\rho \kappa_{\rm P} \left(B -c e_{\rm rad} \right) + Q_{\rm visc} .\label{eq:entropyequation}
\end{align}
where $d/dt = \partial/\partial t  + \vec v \cdot \vec \nabla$ is the Lagrangian time-differential operator.
This equation shows that specific entropy can only be generated through viscous heating or lost through coupling to radiation. The specific entropy is directly proportional to a diagnostic quantity often used in atmospheric physics, namely the potential temperature $\rm \vartheta$ \citep[see e.g.][]{vallis} and it is $S = c_P \ln(\vartheta)$ for a gas of constant specific heat capacity at constant pressure $c_P$. The potential temperature can be thought of as a temperature corrected for effects of compression/expansion and is therefore a quantity derived from the simulation data, not an additional temperature we solve for. Just like the entropy, $\vartheta$ remains constant in regions of active convection, and has a positive gradient vs. gravity in radiative regions. 

As we show in Appendix \ref{sec:appendix_entropy}, the potential temperature can, for an arbitrary reference density $\rho_0$ and a constant adiabatic index of $\rm \gamma=1.4$ which we use, be computed as 
\begin{align}
\vartheta = T \, \left(\rho / \rho_0 \right)^{-2/5}.
\label{eq:pottemp}
\end{align}
It is therefore not only a quantity to assess dynamical instability, but also a useful one to compare envelope structures to one another, without the need to inspect both density and temperature.\\

\begin{figure}
   \centering
   \includegraphics[width=\hsize]{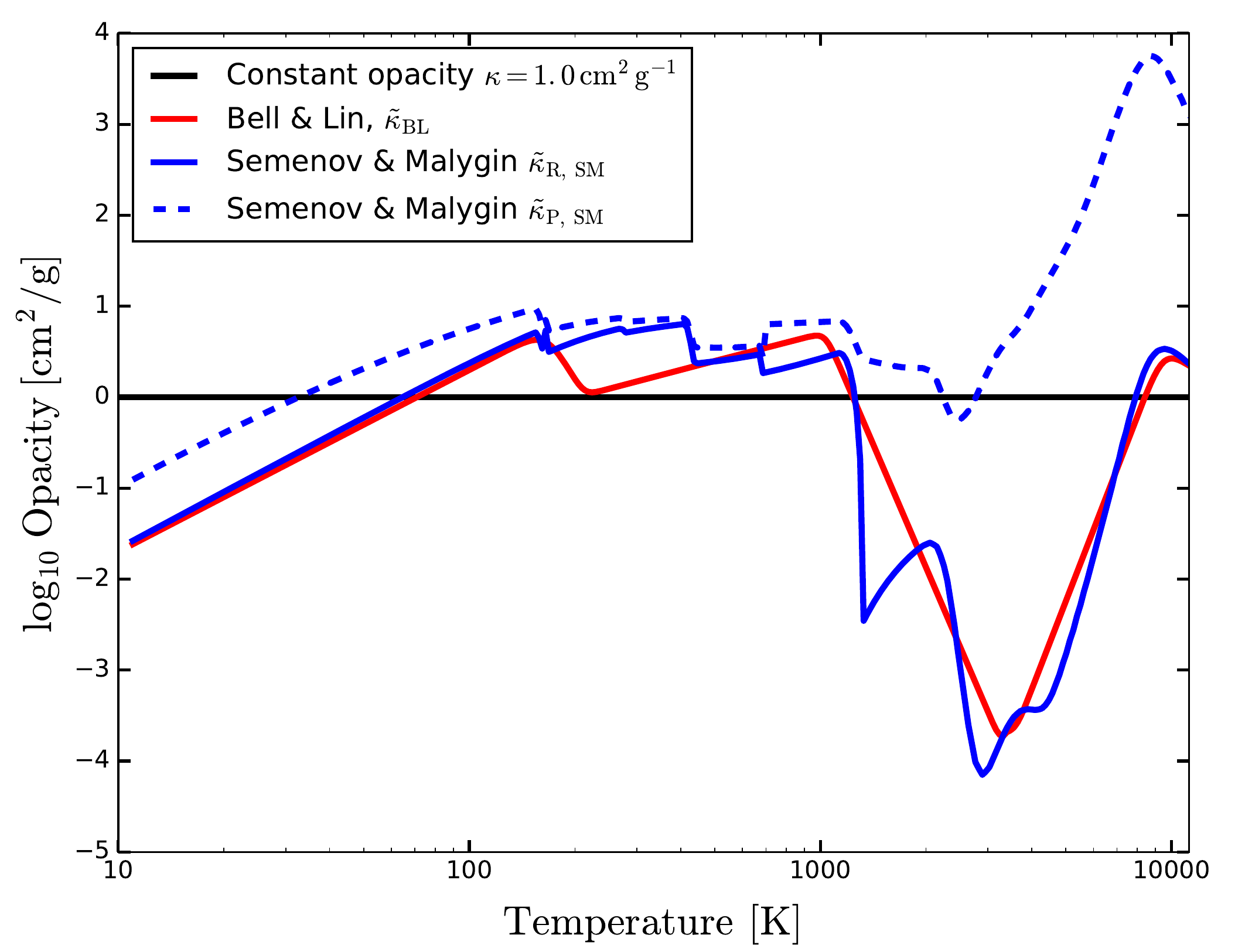}
      \caption{Opacity as a function of the temperature.
			Constant opacities take the three different values $1.0, 0.1, 0.01 \, \rm cm^2 g^{-1}$, where in the plot we indicate only $1.0 \, \rm cm^2 g^{-1}$. The second opacity set from \citet{belllin1994} increases with temperature, shows a transition at the water-iceline at $\approx 200\, \rm K$ and a steep decline of the opacity once dust sublimation become relevant at temperatures of $1000$ K. The third opacity set from \citet{malygin2014} shows a much richer structure in terms of opacity transitions, which is due mainly to the various chemical constituents in \citep{semenov2003} that those authors computed. The large difference between the Planck-opacity and the Rosseland-opacity will play a role mainly in optically thin regions. Midplane temperatures in our simulations at $5.2$ AU without planet range from $30$-$50$K.}
         \label{fig:opacitiesoverview}
\end{figure}

\subsection{Physical parameters - Viscosity and opacities}
\label{subsec:physparam}
In this section we outline our parameter choices for the viscosity, which determines the disc structure, and the opacities, which control the cooling of the planetary envelope and the disc structure. Another important heating source, the compressional work, does not depend on the choice of any free parameters and is therefore not discussed here. 

The physical viscosity $\nu$ is kept throughout our work at $\rm \nu = 10^{-5}$ in code units, c.u., where the planetary distance $r_{\rm p}$ and the Keplerian speed of the planet $v_{\rm K}$ are both unity.
This corresponds for a planet at 5.2 AU around a solar-mass star to a physical viscosity of $\rm\nu = 10^{15} \, \rm cm^2 s^{-1}$. With the standard relation between physical and $\alpha$-viscosity of $\rm \nu = \alpha \mathit{H}^2 \Omega$, our simulation values of $H/r \,=\, 0.021 \,\ldots\, 0.036$, correspond to $\alpha \,=\, 3.3 \,\ldots\, 1.1\cdot 10^{-2}$ at Jupiter's orbit. Thus our adopted nominal viscosity value is comparatively large. We plan to study lower values of the turbulent viscosity in a future publication. 
It is likely that the angular momentum in protoplanetary discs is transported by laminar stresses due to the large-scale magnetic field that penetrates the disc \citep[e.g.][]{turner2014}. Some works therefore adopt a two-alpha approach, with a disc $\alpha$ setting the angular momentum transport and a turbulent $\alpha$ of much lower magnitude setting the turbulent viscosity \citep[e.g.][]{ida2018}. However for reasons of simplicity we adopt here a constant $\nu$ model. 

In order to obtain correct post-shock entropies, we employ a standard method for capturing shocks \citep{neumann1950}, which uses an artificial viscosity to smear out any shock over $\sqrt{2}$ cells. In our work this is mainly relevant for the spiral arm shocks triggered in the disc by the planet. Because our envelopes feature quite substantial pressure support radially as well as vertically, no accretion shock is established for the investigated parameter values.

The opacity, however, is the main physical parameter of interest for this study. 
From equations \ref{eq:energyequation2} and \ref{eq:fluxesfld} it is evident that the Rosseland-mean opacity $\kappa_{\rm R}$ controls the diffusive properties of radiative energy. It is this opacity that determines whether the transition from one cell to the next is optically thick or thin. The Planck-mean opacity $\kappa_{\rm P}$ serves to couple internal and radiative energy. If $\kappa_{\rm P} > \kappa_{\rm R}$ in an optically thin simulation region, then the sourcing term in eq. \ref{eq:energyequation2} proportional to $\rho \kappa_{\rm P}$ will be larger than $\rho \kappa_{\rm R}$ and the gas will lose internal energy more readily. In an optically thick simulation regoin, $\kappa_{\rm P} \neq \kappa_{\rm R}$ will not matter, as the sourcing term is zero anyway in equilibrium.

The two different opacities $\kappa_{\rm P}$ and $\kappa_{\rm R}$ from equations \ref{eq:energyequation2} and \ref{eq:fluxesfld} are taken from one of the following sets: 

\begin{enumerate}
\item Constant opacities $\rm \kappa_{\rm P} = \kappa_{\rm R}  \in \{1.0, \, 0.1, \,0.01\} \, cm^2 g^{-1}$. Constant opacities are probably not realized in nature, but provide a basis for a relatively controlled environment to study the envelope structure and accretion process.\\
\item The Bell \& Lin opacities \citep{belllin1994}. Here the opacities are $\rm \kappa_{\rm P} = \kappa_{\rm R}  = \epsilon \, \tilde{\kappa}_{\rm BL}$ and $\rm \epsilon \in \{1.0, \, 0.1, \, 0.01\}$. A cut through the function $\tilde{\kappa}_{\rm BL} = \tilde{\kappa}_{\rm BL}(\rho, T)$ is plotted in red in Fig. \ref{fig:opacitiesoverview}.\\
\item The Malygin opacities from \cite{malygin2014}. Those include updated Planck gas opacities together with the dust opacities from \cite{semenov2003}. It is $\rm \kappa_{\rm P, \; SM} = \epsilon \, \tilde{\kappa}_{\rm P, \; SM} \neq \kappa_{\rm R, \; SM} = \epsilon \, \tilde{\kappa}_{\rm R,\; SM}$ and $\epsilon = 0.01$. A cut through the functions $\tilde{\kappa}_{\rm P,\; SM}(\rho,T), \; \tilde{\kappa}_{\rm R,\; SM}(\rho, T)$ is plotted in blue in Fig. \ref{fig:opacitiesoverview}. 
\end{enumerate}
\vspace{0.25cm} 

For the latter two sets, where the opacity has a non-constant form, one can think of the parameter $\epsilon$ as reduction factor relative to the ISM opacities, such that $\epsilon \in \{1.0, \, 0.1, \, 0.01\}$ correspond to $\{1.0, \, 0.1, \, 0.01\}$ ISM opacities and thus dust-to-gas ratios of $\{1.0, \, 0.1, \, 0.01\} \%$. The order of magnitude of the constant opacity set and the opacities from \citet{belllin1994} then coincide roughly at the water-sublimation iceline.
In set 3 one finds at higher temperatures a significant deviation of $\kappa_{\rm P}$ from earlier estimates (see Fig. \ref{fig:opacitiesoverview}). \cite{malygin2014} show that this is because $\kappa_{\rm P}$ is mostly sensitive to the spectral line wings of the underlying sampled linelist, and this was undersampled in earlier work. 
The main simulations in our work have been performed using the first two opacity sets, while the third set serves mainly to inform us about the simulation behaviour when $\kappa_{\rm R} \neq \kappa_{\rm P}$.

Although in the third opacity set the difference between $\kappa_{\rm P}$ and $\kappa_{\rm R}$ becomes very large above $T$$\sim$$2000$K, in our work its influence is negligible, as our limitations in resolution prohibit gravitational potentials deep enough for those temperatures. In general however one would expect there to be a visible effect when $\kappa_{\rm P} \neq \kappa_{\rm R}$:
The function of $\kappa_{\rm P}$ is to couple the evolution of $e_{\rm int}$ and $e_{\rm rad}$, while the fluxes that cool the envelope are a function of $\kappa_{\rm R}$. The same is true for the optical thin-thick transition, or the $\tau=1$-surface, defined by $- \int_{\infty}^z \rho \kappa_R dz=1$.
If it is $\kappa_{\rm P} = \kappa_{\rm R}$, then at that same height where $\tau=1$, the coupling between radiation and matter also stops being efficient. When it is however $\kappa_{\rm P} > \kappa_{\rm R}$, then matter keeps transferring internal energy into photons even above the $\tau=1$-surface. This allows for more efficient cooling and thus higher values of accretion rates are expected for set 3 compared to set 2.

The most prominent differences between the two non-constant sets can be seen in the potential temperature structure of the envelope (see figure \ref{fig:sm_entropa}). The multibump structure in the curve for the \cite{malygin2014} opacities is due to the multitude of chemical ingredients used to compute the opacities, each having their own sublimation temperature. The opacity jump then results in differing temperature gradients, which impact the potential temperature.

\section{Numerical procedures and parameters}
\label{sec:methods}

Here we outline the most important properties of the code used and the individual simulation steps. We employ three such steps \citep{lambrechts2017}, which separate physical effects according to timescales and ascertain that our final results are not influenced by, for example, the process of forming gaps or spiral arms.
We also present the simulation runs that we have performed, characterized by their parameter values. The parameters that we have varied are the opacity sets, the dimensionless gravitational smoothing length $\tilde{r}_{\rm s} = r_{\rm s}/r_{\rm H}$ and the number of grid-cells per Hill-radius $N_{\rm c}$. We also use $N_{\rm c}$ as a measure for the numerical resolution in the simulation domain.

The steps we employ are connected via restarting and interpolating onto a new grid. Step A consists of a 2D-radial-vertical run without planet to find the radiative equilibrium structure of the disc, which gets its structure from viscous heating alone. Step B restarts from the end of Step A in a full 3D-low-resolution grid of the same domain dimensions and plugs in the planet smoothly, in order to generate the planetary gap and spiral waves. Once the gap depth satisfies a convergence criterion, we cut a global ring out of the disc and use this to run a 3D-high-resolution grid on it to measure accretion rates, which is step C.

\begin{table*}
\caption{Summary of grid parameters for individual simulation steps}
\label{tab:griddata}
\centering
\begin{tabular}{c c c c}        
\hline\hline                 
Step & Simulation domain  & Number of cells & Grid type\tablefootmark{***}  \\
 & $r\,\times\,\phi\,\times\,\theta \in$ & $N_r \,\times \, N_{\phi} \, \times \, N_{\theta}$ &  \\ 
\hline
	A & $[0.4;2.5]\times[90^{\circ}; 83^{\circ}]$\tablefootmark{*} & $288 \times 32$ & $\rm U \times U$ \\
	B & $[0.4;2.5]\times[90^{\circ}; 83^{\circ}]\times[-\pi;+\pi]\;$ & $288 \times 32 \times 768$ & $\rm U \times U \times U$ \\
	C & $[0.7;1.3]\times[90^{\circ}; 83^{\circ}]\times[-\pi;+\pi]\;$ & $300 \times 72 \times 1486$\tablefootmark{**} & $\rm P \times P \times U/G$\tablefootmark{****} \\
\hline                                   
\end{tabular}
\tablefoot{\tablefoottext{*}{We give code units here, with planetary position and velocity set to unity, to simplify numbers. In this work distances are given in $r_{\rm H}$ or $\rm AU$ when convenient.}
\tablefoottext{**}{Example values for $N_{\rm c}=25$. Higher resolution in $r_{\rm H}$ requires more cells overall}
\tablefoottext{***}{Grid types: U=Uniform, P=Parabola, G=Gaussian}
\tablefoottext{****}{The azimuthal direction uses U with the fargo-algorithm and G without fargo at very high resolutions, depending on the runtime. We made sure both choices give consistent results.}
}
\end{table*}

\begin{figure}
   \centering
   \includegraphics[width=\hsize]{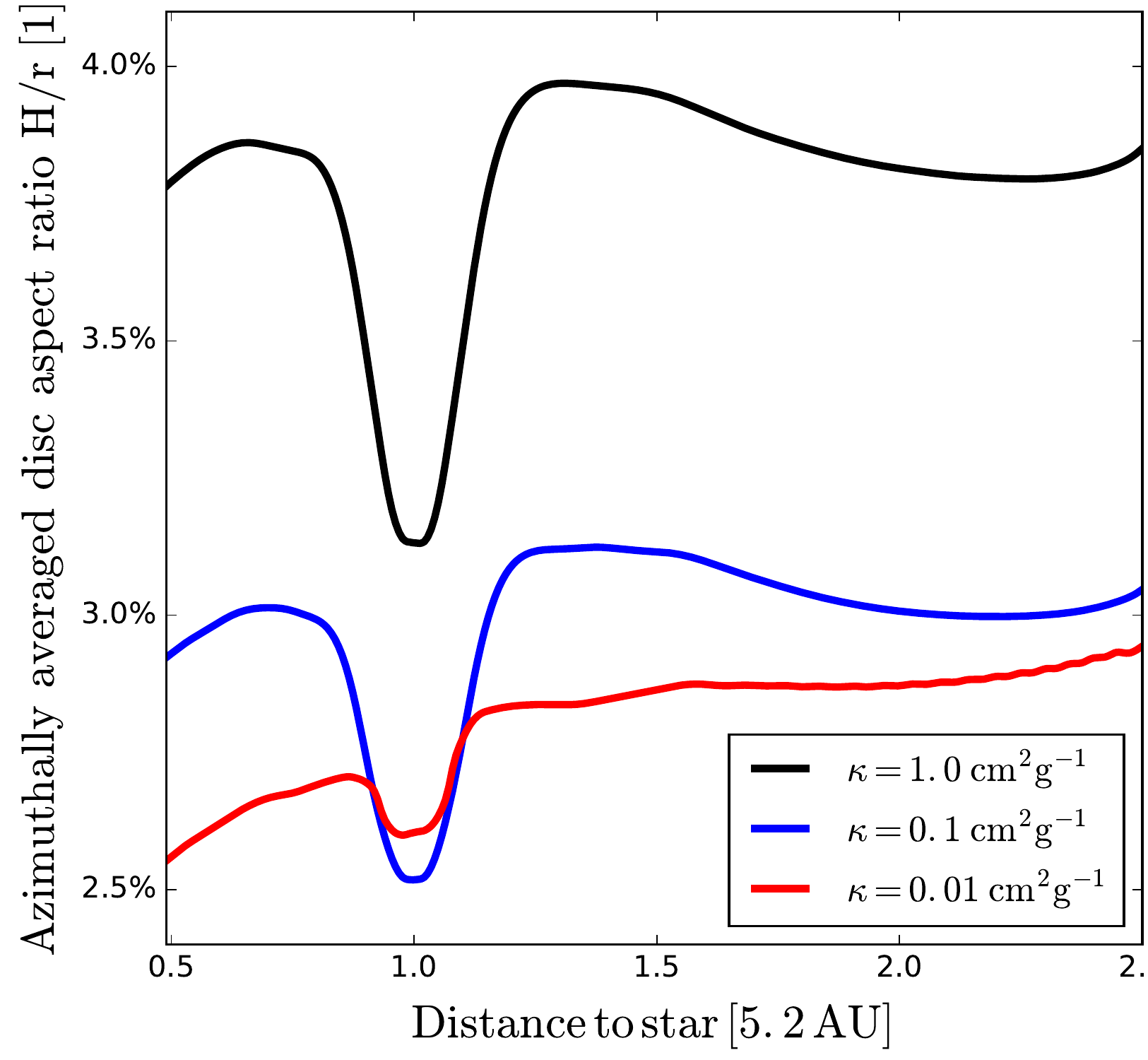}
      \caption{Profiles of the average aspect ratio in the disc midplane for runs with constant opacity after 400 orbits in step B. The planet is kept fixed at $5.2$AU. The three different opacities that we use for later accretion rate measurments are used as running parameter. For $\kappa=0.01 \rm \,cm^2 \,g^{-1}$ the gap becomes optically thin, and is then directly heated by the gap walls, which explains the higher gap temperature compared to $\kappa=0.1 \rm \,cm^2 \,g^{-1}$.
			 }
   \label{fig:gapaspectratio}
\end{figure}

\subsection{Numerical parameters - grid, resolution and smoothing length} \label{sec:numericalparameters}

We employ a grid in spherical coordinates centered on the star, $ [ r,\phi,\theta] $ where r is the radial distance from the star, $\phi$ the colatitudinal angle and $\theta$ the azimuthal angle, varying along the disc orbit.
This results in quasi-cartesian coordinates surrounding any single point of interest around the planetary orbit, when sufficiently high resolution around this point is used. For flows orbiting the star this is a natural coordinate frame, since it minimizes numerical diffusion. However flows that orbit a planet in the disc, and thus in a cartesian region, are then subject to strong numerical diffusion.
To remedy this problem, \citet{lambrechts2017} introduced non-uniform grid spacing between cells along each spherical coordinate in order to resolve the Hill-radii of low-mass protoplanets, similar to the grid with quadratic grid-spacing used in \citet{fung2015}. In this work, we introduce a modified version of this grid in the azimuthal direction, one that follows a gaussian grid-spacing, so that we achieve high resolution in the planetary Hill-sphere and uniform grid-spacing at the simulation boundaries.

The application of the quadratic non-uniform grid is implemented in radius and co-latitude with a functional dependence
\begin{align}
dr \propto (r-r_p)^2 
\end{align}
and
\begin{align}
d\phi \propto (\phi - \frac{\pi}{2})^2.
\end{align}
In the azimuthal direction we use two different strategies: Equidistant spacing for moderate resolutions of up to 65 grid-cells per Hill-radius to allow for the application of the FARGO-algorithm for the orbital advection around the star.
For a number of runs that correspond to our highest achievable resolution, 100 and 150 grid-cells per Hill-radius, we apply a non-uniform grid in azimuth. For the latter cases, we applied azimuthal grids that vary gaussian with quasi-equidistant spacing $d\theta_{qe}$ far away from the planet. The gaussian grid has a spacing 
\begin{align}
d\theta \propto ( d\theta_p - d\theta_{qe}) \left\{ \exp \left[- (\theta-\theta_p)^2/W \right] - 1 \right\} + d\theta_p ,
\end{align}
with $\theta_p$ being the planet's azimuthal position, $d\theta_p$ the azimuthal resolution at the planet's position, and $W$ the gaussian width that is found in accordance with requiring a certain number of total cells.

The simulation domain is global in azimuth for all runs, so that $\theta \in [-\pi;+\pi]$. The extent of the simulation box in co-latitude is chosen such that it represents $\approx 3 r_{\rm H}$ in physical space which results in $ \phi \rm \in [90^\circ;83^\circ]$. We do not extend the simulation box below the midplane, but use a half-disc approach instead that mirrors the hydrodynamic quantities in the cell above the midplane into the first ghost-cell below it. Thus, we assume symmetric physics in all variables below and above the midplane in order to speed our code up.

Radially, we choose a ring of $r \rm  \in [0.4;2.5] \cdot 5.2\, AU$ for the low-resolution Steps A and B and $r \rm  \in [0.7;1.3] \cdot 5.2 \, AU$ for the high-resolution step C, and the planet is kept fixed at $r_{\rm p} = 5.2\, \rm AU$.
Numerical resolution -- the number of grid cells between the box limits and their individual spacing -- varies between the simulation steps and is thus documented individually below.

The temperature near the planet in Step C (discussed in Sect. \ref{sec:stepc} below) where the accretion takes place, is mainly regulated by the opacity and the gravitational smoothing length $\rm r_s $\citep{klahrkley2006}. In our work the latter takes values of $ r_{\rm s} \, \rm \in [0.5;0.05] \,$ $r_{\rm H}$ and we define a convenient quantity, the Hill-sphere normalized smoothing length $\tilde{r}_{\rm s} = r_{\rm s}/ r_{\rm H}$. 

\begin{figure}
   \centering
   \includegraphics[width=\hsize]{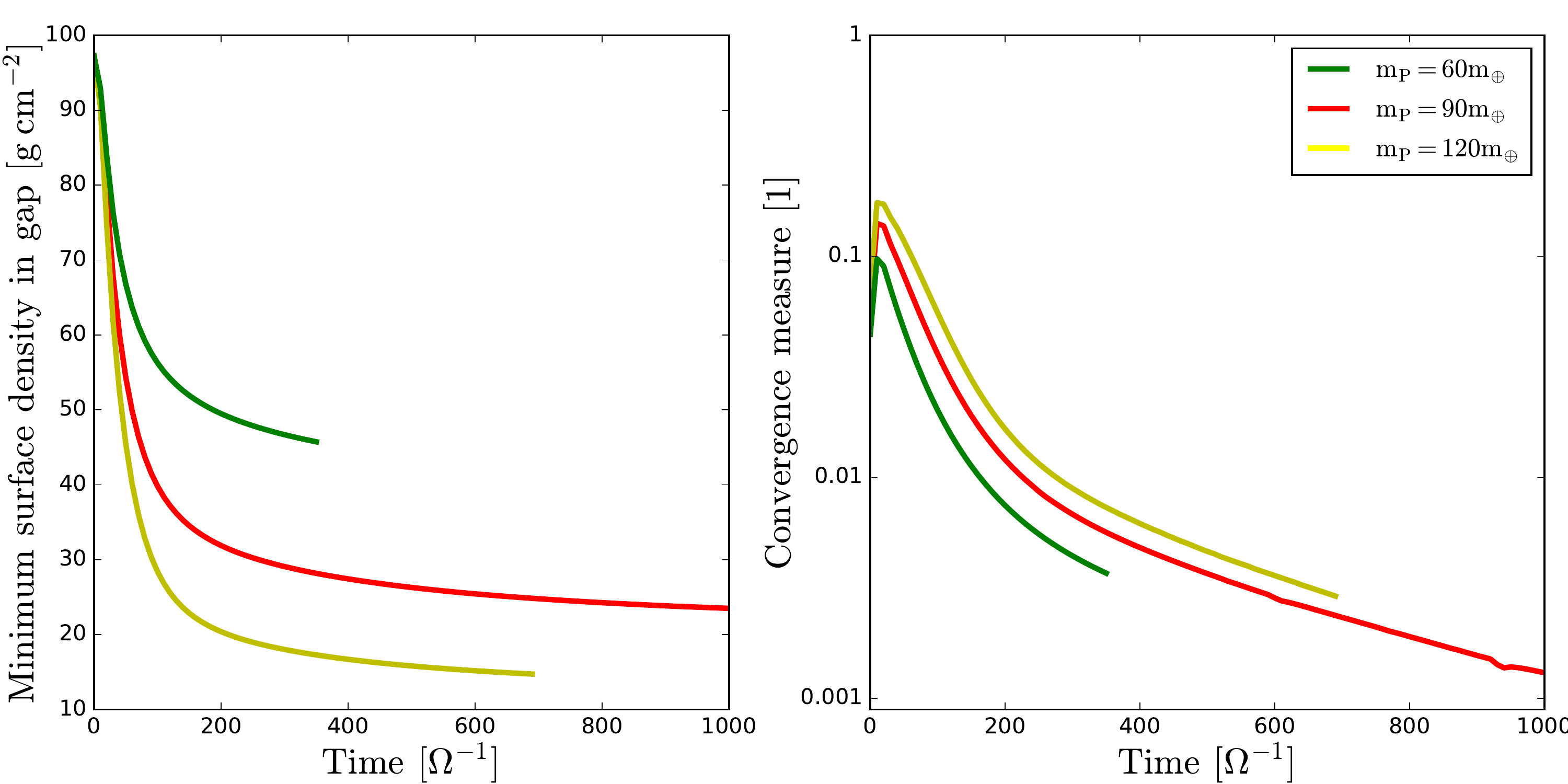}
      \caption{Gap depth vs. time and convergence ratio: For three different opacities with all other numerical and physical parameters being constant, we compare the resulting gap depth as a function of orbits and the convergence measure $(\Sigma(t)-\Sigma(t+dt))/\Sigma(t)$. We see that after 400 orbits the gap depth has essentially reached its final value (up to 10-20\%). The convergence rates for the gap depths are independent of opacity after a few orbits and thus their relative depth will not change. }
         \label{fig:gapdepthconvergence}
\end{figure}

\begin{table}
\label{tab:simdata}
\centering
\caption{Simulation runs as characterized by their parameter values in step C.}
\begin{tabular}{c c c c c}        
\hline\hline                 
Opacities & $\kappa$ or $\epsilon$ & $\tilde{r}_{\rm s}$ & $N_{\rm c}$  \\    
\hline                        
  1 & 0.01 & 0.5 & 25 \\      
  1 & 0.01 & 0.2 & 25,35,50,65,100  \\
\hline
	1 & 1.0,0.1,0.01 & 0.2 & 100  \\
	1 & 1.0,0.1,0.01 & 0.1 & 100  \\
  1 & 1.0,0.1,0.01 & 0.05 & 150 \\
\hline
	2 & 1.0,0.1,0.01 & 0.2 & 100 \\
	2 & 1.0,0.1,0.01 & 0.1 & 100 \\
	3 & 0.01 & 0.2, 0.1 & 100  \\
\hline                                   
\label{tab:table2}
	\end{tabular}
	\\[5pt]
	\caption*{The first set of simulations served mainly to investigate the influence of the numerical resolution. The second set is the main data set for our results. There, for each smoothing length, three different values of constant opacity were investigated. In the third set we also investigated runs with non-constant opacities, as introduced earlier.}
\end{table}

\subsection{Step A - heating/cooling Equilibrium for the disc alone in 2D}
\label{sec:stepa}

The first step places an initial mass distribution of $\rm \Sigma = 120 \,g \,cm^{-2} \; (\mathit{r}\,/\,5.2\;AU)^{-0.5}$ with $\rm \Sigma_0 = 120 \, g \, cm^{-2}$ being the gas mass surface density at $\rm 5.2 \, AU$ and gaussian vertical structure in the simulation domain. According to classical literature \citep[eq. 5.9]{accretionpower}, i.e. $v_r = 3 \nu /(\Sigma r^{1/2}) \partial_r (\Sigma r^{1/2}) $  this is the density profile required for zero accretion, which has also been used previously in \cite{bitsch2013a}.
The resolution is $N_r \times N_{\phi} = 288 \times 32$ with equidistant spacing in both dimensions. No planet is present in this step. 
Thereafter, the density and temperature structure of the disc is self-consistently calculated via solving eqs. \ref{eq:massconservation} - \ref{eq:energyequation2} and evolved until steady-state is reached. This steady-state solution is then taken as initial condition for step B.

\begin{figure*}
	
	\hspace*{-0.2cm}
  \begin{subfigure}{0.5225\textwidth} 
	\centering
	\includegraphics[width=\textwidth]{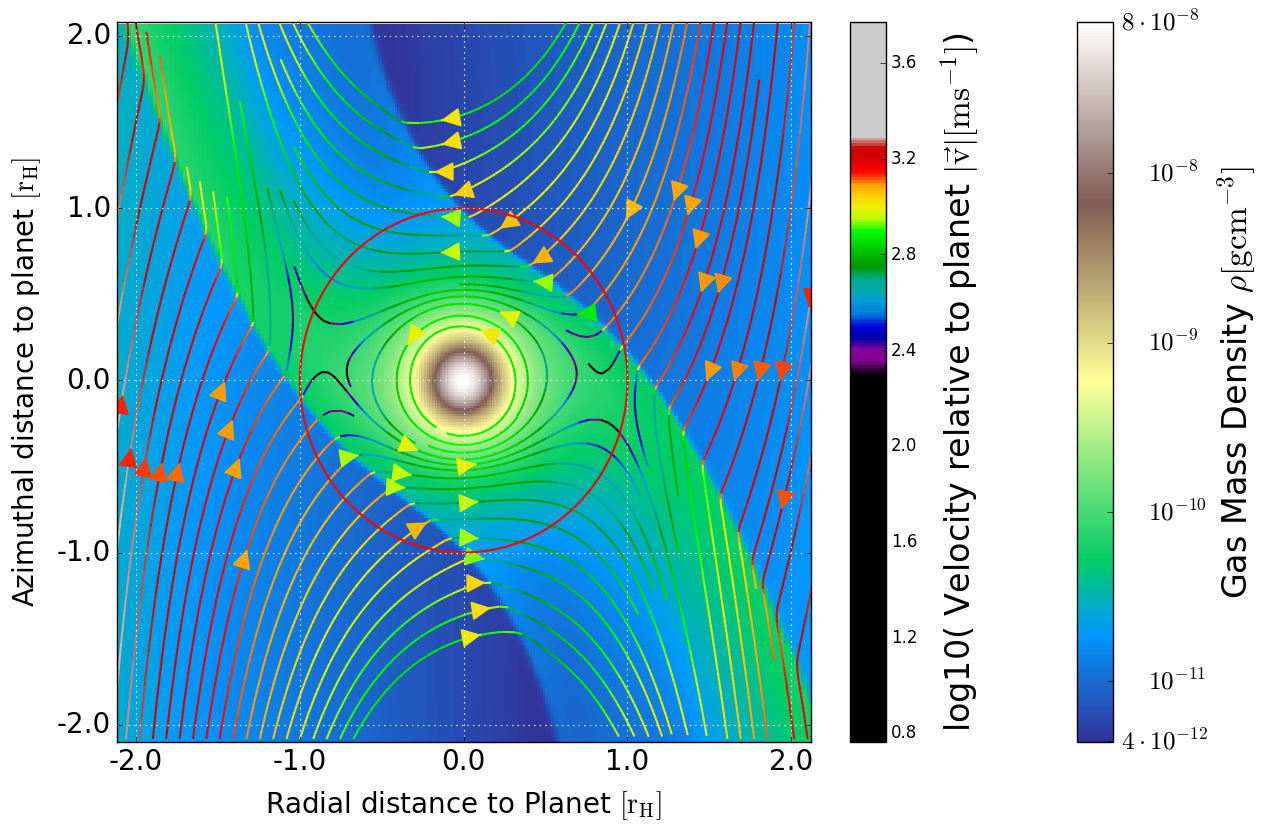}
      \caption{Gas mass density in the midplane and streaklines overplotted with their respective momentary velocities. Spiral arm shocks launched near the planet's Hill-sphere are clearly visible. The horseshoe flows interact with the shocks and are partially deflected. Colours indicate the streakline velocities and highlight flow deceleration at the shocks. Judging from streak-lines only it would be impossible to say whether there is ongoing gas accretion, because most of the horseshoe mass merely passes through the planetary envelope, while only a minor fraction is accreted. The planetary gap is approximately $\sim$$2 r_{\rm H}$ wide, as wide as the plot and attains its unperturbed disc value at $\sim$$3 r_{\rm H}$. For comparison, the simulation domain in step C extends to $ \pm 6 r_{\rm H}$ radially.   }
         \label{fig:overview_1}
   \end{subfigure}%
	\hspace*{+0.5cm}	
	  \begin{subfigure}{0.43\textwidth} 
	\centering
	\includegraphics[width=\textwidth]{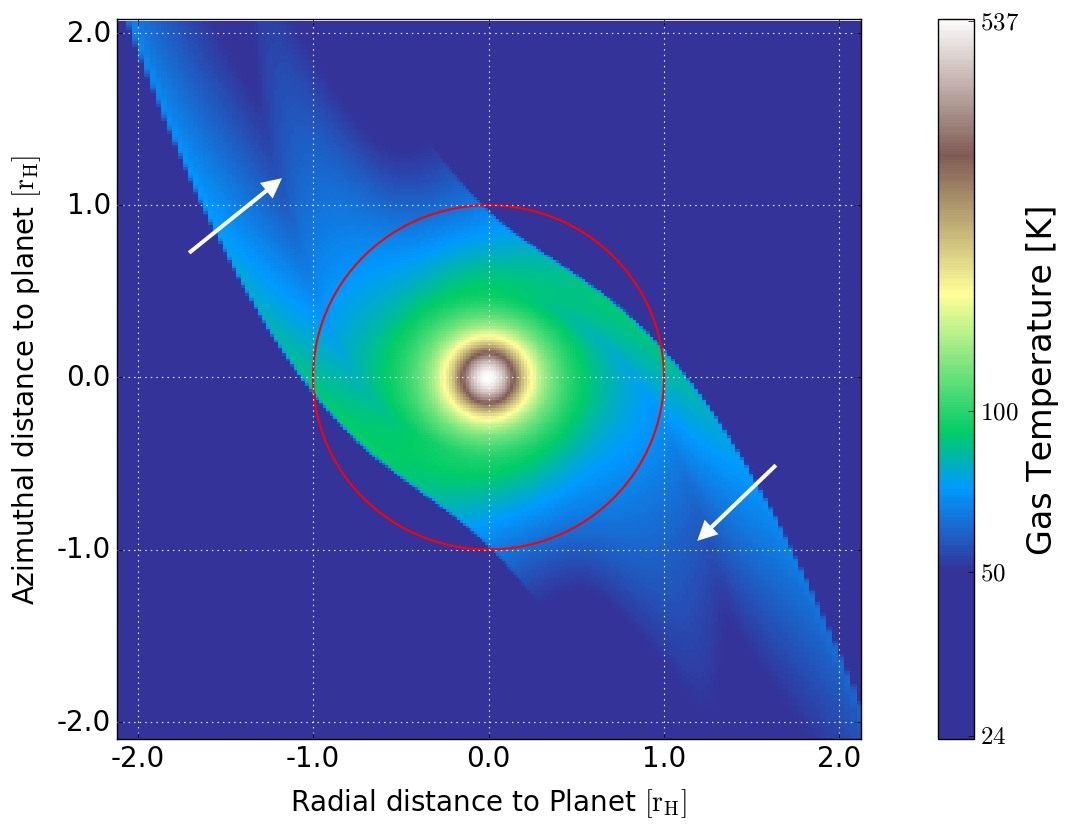}
      \caption{Temperature structure in the midplane. The cold gas in the gap and the surrounding disc is clearly visible as dark background colour, while the planetary envelope and beyond is heated by compressional heat from accretion and the spiral shocks. A small fraction of internal energy is advected along the horseshoe orbits, visible in separate moderately warm ($\sim$$80\rm K$) arm structures marked with arrows.\\ \\ \\}
         \label{fig:overview_2}
   \end{subfigure} 

  \hspace*{+0.2cm}
  \begin{subfigure}{0.425\textwidth} 
   \centering
   \includegraphics[width=\textwidth]{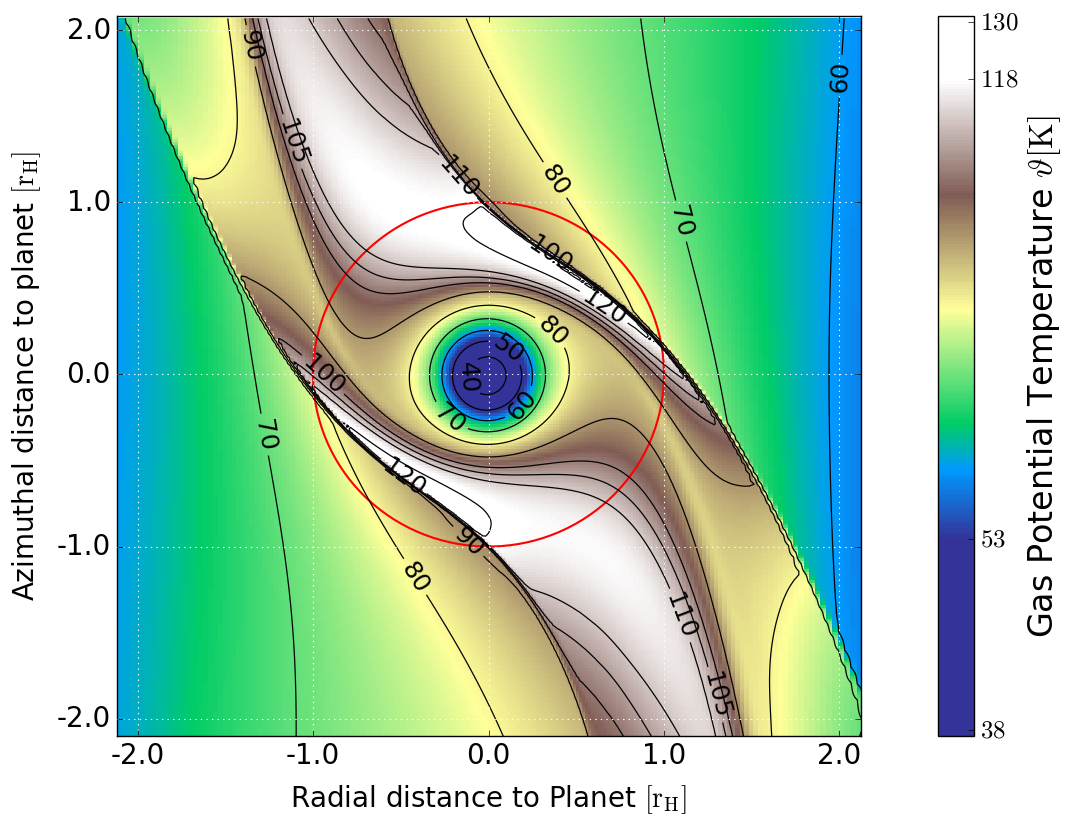}
      \caption{Potential temperature $\vartheta$ overlaid with iso-$\vartheta$ contours. Entropy generation at the spiral arm shocks is strongest where horseshoe-orbits hit the envelope radially. The central region cools most efficiently into the vertical direction, and develops thus a drop in $\vartheta$, indicating a radiative region. The innermost few pixels are subject to a flattening of the $\vartheta$-profile, as here the smoothing length provides a region essentially free of gravity. \\}
         \label{fig:resolutioncontrols_3}
	\end{subfigure}%
	\hspace*{+2.00cm}
	\vspace*{-0.2cm}
	\begin{subfigure}{0.415\textwidth} 
   \centering
   \includegraphics[width=\textwidth]{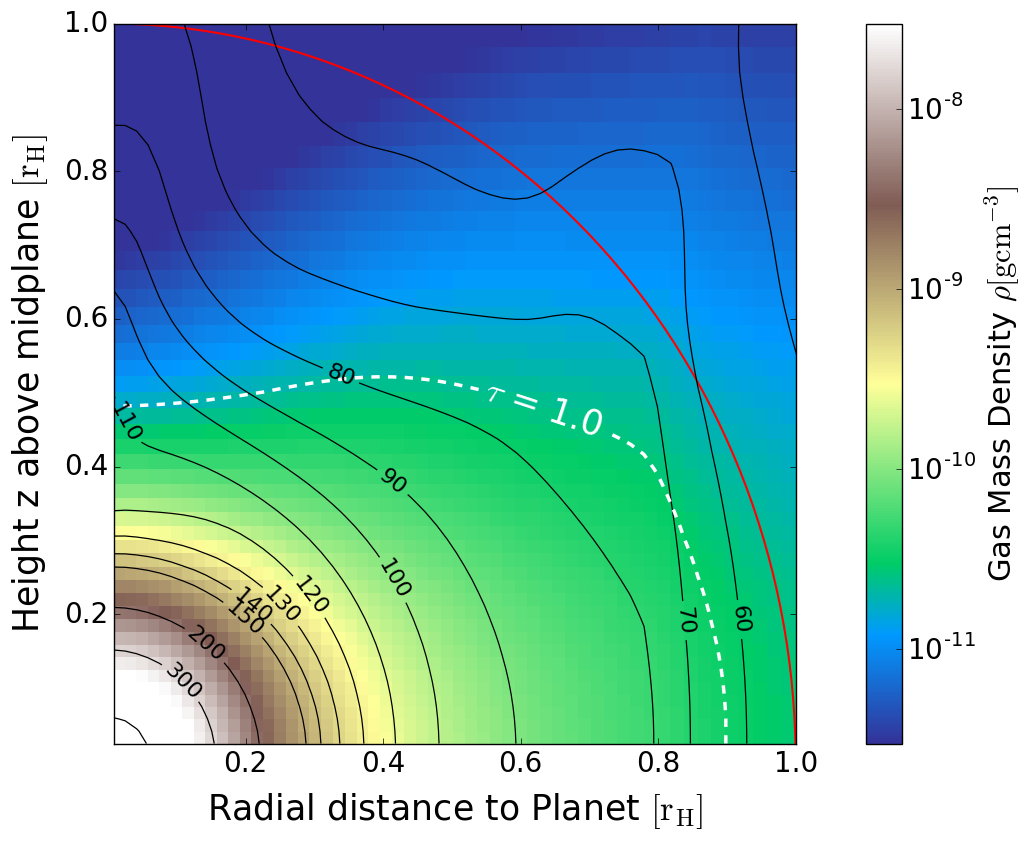}
      \caption{Cylindric average of density as background colour and temperature in contours, as well as the $\tau=1$ surface, denoting the optically thin-thick transition. As this run has a constant opacity, the $\tau=1$ surface follows essentially an iso-density surface. This surface is bent towards the planet at small $r$, due to heavy cooling into the disc atmosphere and the consequencial contraction. This plot only extends out to $1 \, r_{\rm H}$ as otherwise averaging effects smear out the physical properties outside the Hill sphere.}
         \label{fig:resolutioncontrols_4}
	\end{subfigure} 

\vspace*{+0.20cm}
\caption{Overview and snapshot of typical midplane variables during gas accretion in step C at the time of 10 orbits for our nominal simulation run with constant opacitiy of $\rm \kappa = 0.01 \, cm^2 \, g^{-1}$, $\tilde{r}_s = 0.2$ and $N_{\rm c} = 100$. The planet's Hill surface is approximated as a sphere, marked with the red circle.}
	\label{fig:overview}
\end{figure*}

\subsection{Step B - Gap opening and spiral arms in low resolution 3D}
\label{sec:stepb}

With the initial conditions being the end-state from step A, we extend the disc into 3D with the azimuth $\theta \in [-\pi;+\pi]$, $N_{\theta} = 768$, equidistant angular spacing and plant a seed planet into it. The planetary potential is centered at the corners of the 8 grid cells around the point ($r$$ \rm, \phi, \theta) = (5.2\, AU, 90^\circ, 0^\circ)$ and is ramped up slowly over 10 orbits from $\rm 0$ to its final mass of $90 \, m_{\oplus}$ with $\tilde{r}_s = 0.5$ in order to avoid unphysical shocks in the disc and the planetary envelope. Subsequently we let the planet continue for $\rm 400$ orbits until a steady-state gap has been opened. For our relatively large viscosity the $\rm 90 \, m_{\oplus}$ planet opens a partial gap of about 30\% of the unperturbed gap depth (see fig. \ref{fig:gapdepthconvergence}). 

We compare our gap to the semi-analytical predictions from \citet{crida2006},  where we have taken the value of the gap depth after half a viscous time
 $t=t_{\nu}/2$ and $t_{\nu}=r_p^2 /\, \nu = 10^{4.2} \Omega^{-1}$.
Our viscosity corresponds to a Reynolds number of $Re \equiv t_{\nu} \Omega = 10^{4.2}$ 
and with this we can compute the expected gap depth from the Crida-formula, resulting in a predicted 40\% gap. It is interesting that our gaps are slightly deeper for the same $h/r$, possibly hinting at some additional physical processes contributing to the gap formation process.

In fig. \ref{fig:gapaspectratio} we show the aspect-ratio of the disc as function of the distance to the star for the constant opacity cases. As the locally isothermal midplance scale height $H$ depends on the local temperature $T$ through $ H = \rm{k_B}$$T / \mu g$, it is evident that one finds colder temperatures with decreasing opacity in the optically thick disc mid-plane. For $\rm \kappa=0.01 cm^2 g^{-1}$ the planetary gap becomes optically thin and is then directly heated up by the gap walls. Thus, the material inside the gap becomes hotter for $\rm \kappa=0.01 cm^2 g^{-1}$ than for $\rm \kappa=0.1 cm^2 g^{-1}$. Surface densities follow the temperature rather than the opacities, which indicates the importance of radiative transfer for predicting correct gap depths.

We use the end state of this simulation step again as initial conditions for step C. Therefore it is important to check whether the planetary gaps are converged before we do so. We confirm this in fig. \ref{fig:gapdepthconvergence}.

\subsection{Step C - High resolution step in 3D}
\label{sec:stepc}

The previous steps ascertain that the disc around the planet is in radiative equilibrium once we apply the non-uniform grid (see sec. \ref{sec:numericalparameters}). To limit computational cost, we now shrink the simulation domain in radius to $ r \rm \in [0.7;1.3]  \cdot 5.2 \, AU$. 
We keep the information in the physical variables from step B and apply boundary conditions where the boundaries are relaxed to the values obtained in Step B. 
This approach is valid if the new boundaries are far enough from the planet to not be affected by the physics inside the Hill sphere. The planet sits at $r_{\rm p} = 5.2\, AU$ and has a Hill radius of $r_{\rm H} = 0.25 \,\rm AU$, thus the simulation domain radially extends $\sim$$6 \, r_{\rm H}$ radially to both sides, and about $3 \, r_{\rm H}$ in co-latitude. \\

We then interpolate the data from step B onto the step C non-uniform grid, deepen the potential depth during a time of $t_{\rm ramp}=2\,\Omega^{-1}$ from $r_{\rm s}=0.5 r_{\rm H}$ to smaller values of $\rm r_{\rm s} = 0.2, \,0.1, \, 0.05 \, r_{\rm H}$.

\subsection{Summary of simulation parameters}

After explaining the simulation steps in detail, we briefly list the most important numerical parameters in table \ref{tab:griddata}.
When step C is reached, accretion rates can be measured; we explain this in more detail in the next subsection.
Various simulation runs have been performed in order to ascertain numerical convergence and investigate parameter dependencies of those accretion rates. We list them in table \ref{tab:table2}.

\subsection{Measurement of mass accretion rates}

The central aim of this study is the measurement of accretion rates in step C and to obtain an understanding of the influence of resolution, gravitational smoothing length and opacity on the results.
The influence of those numerical parameters will be presented in the results section, and we now turn to describing the two methods with which the mass accretion is calculated. 

1. Direct measurement of mass per snapshot in time. Two successive snapshots can be subtracted to calculate a mass accretion rate. Thus it is
\begin{align}
\dot{m} = \frac{m_{\rm H}^{\rm n+1}-m_{\rm H}^{\rm n} }{t^{n+1}-t^{n}} .
\label{eq:mass1}
\end{align}
where $m_{\rm H}^n$ is the mass inside the Hill sphere at time-step $n$.

2. Compute a net mass accretion rate via average mass fluxes in/out of the Hill-sphere 
from only one snapshot. With this, we get the accretion rate
\begin{align}
\dot{m} = - \int_{\partial \Omega} d \vec n \cdot \vec u \rho .
\label{eq:mass2}
\end{align}
where $d \vec n$ is the normal vector on the Hill-surface $\partial \Omega$. We find that both these methods agree reasonably well within 10\%, thus to simplify the presentation of our results, we only show results given by \ref{eq:mass1}. 
We will now proceed to introduce general results from the measurements of mass accretion rates, and their dependencies on numerical parameters.

Our disc accretion rate on the other hand is set to zero. This is implemented self-consistently by chosing a surface density profile of $\Sigma\sim r^{-0.5}$ in Step A that corresponds to an equilibrium disc.

\begin{figure}[]
   \centering
   \includegraphics[width=0.50\textwidth]{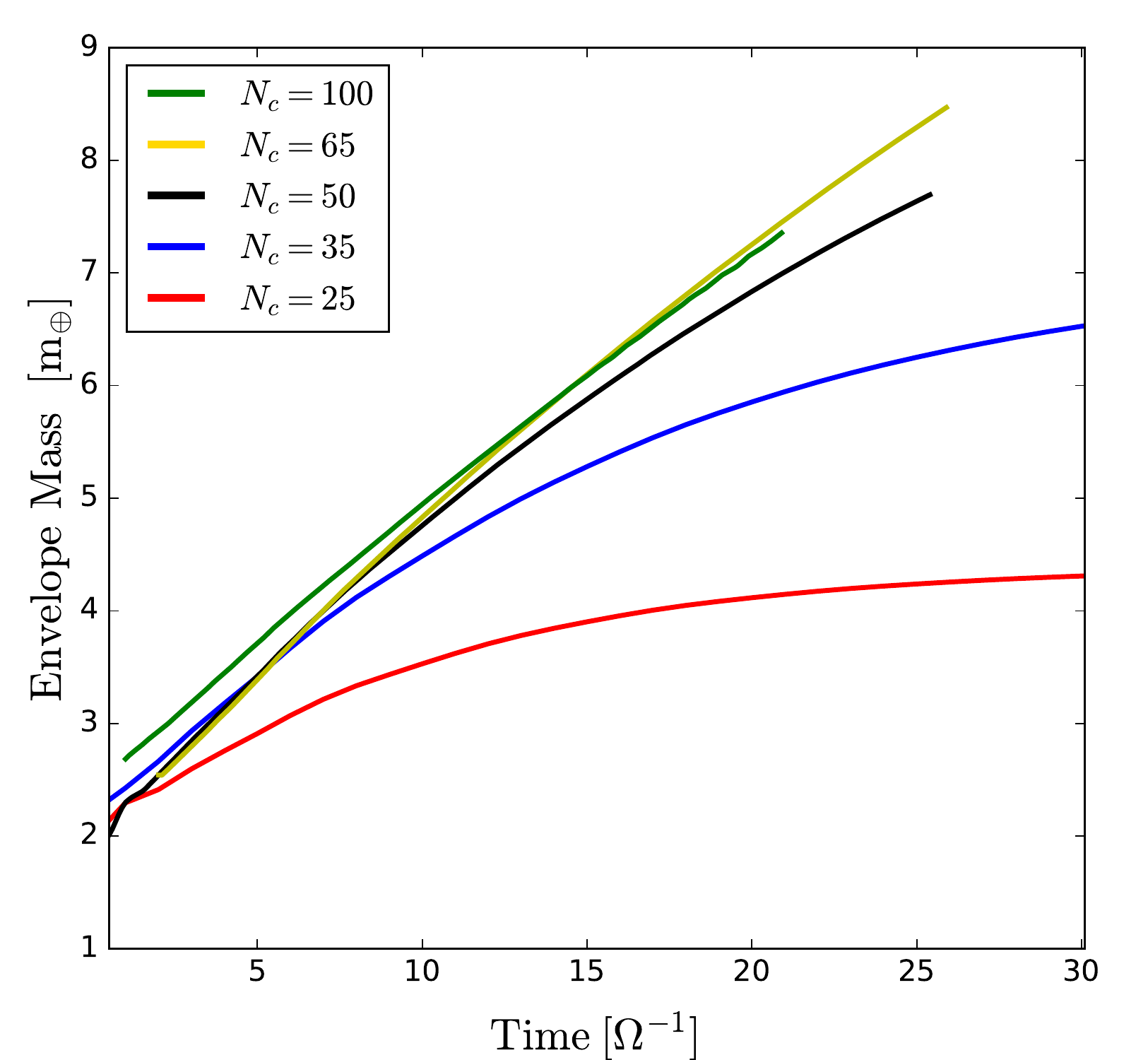}
      \caption{Total envelope mass within the Hill radius as function of time, just after the potential ramp-up time of 2 orbits, plotted against the numerical resolution of the envelope. We show the runs with cells per Hill-radius $ N_{\rm c} \in \left\{25,35,50,65,100 \right\}$ and smoothing length $\tilde{r}_s = 0.2$ for a constant opacity of $\kappa = 0.01 \rm cm^2 \, g^{-1}$. Under-resolving the Hill-radius leads to a gradual decrease in the measured accretion rate, which results in that the envelope as a whole achieves a dynamic steady state.
Increasing the resolution step-by-step we find numerical convergence of accretion rates and a less pronounced drop in accretion rates over time. We note that our highest resolution ($N_{\rm c}=100$) drops slightly in accretion rate compared to the case with $N_{\rm c}=65$.       	 }
\label{fig:envelopemass}
\end{figure}

\section{Dependence of the mass accretion rate on the numerical resolution}
\label{sec:numericalresolution}

This section gives an overview of the accretion rates that we measure in step C, without going into much physics detail at first; we are interested in a clear picture how numerical parameters influence our results first, before the dependence of the accretion physics on the opacity and interpretations of the flow structure inside the Hill sphere are presented in the following two sections. In the overview-plot fig. \ref{fig:overview}, we show the structure of density $\rho$, streamlines and temperature $T$ in the midplane. 
Typical features for gas flows around gas giants are visible: The gas gap along the azimuth, spiral arm shocks and horseshoe-flows feeding gas into the planetary envelope. Two separatrices define an inner and outer envelope region. One might expect there to be a simple relationship of the outer region feeding the inner one, but the 3D-structure is more complicated than that, with some flows coming from the vertical direction.
The temperature distribution is mostly determined by the compressional heating from shocks and accretion balanced by radiative cooling and weak heat advection effects.

To help interpret the mass accretion rates we measure the potential temperature $\vartheta$, as defined in equation \ref{eq:pottemp}. The potential temperature shows a marked drop in the central regions of the Hill sphere. This is a result of entropy reduction by radiative cooling and is a clear evidence that the gas envelope is slowly accreted in our simulations.

\begin{figure}
   \centering
   \includegraphics[width=0.5\textwidth]{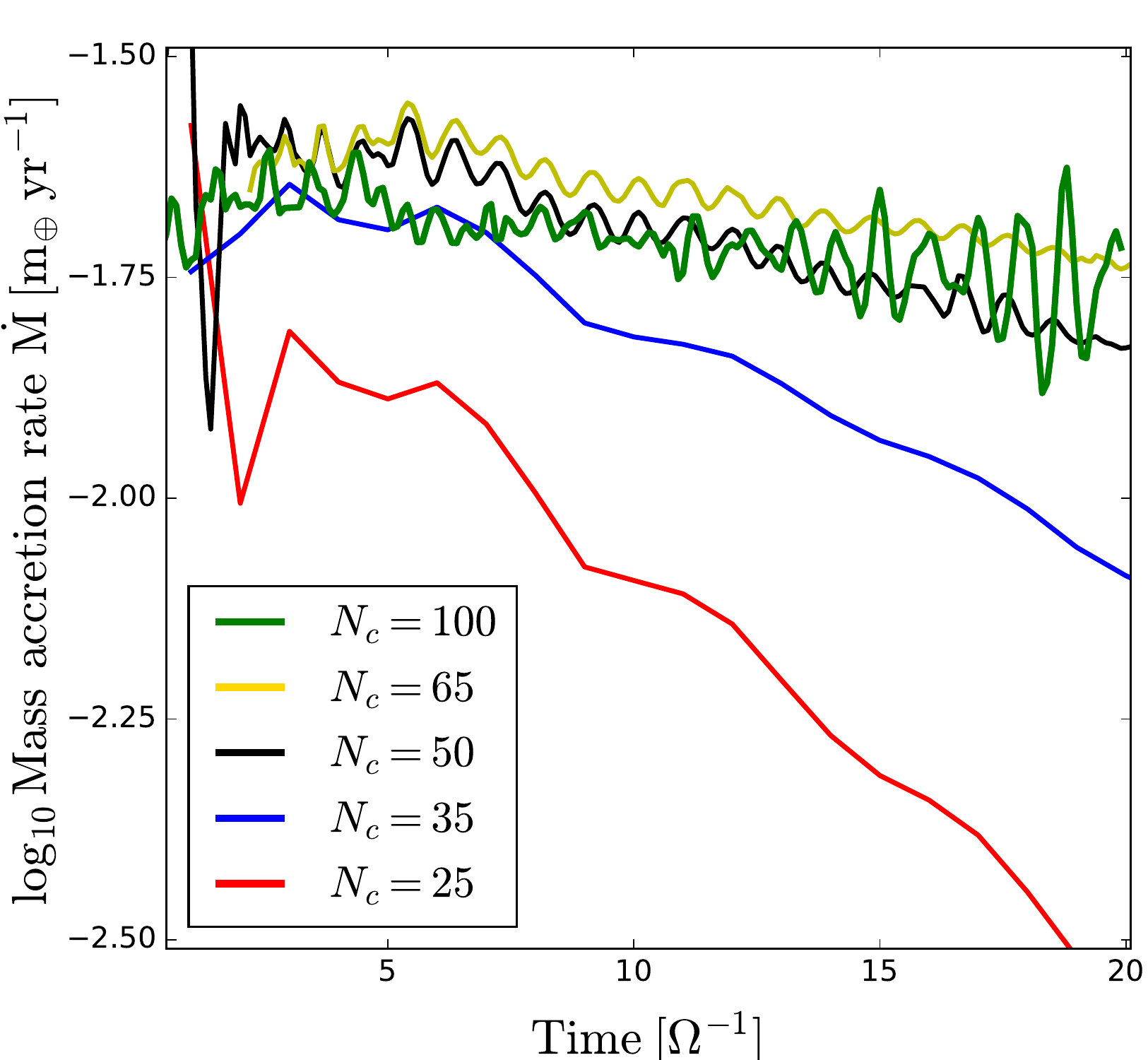}
      \caption{Mass accretion rate as a function of time.
							We present results of runs with cells per Hill-radius $\rm N_c \in \left\{25,35,50,65,100\right\}$ and smoothing length $\tilde{r}_s = 0.2$ for an opacity of $\kappa = 0.01 \rm cm^2 \, g^{-1}$. Those values are two orders of magnitude higher than the necessary accretion rate to form Jupiter within three million years, approximately $\dot{M}=10^{-4} \, m_{\oplus} \, yr^{-1}$. }
         \label{fig:accretionrates}
\end{figure}

When measuring the increase in mass in the planetary envelopes we first noted a strong dependence of the accreted mass on the resolution. This dependency led us to perform a convergence study first, before we discuss the physical state of the envelope further. Studying this it became clear that underresolving the smoothing length, not the Hill sphere, is the responsible parameter for this strong dependence. At low resolution of the smoothing length one finds only weak entropy gradients in the envelope. The envelope then finds a steady-state and constant mass quickly. In this steady-state, the envelope still features a strong flow of horseshoe-gas through the itself, but none of this gas actually gets accreted. 

Only when increasing the resolution, the entropy drops significantly and a part of the mass from the horseshoe gas starts being accreted. Finally, at sufficiently high resolution we find that the mass in the envelope and the accretion rates converge, as one would expect from a physical result. 

Interestingly, we have found that an underresolved simulation in steady-state can be restarted at higher resolution, whereafter that simulation will be accreting again. 
Thus we conclude that 1.) the steady states found must be a numerical artifact 2.) while the decrease in entropy can be used to describe the accretion process, the entropy loss is not guaranteed, as it is rather a change in heating and cooling physics at high resolution that allows the entropy to drop further and let accretion continue.

In fig. \ref{fig:envelopemass} we show that the mass in the envelope compared to the core mass ($90 \, m_{\oplus}$) is small at all times. The initial value of the envelope mass is a result of the midplane density in the disc into which we relax the planetary core mass. Growth continues only for a short time of a few tens of orbits, as this is limited by our computation time. As $m_{\rm env} \ll m_{\rm p}$ our assumption to neglect self-gravity in our model is validated for all runtimes. The neglect of self-gravity in our model is thus self-consistent inside this model, but a realistic core of $90 \, m_{\oplus}$ would probably contribute a significant portion of this mass to the inner envelope, which we do not resolve. We discuss this further in Sect. \ref{sec:envelope}.

A look at the accretion rates in fig. \ref{fig:accretionrates} allows a more detailed study of the numerical convergence behaviour, as well as a comparison with necessary accretion rates to build the solar system gas giants. With increasing resolution, accretion rates quickly converge during the first $\approx$$5$ orbits, but then rapidly decline if the smoothing length is underresolved. However, even for our best-resolved simulations there is a slight decrease in mass accretion rate, the non-convergence saturating at a level that converges. We discuss this decrease in more detail in Sect. \ref{sec:envelope}. If taken at face value, the accretion rates that we see in fig. \ref{fig:accretionrates} would be enough to form a Jupiter-mass planet in $1\%$ of the nominal disc lifetime of 3 Myr. This is further discussed in Sect. \ref{sec:discussion}.

\begin{figure}
\centering
  \includegraphics[width=0.46\textwidth]{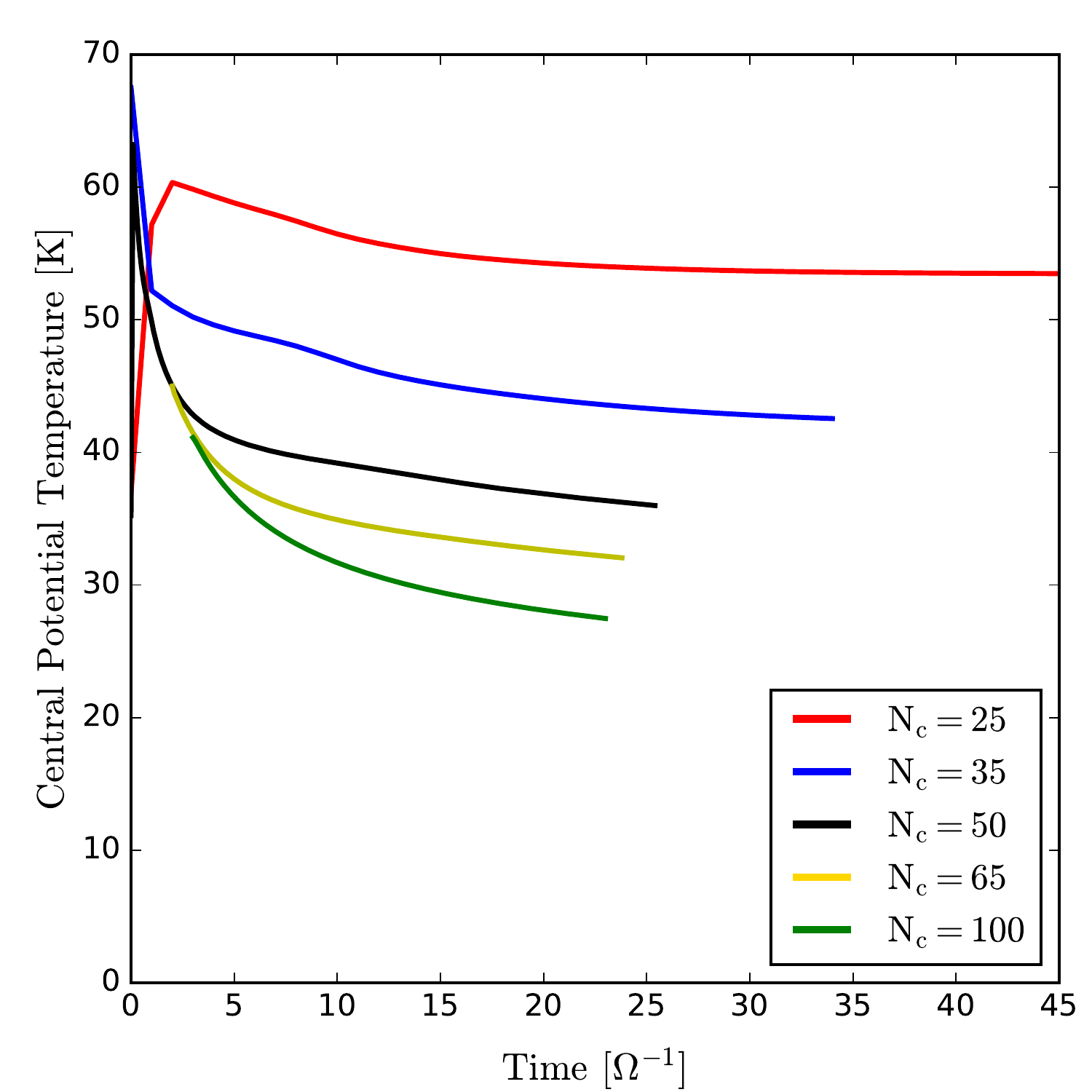}
  \caption{Central potential temperature for the same simulation parameters as in figs. \ref{fig:envelopemass} and \ref{fig:accretionrates}, $\rm \kappa = 0.01 \, cm^2 \, g^{-1}$, $\tilde{r}_s = 0.2$. 
	The steady decrease in potential temperature confirms that the accretion is final and irreversible. Just to be sure, we checked and confirmed that in fact the entropy in the rest of the envelope decreases as well.}
  \label{fig:centralentropy}
\end{figure}

In order to investigate whether the gas accretion into the envelope is irreversible, one needs to compute the evolution of the specific entropy of the gas. As introduced earlier, we investigate the related potential temperature instead, which is simple to compute directly from our simulation data. 
We plot the central potential temperature $\vartheta_{\rm c}$ in fig. \ref{fig:centralentropy}, where we emphasize that $\dot{\vartheta} > 0$ can only happen through the action of viscosity, and $\dot{\vartheta} < 0$ indicates cooling (see our earlier mention of eq. \ref{eq:entropyequation}), possibly with a small positive contribution from viscosity. We find $\dot{\vartheta}_{\rm central} < 0$, thus the mass accretion is irreversible and limited by cooling. 
Interestingly, the evolution of $\vartheta_{\rm c}$ with increasing resolution indicates that there are still physical effects that are insufficiently resolved, although the accretion rates converge. The central temperatures converge as well, thus the behaviour of $\vartheta_{\rm c}$ indicates that $\rho_{\rm c}$ increases as resolution goes up, or differently said, the mass inside the envelope concentrates more and more in the center as resolution increases.

\section{Dependence on smoothing length and opacity}
\label{sec:opacsmoothing}

Besides the numerical resolution, there is also the gravitational smoothing length $r_{\rm s}$ as free numerical parameter. In this section, we investigate the influence of the value of $r_{\rm s}$ on the accretion rates. 
When varying the smoothing length it is necessary to clarify how much resolution is needed for convergent results. After all, it is possible that more resolution is needed than in the cases presented before, as the gravitational potential is deepened.
After analysing the dependence of the measured gas accretion rates on the resolution of the smoothing length, we use the numerical parameters that satisfy convergence in order to investigate the accretion behaviour when changing the opacity. 

\begin{figure*}[]
	
  \hspace*{+0.2cm}
  \begin{subfigure}{0.50\textwidth} 
	\centering
	\includegraphics[width=0.95\textwidth]{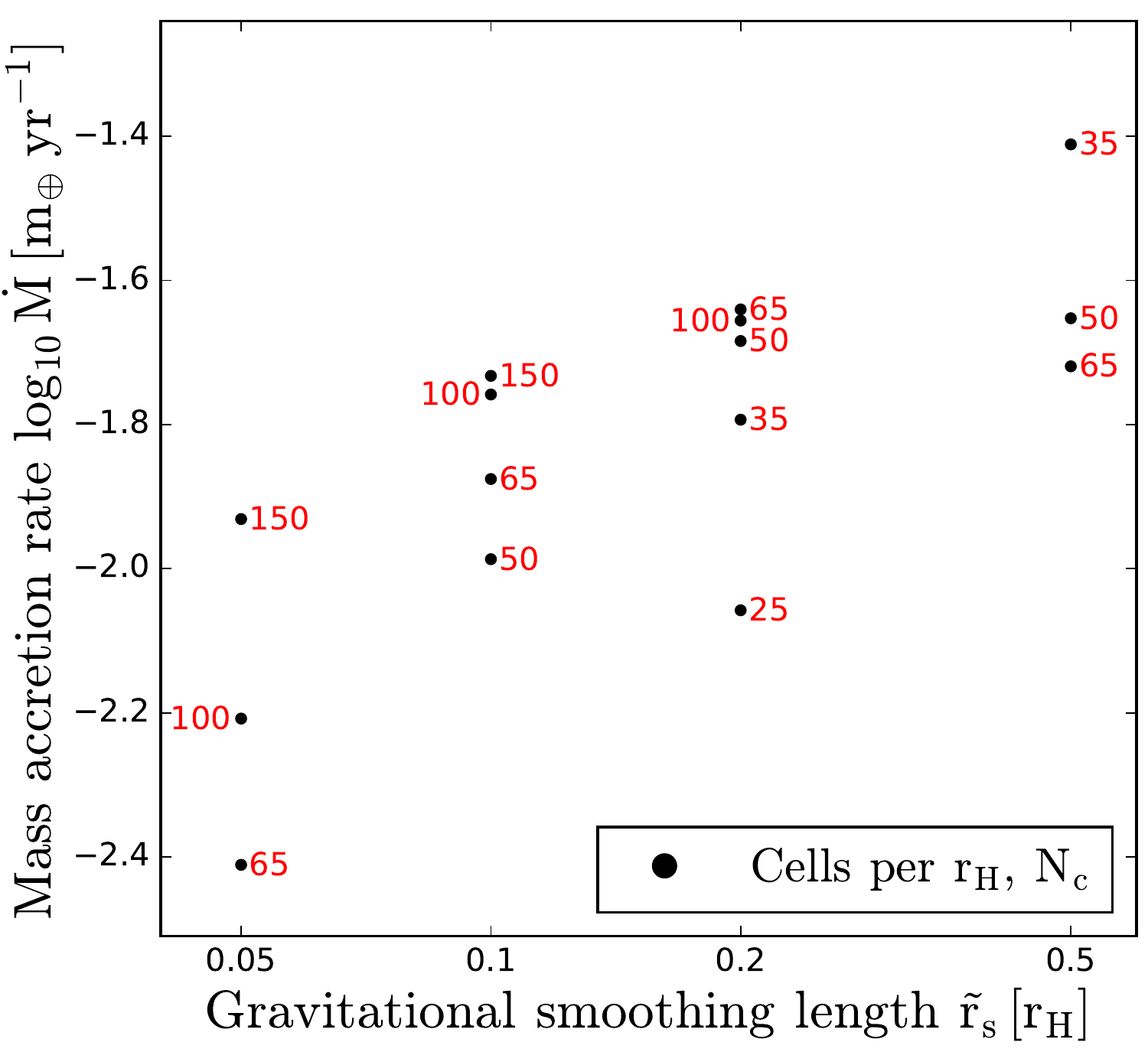}
      \caption{Gas accretion rate inside the Hill sphere as function of the smoothing length, for varying resolution numbers. Each dot is a simulation run, with the resolution of the Hill-sphere indicated as cell numbers. It is evident that as resolution increases, accretion rates cluster closer together, indicating numerical convergence. However more resolution is needed for deeper potential wells. \\ \\ }
   \end{subfigure}%
	\hspace*{+0.2cm}
	  \begin{subfigure}{0.50\textwidth} 
	\centering
	\includegraphics[width=0.95\textwidth]{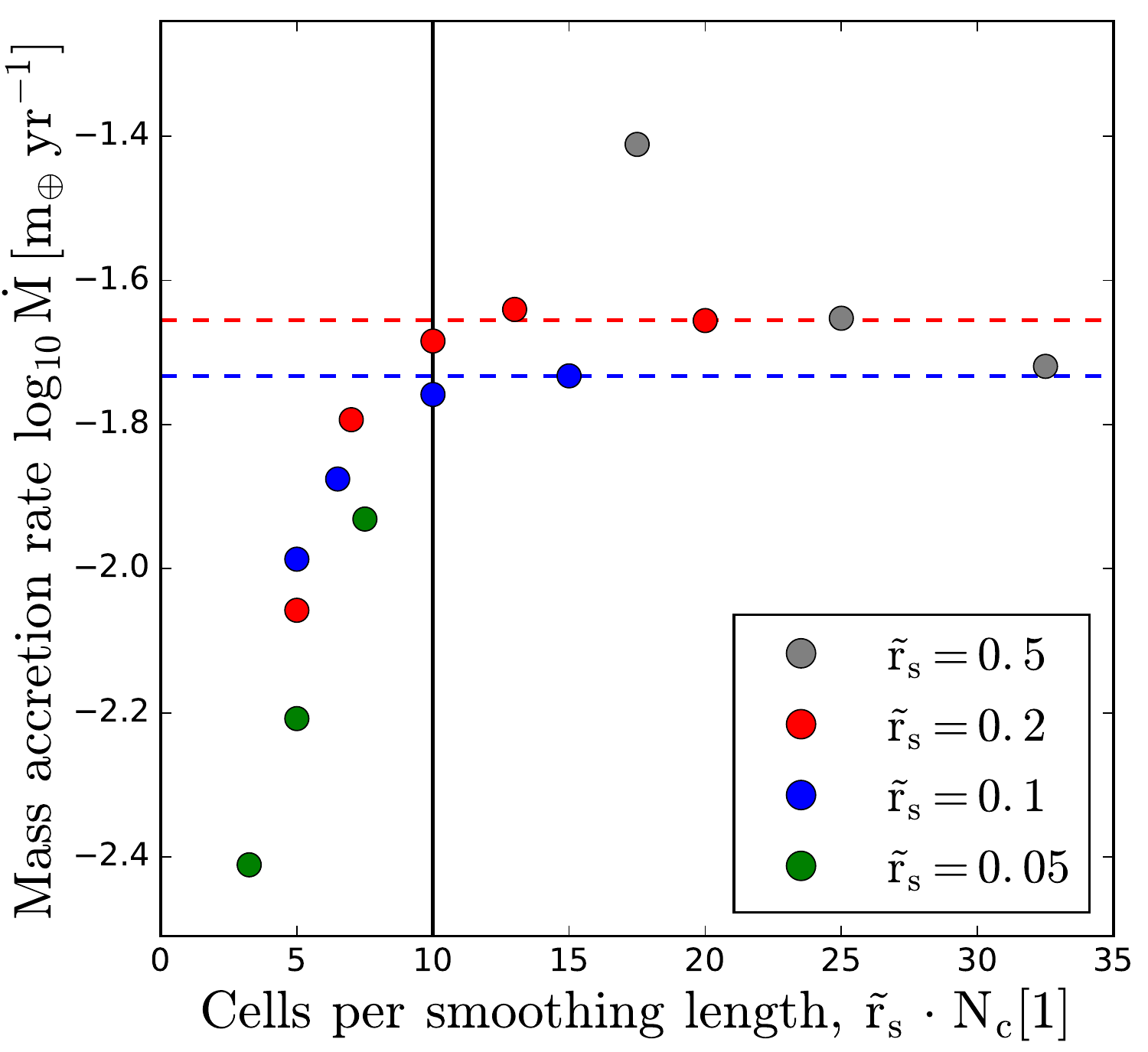}
      \caption{Gas accretion rate inside the Hill sphere plotted against the number of cells per smoothing length. Different colours indicate different smoothing lengths. It is evident that we need to resolve the smoothing length with at least 10 cells, before convergent results can be achieved (vertical lines). Thus, we define a 'well-resolved simulation' as one that hits convergent values of $\dot{M}$ at $\tilde{r}_{\rm s} \cdot N_c > 10$. Once converged, the accretion rates differ only slightly when deepening the gravitational potential, indicated with dashed lines. }
   \end{subfigure} 
	
\caption{Our complete set of all simulation runs for constant $\kappa=0.01 \rm \,cm^2\,/\,g$ where we varied the smoothing lengths as well as the numerical resolution, plotted as accretion rate vs. smoothing length (left) and as accretion rate vs. cells per smoothing length (right). All accretion rates shown are averages of the first 10 accreting orbits after ramp-up of the gravitational potential in Step C, see figs. \ref{fig:envelopemass} - \ref{fig:centralentropy}. }
\label{fig:convergencers}
\end{figure*}

\subsection{Convergence criterion}
\label{sec:convergence}

When performing simulations with smaller smoothing length (see fig. \ref{fig:convergencers}), we note that in fact an even higher resolution is needed before accretion rates converge. This indicates that it is $r_{\rm s}$, not $r_{\rm H}$ that is the relevant length scale to be resolved. In order to quantify how high resolution we need to be achieve convergence, we plot the values of the accretion rates averaged over 10 orbits with 10 samples per orbit against the number of cells per smoothing length. The result can be seen in fig. \ref{fig:convergencers}. There, the accretion rate can be seen to first rise near-linearly with increased resolution for all $r_s$ except for $\tilde{r}_s=0.5$, before leveling off after $r_s$ has been resolved with $\geq$10 cells. We consider the runs with $\tilde{r}_s=0.5$ as anomalous, as the force-free region inside $r_s$ there connects directly with the horseshoe flows, giving rise to strange velocity shears that otherwise would not exist.
The value towards which the simulations converge is slightly different for the individual values of $\tilde{r}_s$ (indicated in fig. \ref{fig:convergencers} see the dashed blue and dashed red lines). Although small, we deem this difference real, as there are physical differences in the envelopes that are introduced by deepening the potential. We discuss this further in Sect. \ref{sec:envelope}.

From fig. \ref{fig:convergencers} we conclude that we need at least 10 cells per smoothing length (\cite{lambrechts2017} already argued for $>5$) in order to obtain convergent accretion rates. Due to runtime considerations this limits our minimum achievable smoothing length to $\tilde{r}_{s} = 0.05$ and with this parameter we unfortunately do not achieve full convergence, but we can already observe a trend that is similar with the simulations of larger smoothing length. The physics that needs to be resolved by those 10 grid-cells resolution will also be discussed in Sect.  \ref{sec:envelope}.

\begin{figure*}
   \centering
   \includegraphics[width=0.95\textwidth]{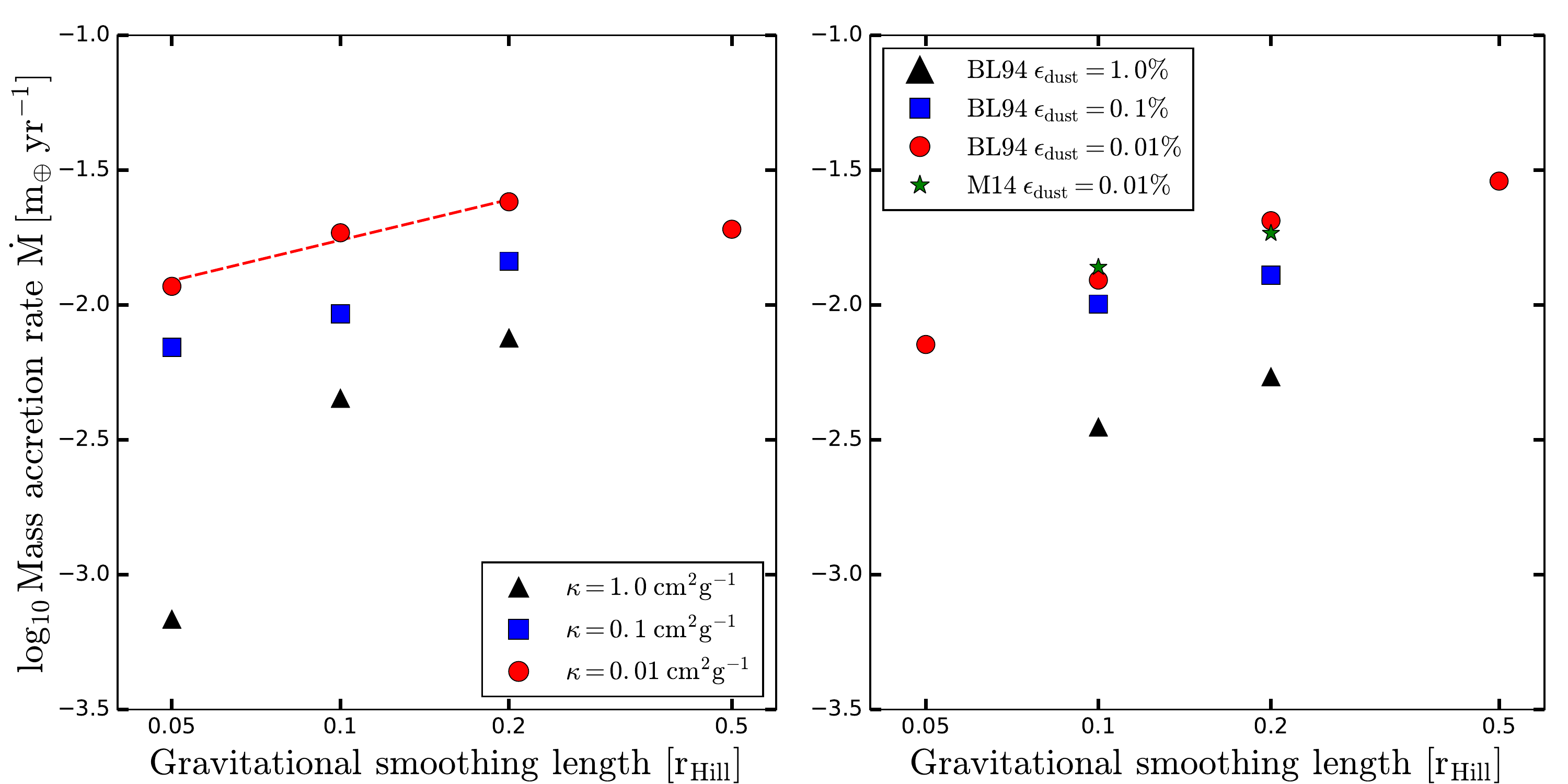}
      \caption{Dependency of the accretion rates on the smoothing length and opacities for simulations with highest available resolution as depicted in fig. \ref{fig:convergencers}.
			\textit{Left:}
							 Averaged accretion rates from 5 to 15 orbits, for $\tilde{r}_s=0.5$ down to $\tilde{r}_s = 0.05$ and our three different opacity values. Simulation runs with $\tilde{r}_s=0.5$ have such a shallow potential that the horseshoe flows, which penetrate down to approximately $\tilde{r}_s=0.5$ and interfere with the envelope cooling process.
			\textit{Right:} Accretion rates with Bell \& Lin opacities, with $1\%$ dust (=solar), $0.1\%$ and $0.01\%$ dust opacities, as well as a run with Malygin opacities and $\epsilon = 0.01 \%$. 
			Simulation runs with $\tilde{r}_s=0.2,\,0.1$ have been shown in fig. \ref{fig:convergencers} to be sufficiently resolved. We therefore interpret the seen downwards trend of $\dot{M}$ with $\tilde{r}_s$ as real (red dashed line), even if runs with  $\tilde{r}_s=0.05$ are only at the verge of fulfilling our convergence criterion. }
         \label{fig:resultssmoothing}
\end{figure*}

\begin{figure}
   \centering
   \includegraphics[width=\hsize]{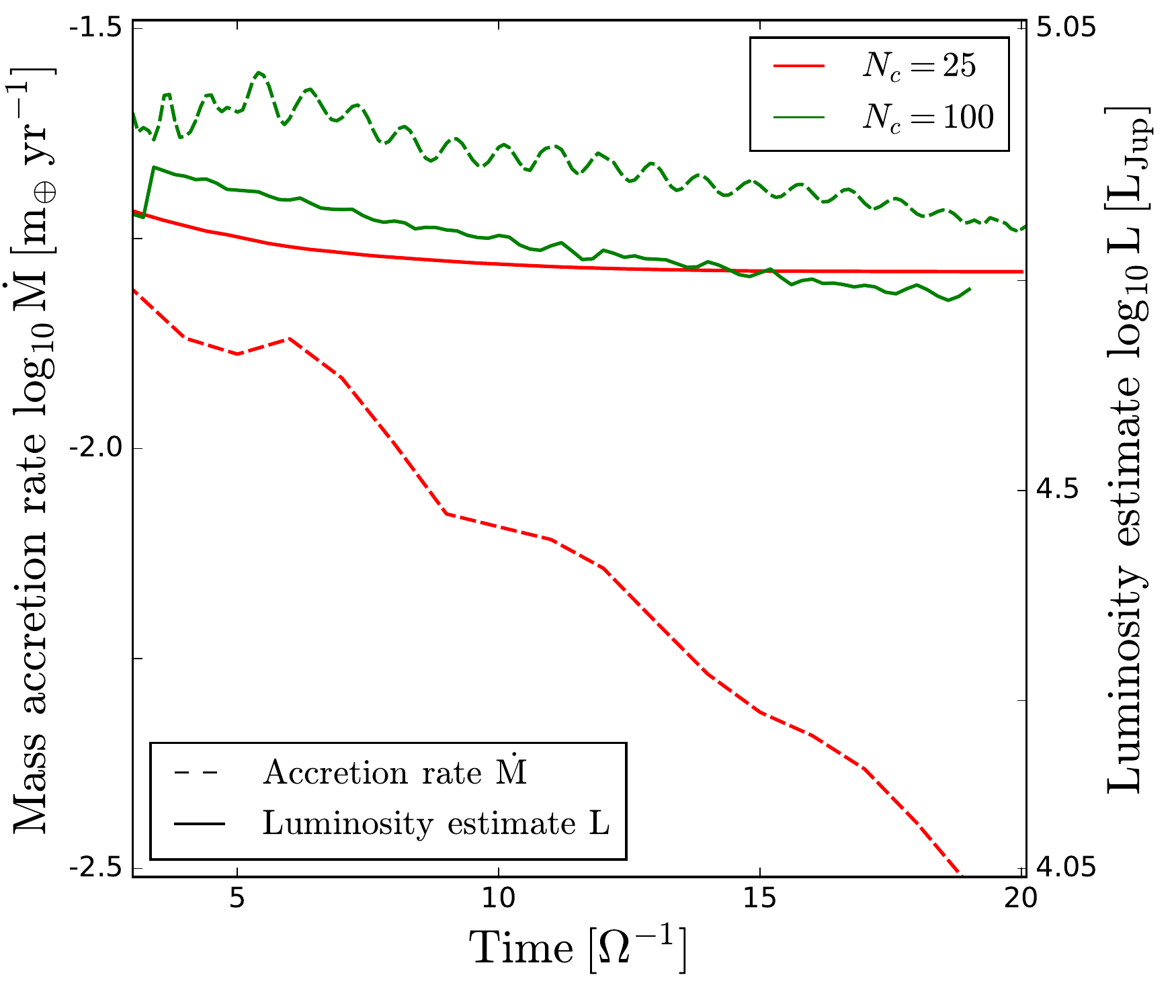}
      \caption{Evolution of measured luminosity vs. measured accretion rate for our $\kappa=0.01 \cmg$, $\tilde{r}_s=0.2$ simulation run. Sufficiently resolved simulations show behaviour such that the measured luminosity and the measured mass accretion rate correlate well. Unresolved simulations however run into a state of zero accretion rate, but finite luminosity, which remains higher than that of the final luminosity in the resolved simulation. This hints at numerical effects or other physics providing an additional energy source in unresolved simulations.}
         \label{fig:lumivsmdot}
\end{figure}

\subsection{Trends with smoothing length and opacity, once converged}

We now analyse the behaviour of simulations at the highest available resolutions as function of the smoothing length and opacities. Those simulations are either fully resolved or very close to it, so that we can ignore effects of resolution in this subsection.
The first set of simulations we introduce here are the constant opacity runs. The second and third set are those with non-constant \cite{belllin1994} and \cite{malygin2014} opacities. We show the measured mass accretion rates at highest resolutions in fig. \ref{fig:resultssmoothing}.

Our simulation runs with values of $\tilde{r}_s=0.2, 0.1, 0.05$ demonstrate that there is a coherent decrease of accretion rates when deepening the gravitational potential for all constant opacities as well as for more complex opacity functions. The runs with $\tilde{r}_s=0.05$ are marginally well-resolved according to our criterion,	and it is conceivable that the converged value for this parameter for $\dot{m}$ is slightly higher.
However our simulation runs are at the moment limited to the maximal resolution of $N_{\rm c}=150$. Furthermore, those runs seem to agree with the trend set by $\tilde{r}_s=0.2, 0.1$. We therefore see no reason to discard this datapoints as insufficiently converged and use them for the extrapolation of the trend of $\dot{m}$ vs. $r_s$. The run with $\tilde{r}_s=0.05$ and $\kappa=1.0 \cmg$ is discussed separately in sec. \ref{sec:discussion_adiabatic}. 
Finally, the runs with $\tilde{r}_s=0.05$ and complex opacities showed stronger irregularities and variability than in their constant opacity counterparts. Strong oscillations prohibited us from executing those runs, and probably require an even higher resolution than we can achieve here, to be executed correctly.

The overall trend of decreasing $\dot{m}$ with $\tilde{r}_{\rm s}$ can be puzzling at first: A smaller gravitational smoothing length, means stronger gravitational forces are acting on the gas close to the planet, which one would associate with a higher accretion rate. 
However a deepened the gravitational potential means first and foremost that there is more gravitational energy available, which is used to heat the gas. Thus when deepening the potential, gravitational forces quickly  compress the gas adiabatically, rising temperatures and increasing pressure support.

However only a fraction of the additional injected internal energy is able to diffuse outward as radiative energy, which results in a disproportionally small increase in envelope luminosity.
When going from $\tilde{r}_s=0.2$ to $\tilde{r}_s=0.1$ and thereafter to $\tilde{r}_s=0.05$, we measure a luminosity increase each time by a factor $\sim$ $1.5$, while internal energy and central temperatures rise by a factor of $\sim$$2.0$, proportional to $GM/\tilde{r}_{\rm s}$.
With the help of eq. \ref{eq:gmmdot} one can then post-predict the accretion rates, and in fact sees that $\dot{m}$ will decline.  

Another interesting feature is the departure of the simulation run $(\tilde{r}_s, \kappa)=(0.05, 1.0 \cmg)$ from the trend seen for the preceeding $\kappa=1.0 \cmg$ (black triangles in fig. \ref{fig:resultssmoothing}). It is conceivable that this might be a simple effect of under-resolving this particular simulation, or that high opacities simply need more resolution than low opacities. Upon testing this, we could however not find a difference in the resolution trend seen in fig. \ref{fig:convergencers} which was done for $\kappa=0.01 \cmg$ and those performed at $1.0 \cmg$. We conclude that this is not an effect of under-resolving that particular simulation run, but a physical effect. We are discussing this further in sec. \ref{sec:discussion_adiabatic}.

The differences between the constant opacities and the more advanced Bell \& Lin and Malygin opacities are not very dramatic. Neither the downwards trend with lowering $\tilde{r}_s$, nor the spacing when changing the opacity prefactors seems to make a huge difference in the general behaviour. 
We analyse the envelope structure for different choices of opacity law in connection with the accretion rates in more detail in the next section.

\begin{figure*}
   \centering
   \includegraphics[width=0.8\hsize]{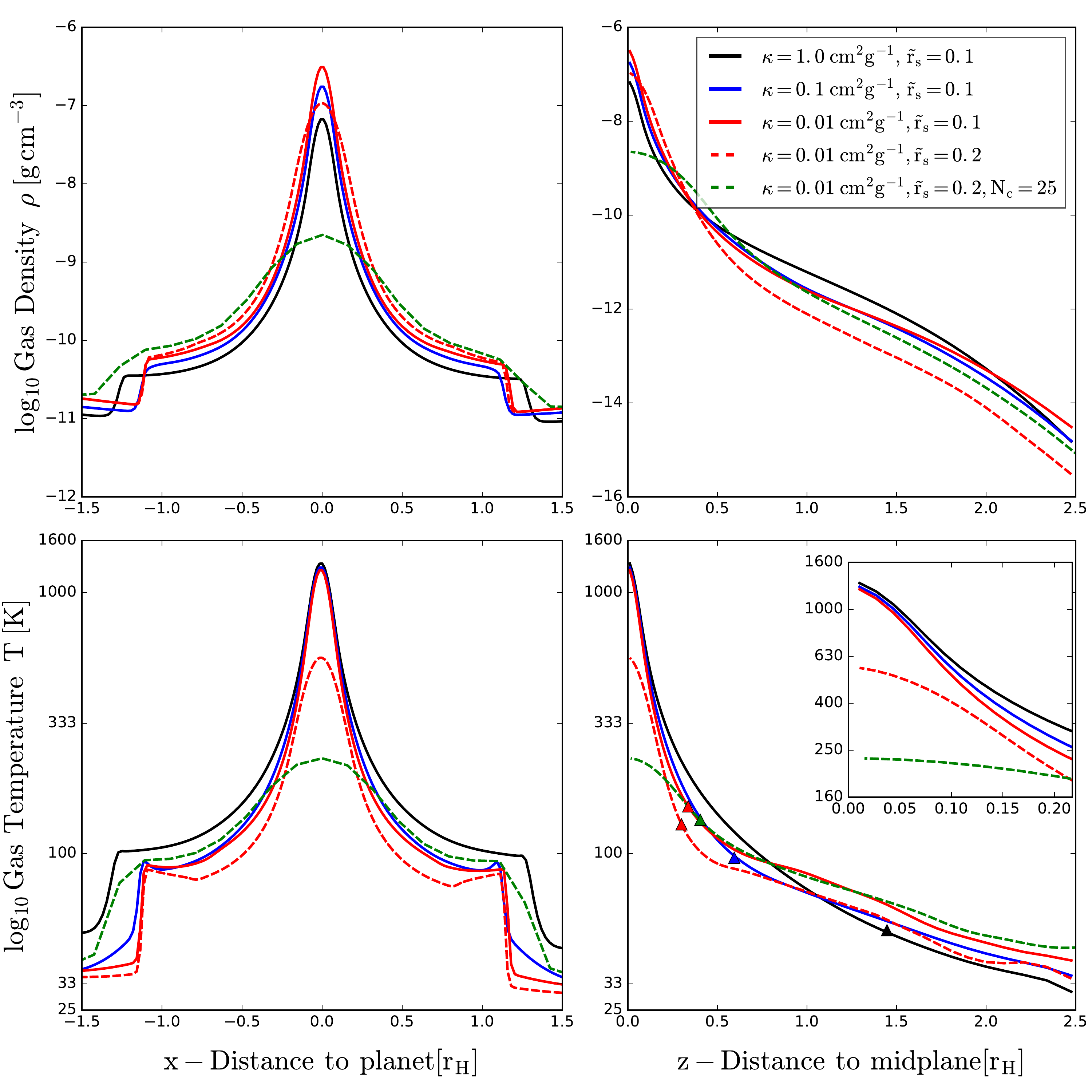}
      \caption{Density and temperature structures for constant opacity runs in radial and vertical slices after 10 orbits: Data is shown for runs with $\tilde{r}_s=0.1$ and the nominal run with $\tilde{r}_s=0.2$ and $N_{\rm c}=100$ for comparison. A deeper potential raises the whole envelope temperature as long as no convective region is formed. Optically thick regions are hotter for higher opacity (see inset), while the optically thin regions approach isothermal structure and their temperature trend reverses according to the position of the $\tau=1$ surface (indicated with triangles). The density structure follows simple expectations: A hotter envelope has lower density in optically thick regions. }
         \label{fig:structures_const_denstemp}
\end{figure*}

\section{Envelope structure}
\label{sec:envelope}

This section aims to link the envelope structure, resulting from the simulation parameters $N_c$, $\tilde{r}_s$ and $\kappa$, with the measured accretion rates. We use the previously introduced theoretical as well as measured luminosities and explain those with the observed density and temperature structures.

\subsection{Comparison to simple 1-D models}
\label{sec:comparison1d}

After establishing numerical convergence, we are interested in understanding whether the solution converged to is also the correct, physical one.

A popular way to estimate mass accretion rates in 1-D models of gas giant formation \citep[see i.e.][]{mizuno1980, pollack1996} is based on the stellar structure literature and connects the energy loss from the envelope, the luminosity $L$, with the mass accretion rate via 
\begin{align}
L = \frac{G M \dot{M}}{r} .
\label{eq:gmmdot}
\end{align}
This relation would be true for slow Kelvin-Helmholtz contraction, and we can check our simulation for consistency with this mechanism. Also this is a slight improvement over the methodology of \cite{lambrechts2017}, where the compressional heating was assumed to be identical to the luminosity.
From our simulations we can measure both $M$ and $L$ and check whether they obey relation \ref{eq:gmmdot}, to answer the question of which physical process we are investigating. To this end, we estimate the 3D-position of the $\tau=1$-surface, integrating all optical depth elements $dz \rho \kappa$ vertically from infinity (the simulation boundary) towards the planetary envelope. An example of the computation of this surface can be seen in fig. \ref{fig:overview}. Once this surface is found, the temperature $T_{\rm gas,\, \tau=1}$ on all surface elements is taken, and it is then $L= \sum \Delta A\, \sigma T_{\rm gas, \, \tau=1}^{4}$, with $\Delta A$ the area of a surface element and $\sigma$ being the radiation constant. Throughout this work we use the unit of a Jupiter-luminosity $L_{\rm J} = 3 \times 10^{24} \; \rm erg \, s^{-1}$. The example surface in fig. \ref{fig:overview} has a luminosity of $7 \times 10^{4} \, L_{\rm J}$ and its evolution in time is shown for a resolved and non-resolved simulation in fig. \ref{fig:lumivsmdot}, confirming that it is consistent with the mass accretion rates. This is because the flux-limited diffusion luminosity quickly becomes the free-streaming luminosity $F_{\rm FS} = c E_{\rm rad}$ when passing through the strong density gradients of the planetary envelope into the disc atmosphere and the planetary gap. This quick transition is also shown in Appendix \ref{sec:appendix_fld}.

This approach is valid only for the simulations for which the $\tau=1$-surface forms an enclosed surface on the planetary envelope i.e. when the planetary gap is optically thin. In fact we find a good agreement between the mass accretion rates and the luminosities in those cases. The downwards trend of luminosity also follows the downwards trend for the mass accretion rate, hinting at the increasingly radiatively inefficient envelope. We checked this downwards trend more thoroughly with a full 3D computation of the luminosity based on the actual fluxes in the simulation. This is shown in appendix \ref{sec:appendix_lumi}, where we conclude that the actual luminosity is 20$\%$ less than the one by our estimate on the $\tau=1$-surface, but that the two luminosities have a relatively constant ratio.

This is interesting for the purpose of our analysis in that we can now link differences in luminosity to differences in accretion rate behaviour when varying opacity and smoothing length. The connection of luminosity to the envelope structure will be discussed further in connection with the averaged 2D profiles of the envelope.

\begin{figure*}
   \centering
   \includegraphics[width=0.8\hsize]{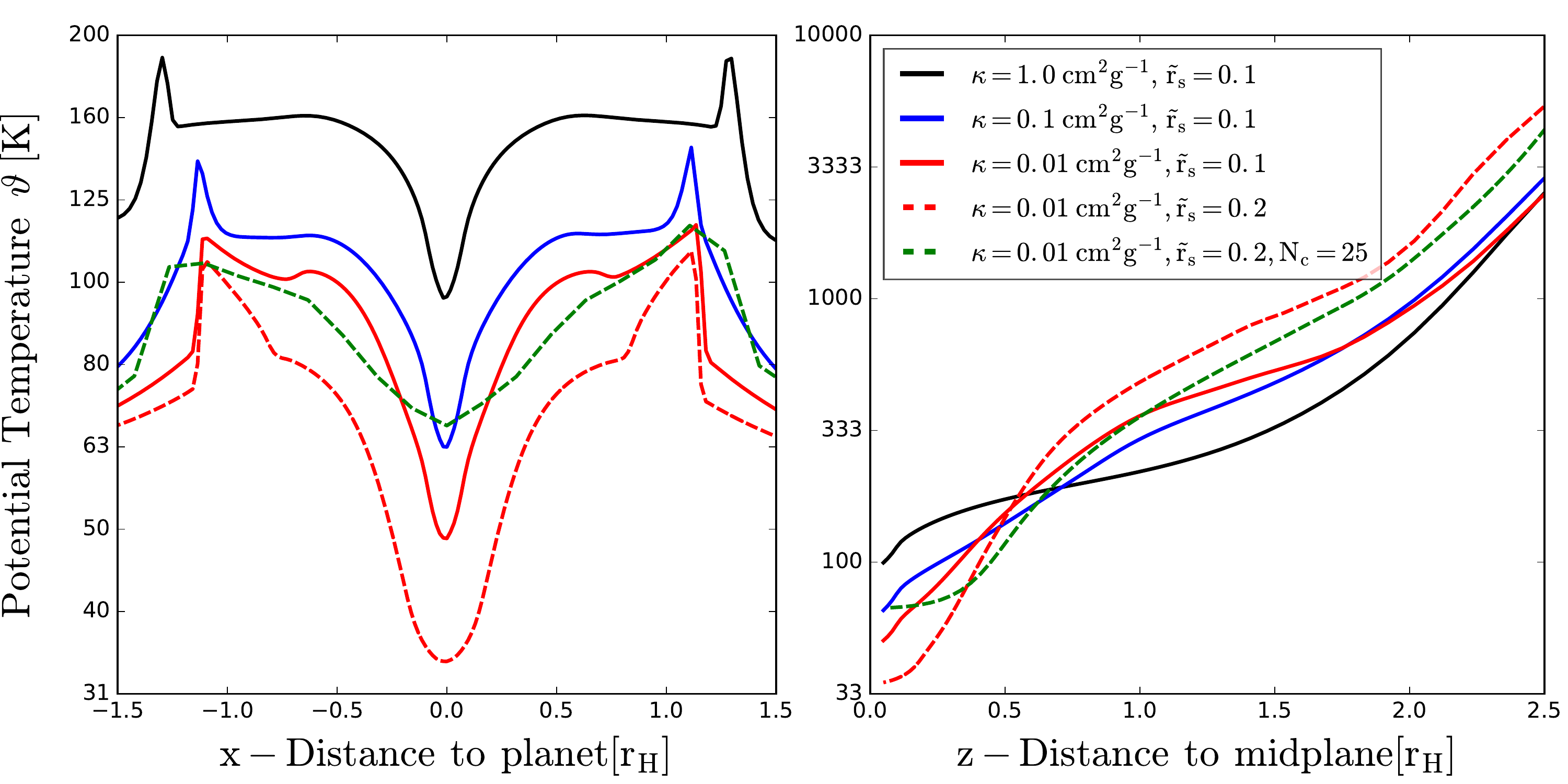}
      \caption{Potential temperatures for opacity set 1 with $N_{\rm c}=100$. 
			Deeper potential and lower opacity leads to a flatter potential temperature profile. Radially, the simulation runs with $\tilde{r}_s=0.1$ have flat $\vartheta$-profiles, this does not denote the onset of convection, but rather the action of the horseshoe-flows that can enter and exit the gravitational well freely. In $z$, a steeper profile corresponds to stronger accretion flows from this direction. The much higher values of $\vartheta$ result from the low density in the disc atmosphere, as we have $r_{\rm H} \cong H_{\rm disc}$ in throughout our work.} 
         \label{fig:const_entropa_real}
				
				\includegraphics[width=0.8\hsize]{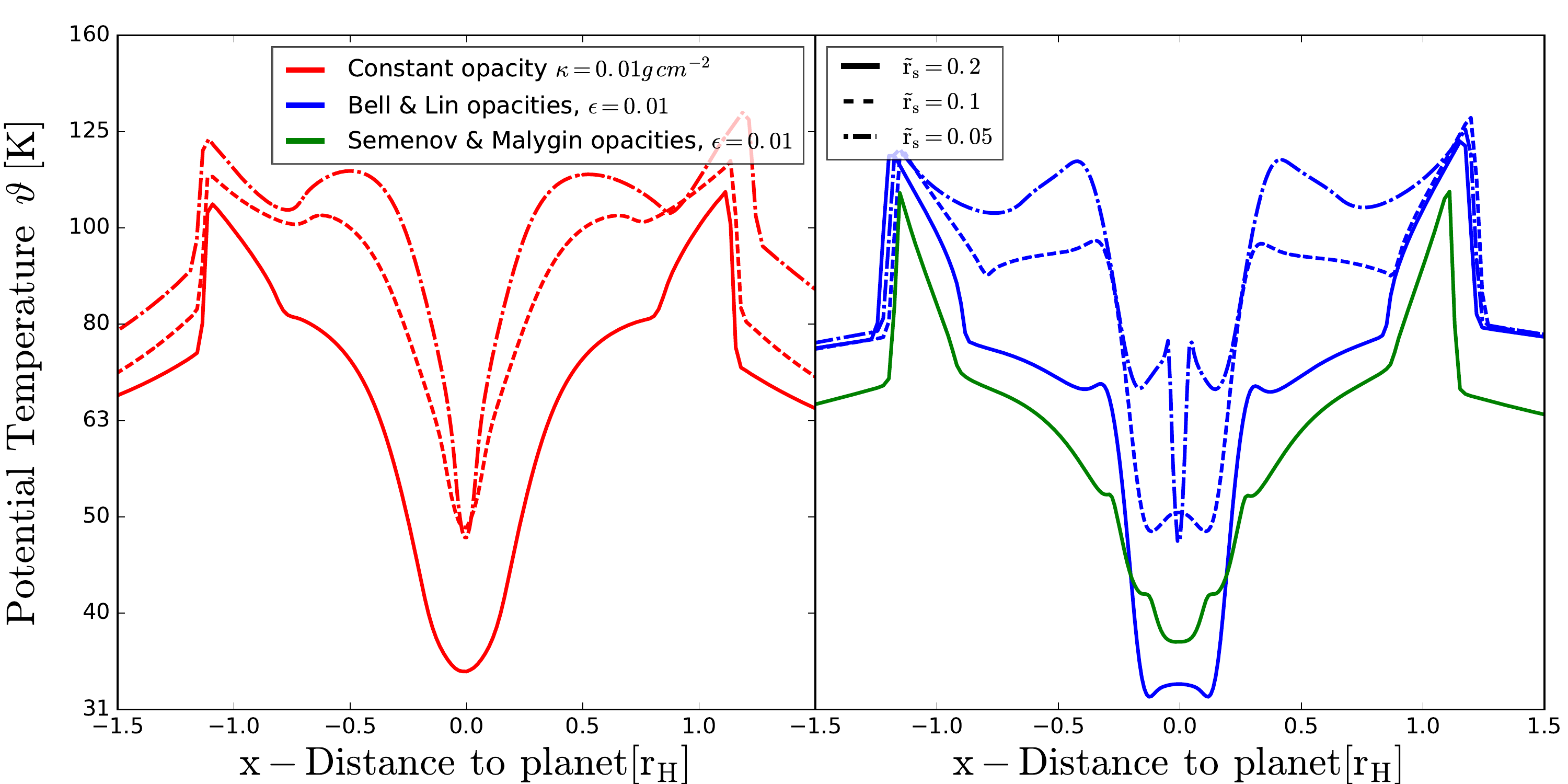}
      \caption{Potential temperatures in simulations with constant (left, red curves), Bell \& Lin (right, black curves), and Malygin opacities (right, green curve) for different values of $\tilde{r}_s$. Note that both plots share the two legends. The $r_{\rm s}=0.1$ have $N_{\rm c}=100$ and correspond to the simulations shown in fig. \ref{fig:const_entropa_real}. A major difference between constant and Bell \& Lin opacities is that close to the water-iceline at around $\tilde{r}=0.25 \, r_H$ there are indications of convective instability. The notion that this is related to the vicinity to the smoothing region is weakened, as the trend of inverted $\vartheta$-profile continues as we deepen the potential in a well-resolved manner.	The comparison to the more realistic \cite{malygin2014}-opacities is done only for $\tilde{r}_s = 0.2$ (compare the green and the straight black line). There, it is evident that the gradients in $\vartheta$ are weakened by a multi-step transition in $\kappa$. }
         \label{fig:sm_entropa}
\end{figure*}

\subsection{Density and temperature}

We now present an overview in fig. \ref{fig:structures_const_denstemp} of 1D slices of density and temperature profiles of planetary envelopes. Five different simulations are presented of which three are identical except for the chosen opacity (the solid red, blue and black curves), two are identical except for the smoothing length (the solid red and dashed red curves) and another two are identical except for the resolution (the dashed green and dashed red curves). In this way we highlight the physical differences in the envelopes when varying parameters.

\subsubsection{Change in resolution}
At first, we compare a well-resolved and under-resolved run ( $N_c=100$ cells per $r_{\rm H}$ red dashed with green dashed $N_c=35$ cells per $r_{\rm H}$) in fig. \ref{fig:structures_const_denstemp}. Those profiles are only snapshots in time. However, they are representative of the differing evolutionary paths that simulations can take, particularly seen in fig. \ref{fig:lumivsmdot}.
The underresolved simulation has low central temperature, but for the accretion rates, this does not play a role. It is the potential temperature (see fig. \ref{fig:radvert_1const}) that helps us interpret the difference between resolved and non-resolved simulations properly: Being closer to a full adiabatic profile, the unresolved simulation must be radiatively less efficient and is thus accreting less. The deeper reasons for this shall be discussed in sec. \ref{sec:boundary}.

\subsubsection{Change in smoothing length}
Secondly, we compare a well-resolved deep potential with a well-resolved shallow potential (straight red with dashed red curve) in fig. \ref{fig:structures_const_denstemp}. 

Both runs start with the same initial envelope mass, so naturally for the deeper potential the central density is higher and the outer envelope densities are lower. This is true for the radial structure of the envelopes. 
This is different in the vertical direction straight above the planet, where at $z > 1.0 r_{\rm H}$ the deep potential has a higher density than the shallow potential. The density structure at those altitudes, in the transition region from planetary envelope to disc atmosphere, is important because the optically thin-thick transition happens at densities of around $\rho \sim 10^{-13} \gcm$ and thus determines the luminosity.

Central temperatures scale linearly with the smoothing length as long as no complex opacity laws are used. This stems from the fact that the energy input into the envelope is $\rho GM_{p}/r_s $ which is then translated into internal energy $c_p \rho T$. The radiative temperature gradient is then followed upwards in $z$ until the optically thin-thick transition is encountered, and photons can escape into space, at which the temperature slope changes. The remaining temperature gradient comes from the radiative coupling with the disc atmosphere, which decreases in efficiency as $\rho \kappa_P$, see eq. \ref{eq:energyequation2}.

An important point to discuss is the difference in envelope temperatures and luminosities. The deeper  potential heats the envelope stronger than the shallow one. The resulting luminosities are higher for the deep potential. But this does not automatically impart a higher accretion rate. One can understand the accretion rate according to eq. \ref{eq:gmmdot} as balance between the luminosity and the energy content of the envelope. For our simulations the energy content of the envelopes rises as $GM/r_{\rm s}$, but the resulting luminosities increase only by a factor of $\sim$$1.5$ for a doubling of the potential depth. Thus the accretion rates decrease with increasing potential depth.

\subsubsection{Change in opacity}
Inside the deep envelope, where the optical depth is larger than one, a higher opacity causes higher temperatures. This is because the radiation diffusion times are proportional to $\kappa$, while all simulations of same smoothing try to radiate away the same initial compressional work. Density is inversely related to temperature in the central region. Temperature gradients in $r$ and $z$ are proportional to $1/\kappa$ while optically thick. However when following the temperature profiles outwards in $z$, low opacities hit the optically thin-thick transition earlier and from there on the temperature remains quasi-isothermal with only weak, opacity independent, temperature gradient. Radially, both $\rho$ and $T$ show signatures of the spiral arm discontinuity and jump by a factor of $3-4$, indicating a strong hydrodynamic shock.

\subsection{Potential temperature and lessons from it}

The potential temperature $\vartheta$ is straightforwardly computed from density and temperature profiles (eq. \ref{eq:pottemp}) and its gradients inform about the energetic state of the solution. Positive gradient vs. gravity indicates a radiative region while a null-gradient indicates convective instability. 
Strong gradients in potential temperature indicate heavy cooling, and must be followed up by contraction. Thus we take potential temperature gradients to explain differences in accretion rates, based on envelope structures.

As is visible in fig. \ref{fig:const_entropa_real} using higher opacities, deepening the gravitational potential and under-resolving flattens the potential temperature profile, corresponding to a decreasing $\dot{M}$. This explains why a deeper potential does not increase the mass accretion rate: The temperature in the central region cannot diffuse away efficiently enough, even for $\kappa = 0.01 \cmg$, and reaches values such that the entropy gradients are overall decreased, which must decrease accretion rates. 

Issues with high temperature in the central region of high-mass planets have been previously reported \citep{judith2016}. 
One might worry that those temperatures are originating in underresolving the gravitational potential. Here we show that underresolving a potential actually lowers the central temperatures. However the density relative to the temperatures is still lower, which results in an increased entropy.
At any given resolution this unfortunately does not help to decide whether the physically correct solution has been found, but it removes doubts pertaining to too high temperatures being the cause of non-accretion.

Comparing now a set of low-opacity simulation runs for different smoothing lengths, we can understand step-by-step the role of deepening the gravitational well on the envelope structure (see fig. \ref{fig:const_entropa_real}, straight red and the dashed red lines). Both simulations start out with the same envelope mass from the step B run, but the deeper potential will have a higher temperature and thus a lower density. With eq. \ref{eq:pottemp} it is clear that $\log \vartheta \sim \log T - 0.4 \log \rho$, so both those factor contribute in increasing the potential temperature. 
While for the constant opacities the relative envelope structure in $\vartheta$ remains very similar for all $r_s$, the effect of opacity transitions on entropy are clearly visible in the black and green curves in fig. \ref{fig:sm_entropa}.

As the horseshoe region penetrates down into $0.5r_{\rm H}$, it is this radius which defines a 'feeding region'. There the local entropy gradient will decide whether a part of the horseshoe mass flows into the inner envelope or not. As we see, when letting any individual $r_s$ constant, the entropy gradients at $r=0.5 r_{\rm H}$ are not very different when switching from constant to complex opacities. This would be a possible explanation for their similarities in accretion rates. 

Therefore, although not dramatically different in terms of mass accretion rates, those simulation runs with different opacities would treat a tracer dust species differently, having interesting consequences for the transport of dust. We address this in a future work.

\begin{figure*}

\hspace*{-1.2cm}
	\begin{subfigure}{.345\textwidth}
				\centering
				\includegraphics[width=.95\linewidth]{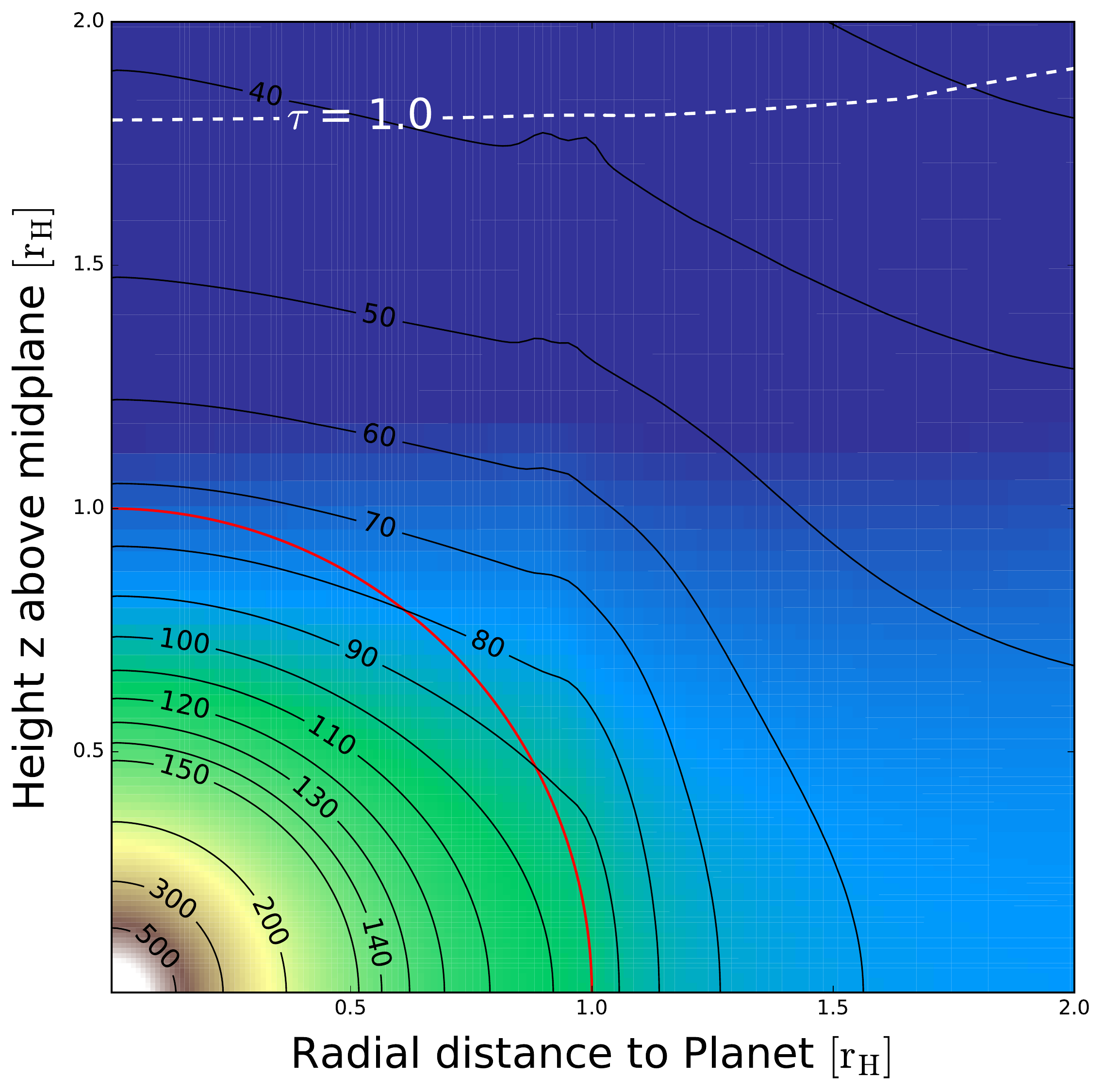}
				\caption{Constant $\rm \kappa=1.0 \cmg$}
				\label{fig:radvert_const1}
		\end{subfigure}%
		\begin{subfigure}{.345\textwidth}
				\centering
				\includegraphics[width=.95\linewidth]{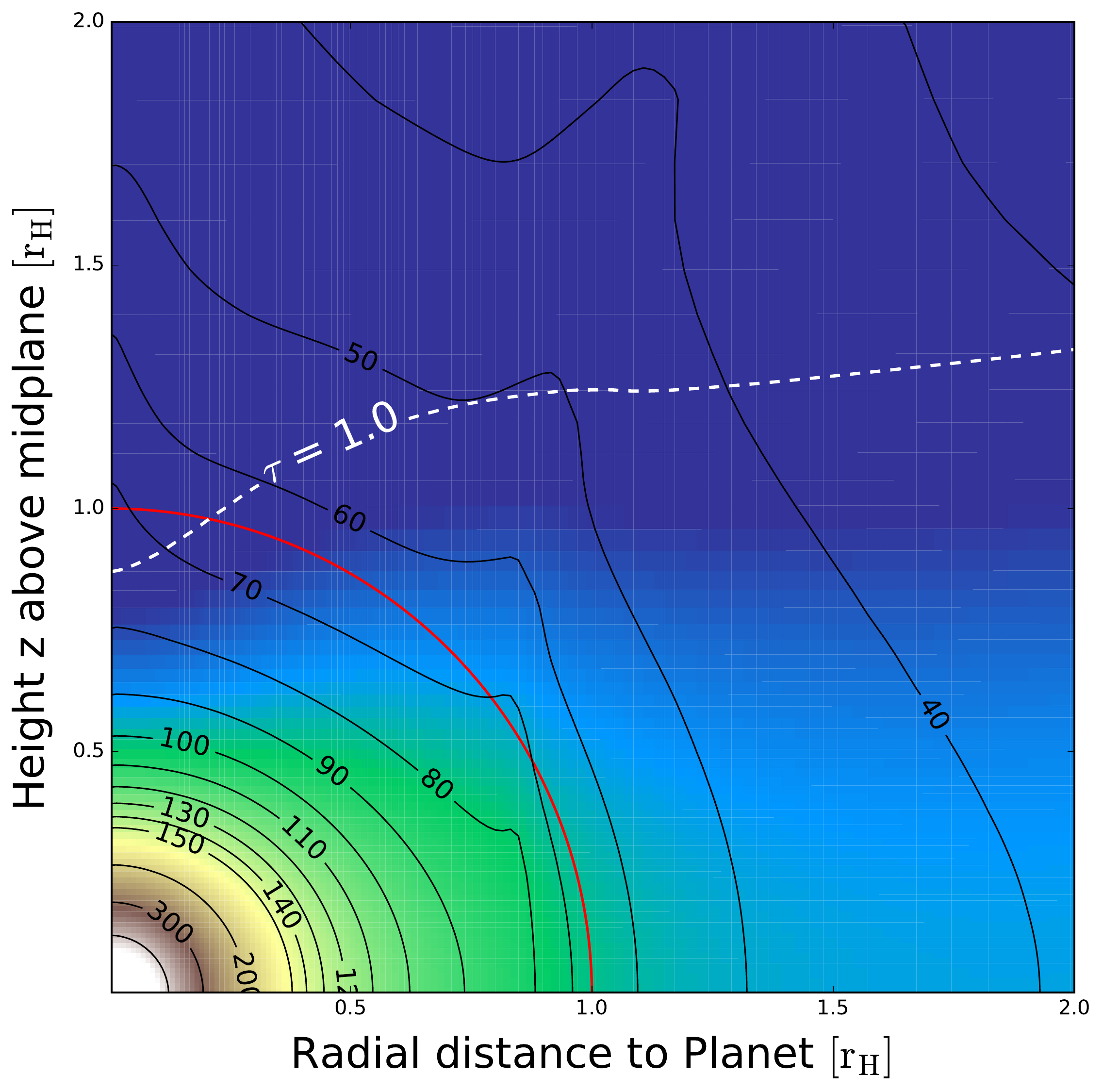}
				\caption{Constant $\rm \kappa=0.1 \cmg$}
				\label{fig:radvert_const2}
		\end{subfigure}%
		\begin{subfigure}{.44\textwidth}
				\centering
				\includegraphics[width=.95\linewidth]{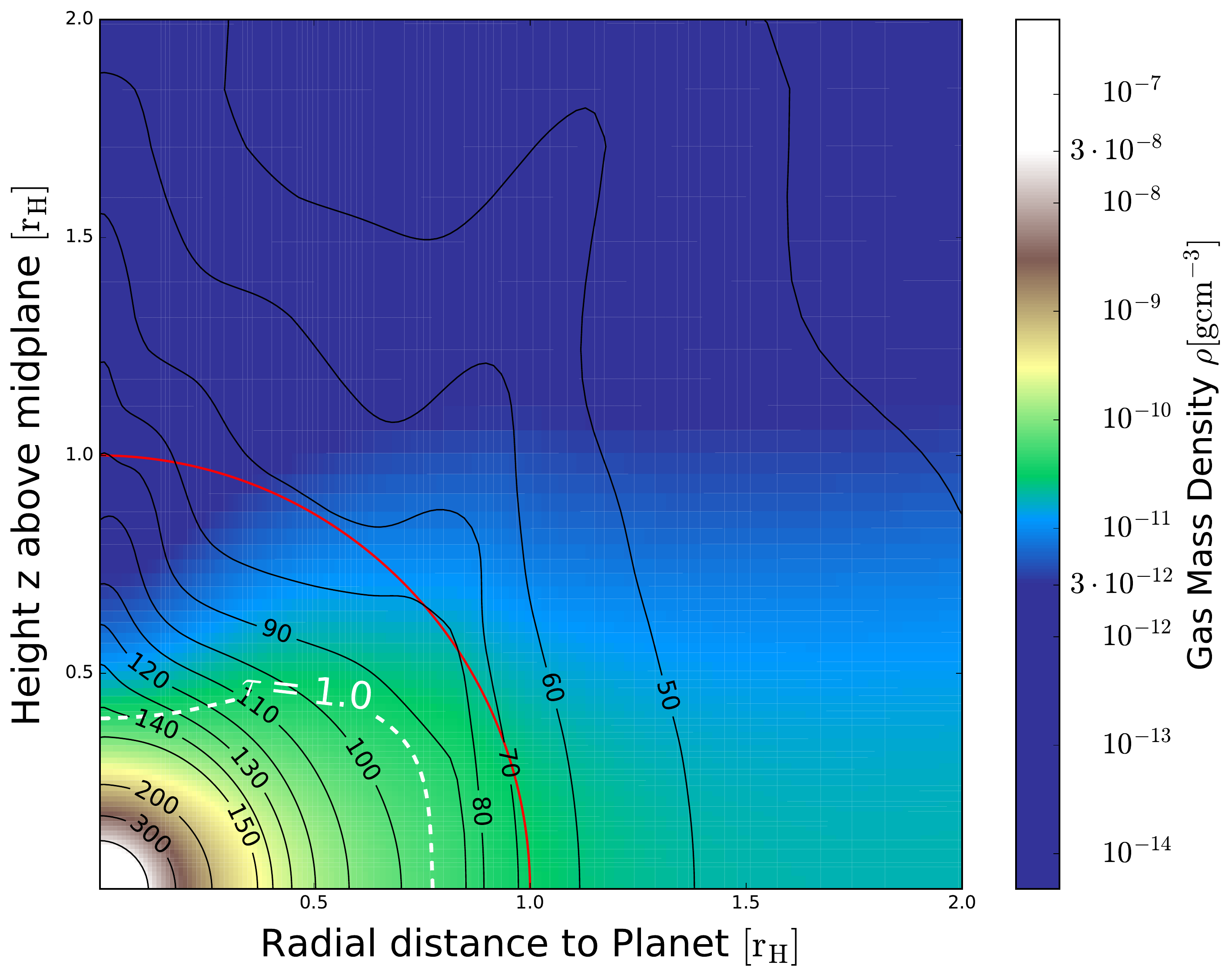}
				\caption{Constant $\rm \kappa=0.01 \cmg$}
				\label{fig:radvert_const3}
		\end{subfigure}%
\caption{Time and cylindrically-averaged density (as background colour) and temperature (in contours) distributions as functions of $r$ and $z$ and the optical thin-thick transition surface measured from the simulation boundary. Parameters are well-resolved ($N_c=100$) runs of opacity set 1 and gravitational smoothing $\tilde{r}_s=0.1$. The time-average ran over 10 orbits with 10 output-samples per orbit. Near the planet, density and temperature are both radially distributed. Farther away from the planet, the density stratifies in $z$ in order to connect to the gap. Moving outward radially in the disc midplane, the temperature becomes quasi-isothermal as $r$ moves into the optically thin gap. In the $z$-direction however, heavy cooling bends the iso-$T$ planes upwards. The luminosity required for eq. \ref{eq:gmmdot} is computed for the $\kappa=0.01 \cmg$ cases in the approximation $\sum dA \sigma T^4$, which is possible as in those cases the planetary gap is optically thin. Luminosities and accretion rates vs. time are shown in fig. \ref{fig:lumivsmdot}. }
\label{fig:radvert_1const}

	\vspace*{0.5cm}
  \hspace*{-1.2cm}
	\begin{subfigure}{.345\textwidth}
				\centering
				\includegraphics[width=.95\linewidth]{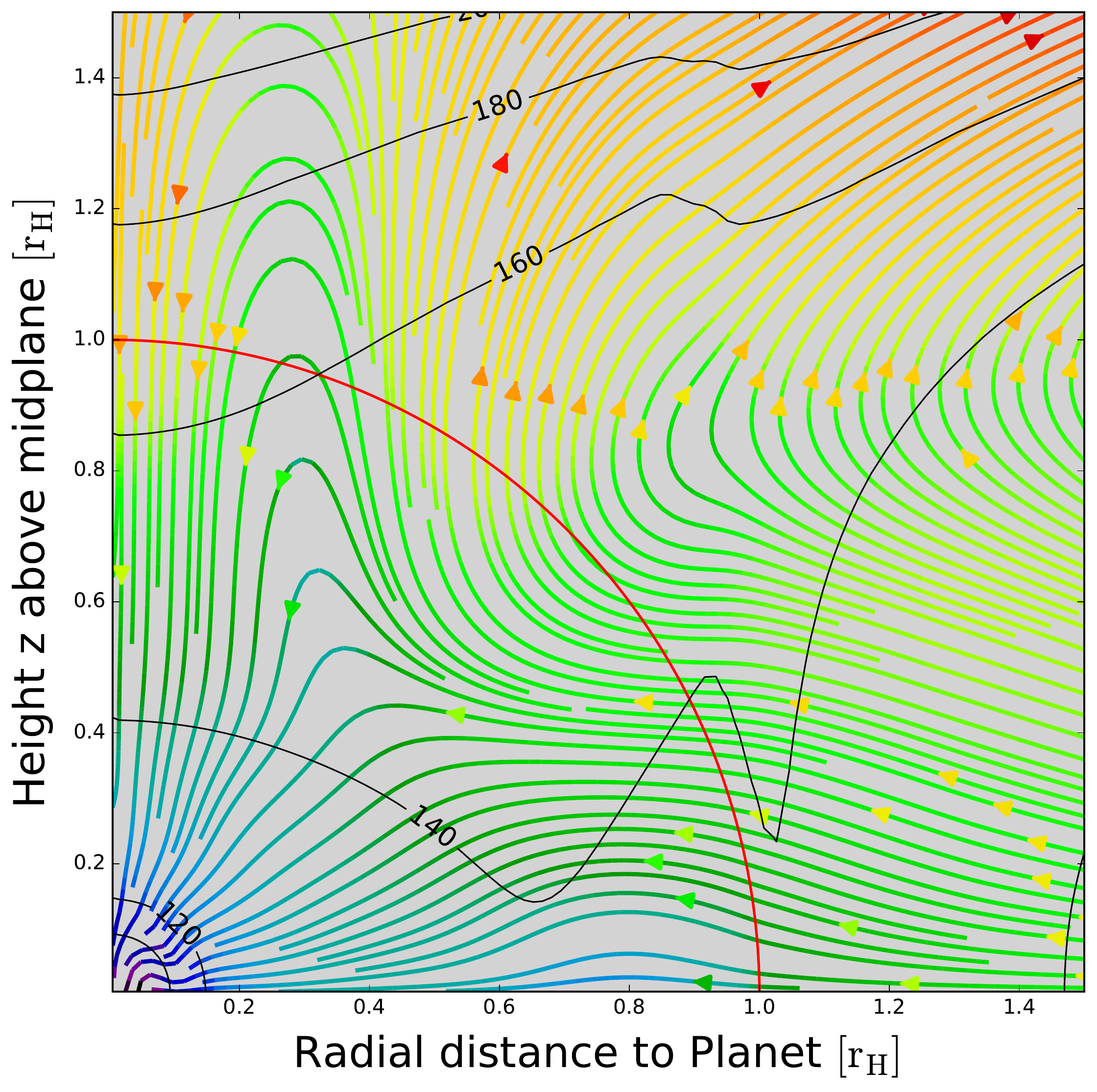}
				\caption{Constant $\rm \kappa=1.0 \cmg$ \\}
				\label{fig:radvert_const4}
		\end{subfigure}%
		\begin{subfigure}{.345\textwidth}
				\centering
				\includegraphics[width=.95\linewidth]{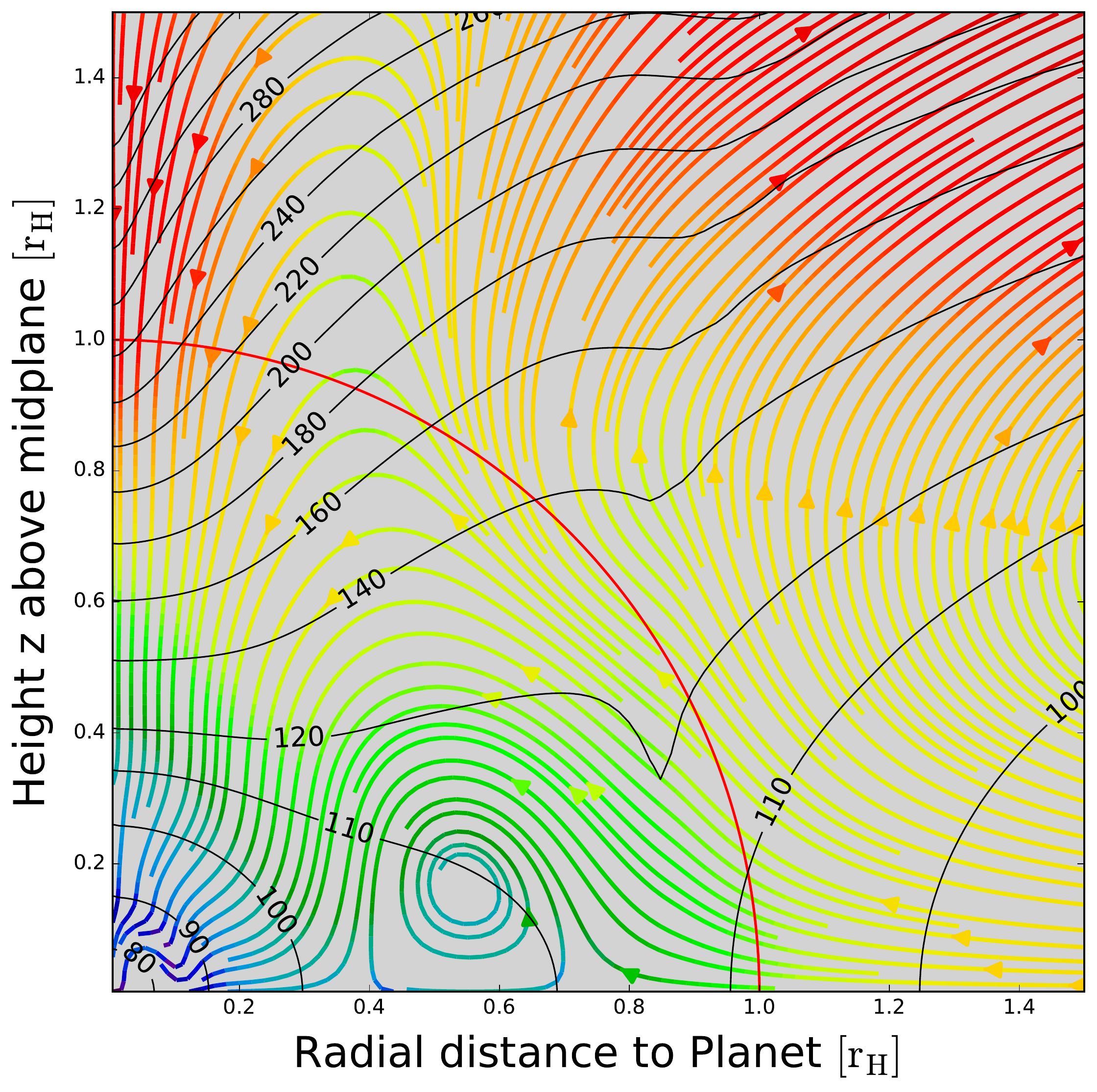}
				\caption{Constant $\rm \kappa=0.1 \cmg$ \\}
				\label{fig:radvert_const5}
		\end{subfigure}%
		\begin{subfigure}{.41\textwidth}
				\centering
				\includegraphics[width=.95\linewidth]{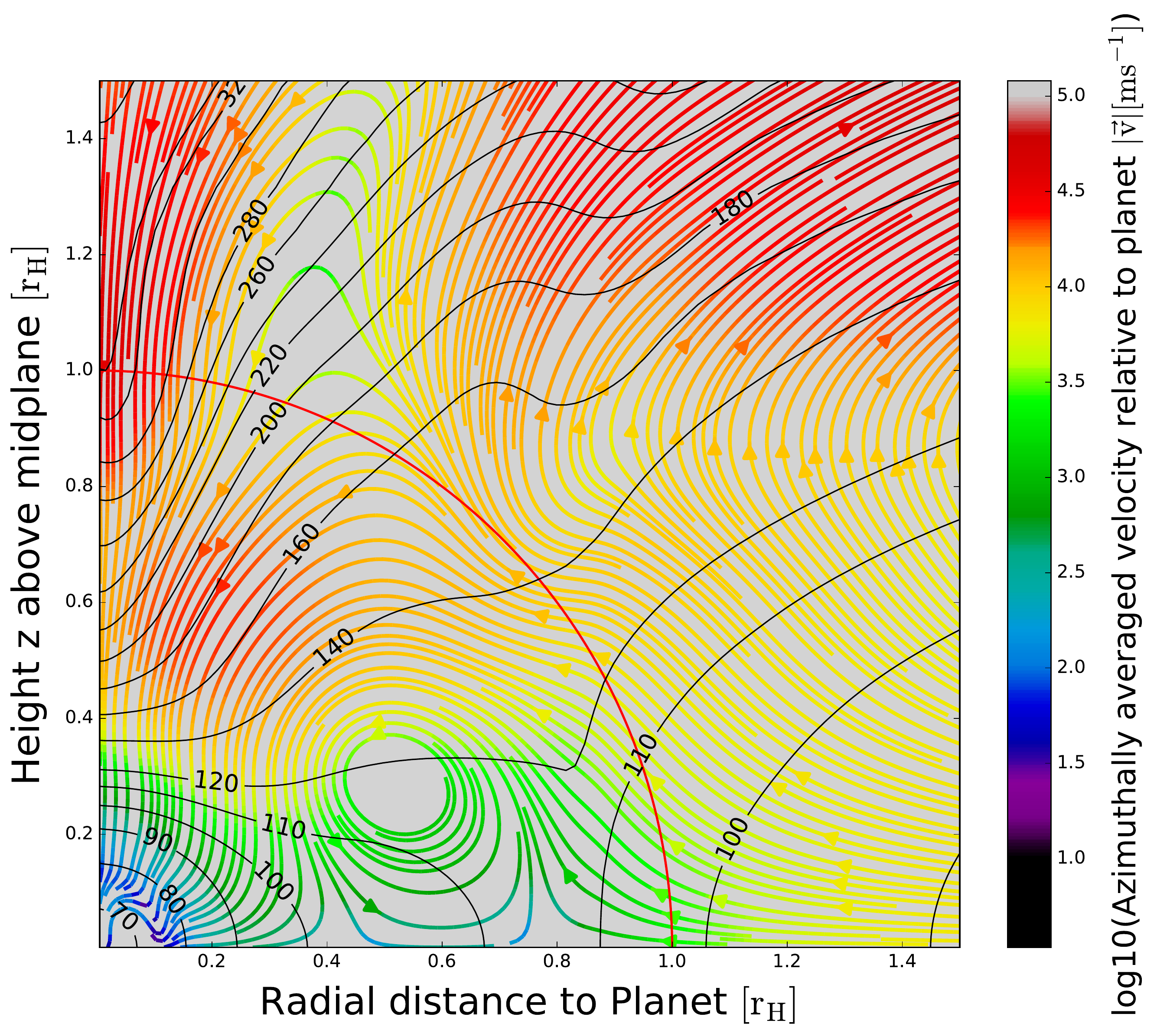}
				\caption{Constant $\rm \kappa=0.01 \cmg$}
				\label{fig:radvert_const6}
		\end{subfigure}%
\caption{ Temporally and cylindrically averaged flow streak lines with their colour-coded velocity relative to the planet and potential temperature $\vartheta$ in contours for well-resolved runs opacity set 1 and gravitational smoothing $\tilde{r}_s=0.1$. The plot is a bit zoomed in compared to \ref{fig:radvert_1const}, to better feature the flow structures.  Correlation of flows and isentropic surfaces is imperfect due to the averaging process. The vortex feature that appears for $\kappa=0.1, 0.01 \cmg$ is a feature that exists in the $x$-$z$ plane, not in the $x$-$y$ plane. This feature stems from families of horseshoe-trajectories that become distorted due to the bend in isocontours of $\vartheta$. This again results from the baroclinicity of the envelope, which can be understood from fig. \ref{fig:radvert_1const} is the angle between iso-density and iso-temperature contours. Those results show that there is roughly a structure of three sections for accreting material: In the midplane material is generally flowing into the Hill sphere, excess material that cannot be cooled is ejected at mid-latitudes. Accretion from the top generally is seen in all cases, but amounts to only 10 \% of the net accretion.}
\label{fig:radvert_const}

\end{figure*}

\subsection{Temporal-2D-averages: $\tau=1$ surfaces, luminosities}

In the previous paragraph we presented a comparison of 1D-$(\rho, T , \vartheta)$-structures for a set of important simulation parameters. We now focus on the comparison of opacities, and show a 2D-averaged version of the full 3D-data of those simulation runs to explain the resulting gas dynamics.

In fig. \ref{fig:radvert_1const} we show the density and temperature in contrast with each other. This is important for understanding the gas dynamics: In regions where density and temperature surfaces are parallel to each other, the 3D simulation can be treated as a 1D problem. From the figures it is evident that the region of spherical symmetry reduces in radial extent as opacity reduces. This is because stronger cooling allows the initially pressure-supported hot gaseous blob to lose pressure, and thus it flattens into a shape that is stronger supported by rotation. Another important non-spherically symmetric feature at low opacities is the efficient vertical cooling that causes upwards bending of the temperature isocontours and downwards bending of density contours, being responsible for the pancake structure in the first place. As is evident from the plots, this region starts at the $\tau=1$ surface, from which photons escape freely.

	The run with $\kappa=0.01 \cmg$ exhibits an enclosed optically thick surface, due to the gap being optically thin. This makes the computation of the energy lost from the envelope particularly simple. We compute the luminosity from this run as $L =\sum dA \sigma T^4$, where the surface elements $dA$ are computed from the data cell-by-cell and $T$ is the temperature of each cell. As example, according to eq. \ref{eq:gmmdot} the measured luminosity of $2.0 \times 10^{29} \rm\, erg/s$, for a Saturn-mass planet with characteristic length being the smoothing length of $0.1 r_{H}$, corresponds to an accretion rate of $0.01 \rm\, m_{\oplus} \, yr^{-1}$, while our actually measured accretion rate is $0.02 \rm\, m_{\oplus} \, yr^{-1}$. 
Thus, although there are complications in the 3D-structure, one can use the measured luminosities to reasonably understand the differences in accretion rates, and follow the envelope evolution, which we shall do shortly.

\begin{figure*}
   \centering
	\begin{subfigure}{.45\textwidth}
   \includegraphics[width=\hsize]{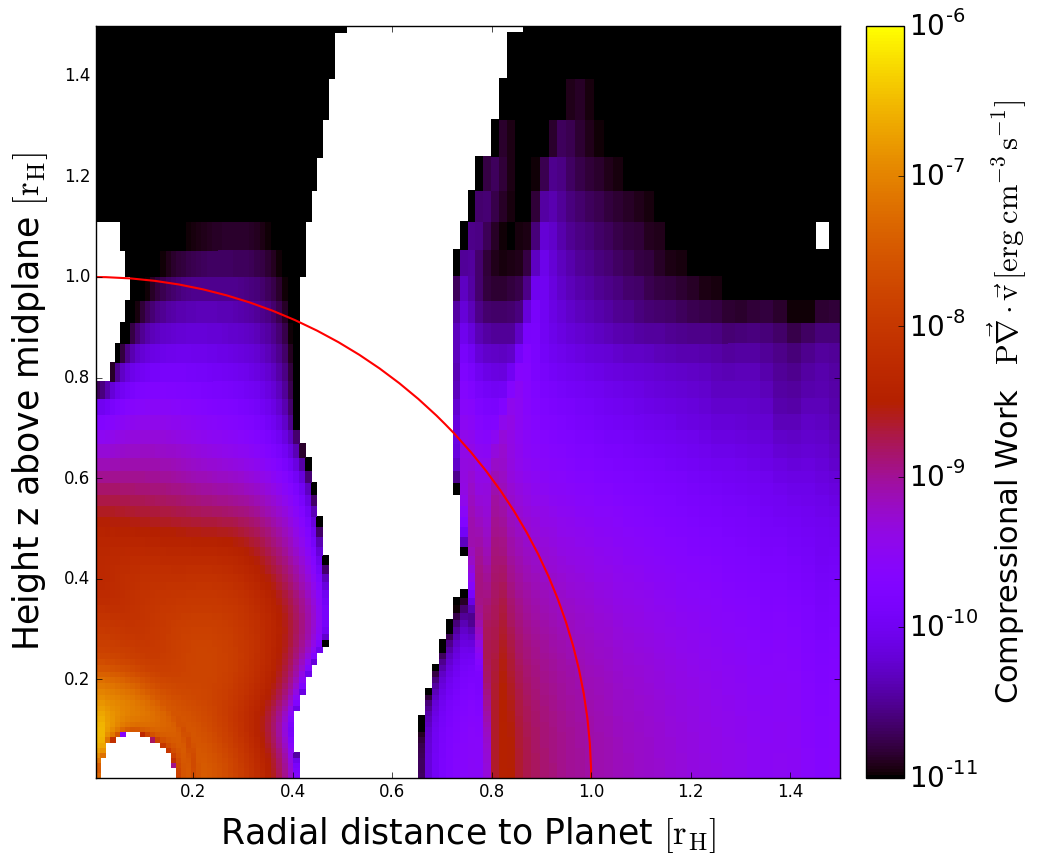}
         \label{fig:pdvwork2}
				\end{subfigure}
				\begin{subfigure}{.45\textwidth}
				\includegraphics[width=\hsize]{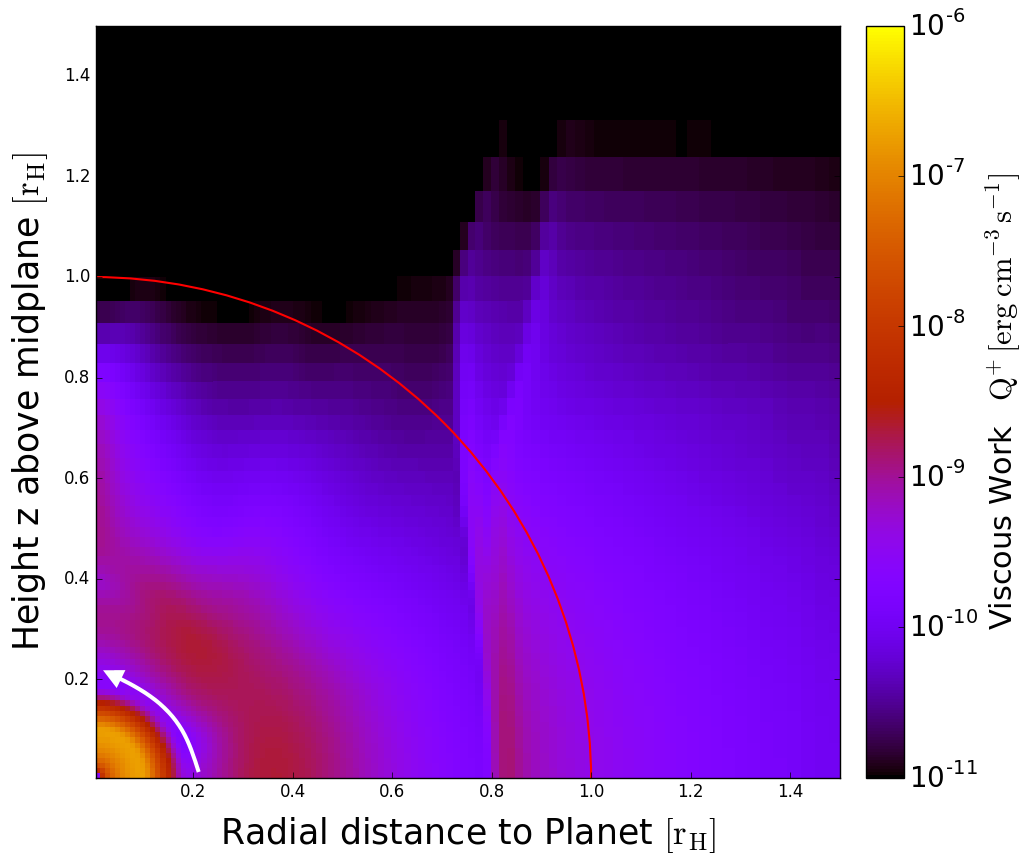}
         \label{fig:viscwork}
		\end{subfigure}
		\caption{Cylindrically averaged compressional (left) and viscous work (right) in the envelope for $\tilde{r}_s=0.2$, $\kappa=0.01 \cmg$, near-identical to the runs with $\tilde{r}_s=0.1$ in figs. \ref{fig:radvert_1const} and \ref{fig:radvert_const}. We chose the run with $\tilde{r}_s=0.2$ because the important physics is better visible.
Most of the heating in the optically thin part of the envelope (for the thin-thick transition compare to fig. \ref{fig:radvert_1const}) is generated at the spiral arm shocks (at $r\approx 0.8 r_{\rm H}$), but radiated away efficiently. 
The optically thick region of the envelope is affected by vertically infalling material, but dominated by contraction of the inner, spherically symmetric region. White regions in the left indicate decompression, which plays only a minor role for the total luminosity in the envelope. 
The decompressing region between $0.4<r/r_{\rm H}<0.7$ is due to two different effects: At $z\leq 0.2$ there is material from the horseshoe flows that decompresses as it falls into the inner envelope. For $z > 0.2$ the decompression is due to the global, meridional flow. The vortical flow does not play a role for the compressional work. Weak decompression is seen above the planet at $r\approx 1.0 r_{\rm H}$ where strong cooling allows for gas to be in free-fall for a very short distance. The viscous work is volume-integrated a factor of $\sim$$10$ times less than the compressional work in the envelope outside of $r_s$, and there is as for the compressional work no heating from the vortical flow visible.
The flow inside of $r_s$ (indicated by the white arrow, also seen in \cite{lambrechts2017}) is probably an artifact that is responsible for the behaviour seen in fig. \ref{fig:lumivsmdot} and further discussed in the text. It is presumably the feature that we need to resolve properly with 10 grid-cells, as found before in fig. \ref{fig:convergencers}.}
		\label{fig:pdvwork}
\end{figure*}

\subsection{Temporal-2D-averages: The baroclinic vortex and gas flows through the Hill sphere}

An interesting phenomenon in our simulations is the appearance of a new circulation feature inside the planetary Hill-sphere, unlike what was observed for gas giants in \cite{lambrechts2019}.
It is well known from classical fluid mechanics that any region with nonzero $\frac{1}{\rho^2} \, \vec \nabla \rho \times \vec \nabla P$ poses a local source for vorticity. Expressed differently, any region where iso-$\rho$ and iso-$P$ contours are inclined towards each other, will produce vorticity. The same argument applies to iso-$\rho$ and iso-$T$ surfaces. We show those surfaces in fig. \ref{fig:radvert_1const}. There it becomes clear, that the stronger the cooling, the sharper the transition from spherical to near-isothermal temperature contours and spherical to vertically stratified density contours is. This poses a stronger vorticity source for stronger cooling, which destabilizes the flow into a vortical feature for $\kappa \leq 0.1 \cmg$. The same feature appears for the Bell \& Lin opacities already at $\epsilon \leq 1.0 \%$. The resulting vortical flows in the outer envelope, can be seen in fig. \ref{fig:radvert_const}. The usage of the opacity set provided by \cite{malygin2014} enlarges the vortex in radial extent by a factor of $\sim$$2$ compared to the vortex in the simulation with \cite{belllin1994} opacities, and thus provides a stronger coupling between the disc and envelope entropies.

There are caveats however to the interpretation of this vortical flow. Firstly, figs. \ref{fig:radvert_1const} and \ref{fig:radvert_const} show time- and cylindrically-averaged views of the flow as well as structure variables. In 2D-cuts in radial-azimuthal direction, the correlation between flow and potential temperature are clearer, but for the sake of simplicity we omit those plots.
It turns out that the vortical feature is mostly a radial feature, i.e. it is strongest in the $r$-$z$ plane, as opposed to the $\phi$-$z$-plane. In the $r$-$z$ plane it also appears centered around $r \approx 0.2 \, r_{\rm H}$ and $z \approx 0.3 r_{\rm H}$, whereas in the cylindrical average it is seen at $r \approx 0.5 \, r_{\rm H}$ and $z \approx 0.3 r_{\rm H}$, so dynamical information extracted from cylindrical averages needs to be taken with caution.
The effect on the 3D gas motion is the following: At $z>0$ horseshoe flows that enter the planetary Hill sphere encounter the baroclinic region. The vortical motion resulting from the baroclinicity gets superimposed onto those flows, being weaker than the forward momentum in azimuth that they already possess. Thus, the vortex causes modified gravitational swing-by trajectories, and does not pose a closed circulation feature.
Secondly, the midplane flows which transport an important part of the mass flux, remain relatively weakly perturbed by this vortical flow. So, the flows as seen in fig. \ref{fig:overview} remain essentially constant, even in the presence of the vortex.
Thirdly, we show in fig. \ref{fig:pdvwork} that the vortex leaves no trace in the overall energetics of the envelope, neither in compressional nor viscous work. This strengthens the notion that this vortical motion is only a passive tracer of the overall baroclinic structure of the envelope, and not a strong feedback mechanism into the envelope energetics.

\subsection{The planetary boundary condition and the energy balance from gas dynamics}
\label{sec:boundary}

Another interesting point in our work is that we are able to quantify the influence of the 3D gas flows on the energetics of the envelope. Because the gas, as it flows through the envelope, compresses and decompresses and is subject to shear which generates viscous heating, one might ask whether this can significantly affect accretion rates. 
We investigate this by averaging the compressional and the viscous work in the envelope. 

The result can be seen in fig. \ref{fig:pdvwork} and we conclude that compressional and shear heating in the outer envelope (at $r > r_s$) are insignificant for the overall energetics. What is dominant is the compressional work in the central region of the envelope, where density and temperature possess spherically symmetric distributions.

However, the amount of central heating might be overestimated. This is because the gravitational smoothing forces a slow circulation (indicated with a white arrow in fig. \ref{fig:pdvwork}) that has non-zero shear. This on average heats the envelope viscously as strong as it does via compressional work, in a pattern that correlates with the smoothing length and thus seems to be a numerical artifact. This is probably the feature that is necessary to resolve with 10 grid cells, explaining our results from fig. \ref{fig:convergencers}.

Fig. \ref{fig:lumivsmdot} showed that in a underresolved simulation, the accretion rate continues to quickly decline, and the luminosity runs into a state of constant non-zero luminosity: If this was a star, we would expect a nuclear burning source to supply energy, but there is no such thing implemented in our simulations. Instead the non-zero luminosity is presumably provided by the viscous work, which is a factor of $\sim$$3$ stronger than the compressional work in the unresolved case, when integrated over the whole Hill-volume. The situation is more complex for the well-resolved cases: There, the ratio of both work integrals over the smoothing length region is $\sim$$1$, but when integrated over the whole Hill-volume this ratio can be as low as $1/10$, giving the wrong impression that viscous work is unimportant overall for the energy balance in the simulation.

That the viscous work is so important in the smoothing length region, even in the well-resolved simulations, could be an artifact of the nonexisting gravity inside $r_s$: Once gas accretes into this region, it still has kinetic energy that needs to be dissipated, as there is no planet to fall on. A region of high shear is generated, until an energy balance is found. 

This problem seems to be avoidable by the usage of a lower viscosity. However it seems possible that the flow could also find a solution with even higher shear, amplifying the viscous heating again. A careful future investigation into this matter is needed, or viscous simulations be avoided altogether.

\begin{figure*}

  \hspace*{-1cm}
	\begin{subfigure}{.35\textwidth}
				\centering
				\includegraphics[width=.95\linewidth]{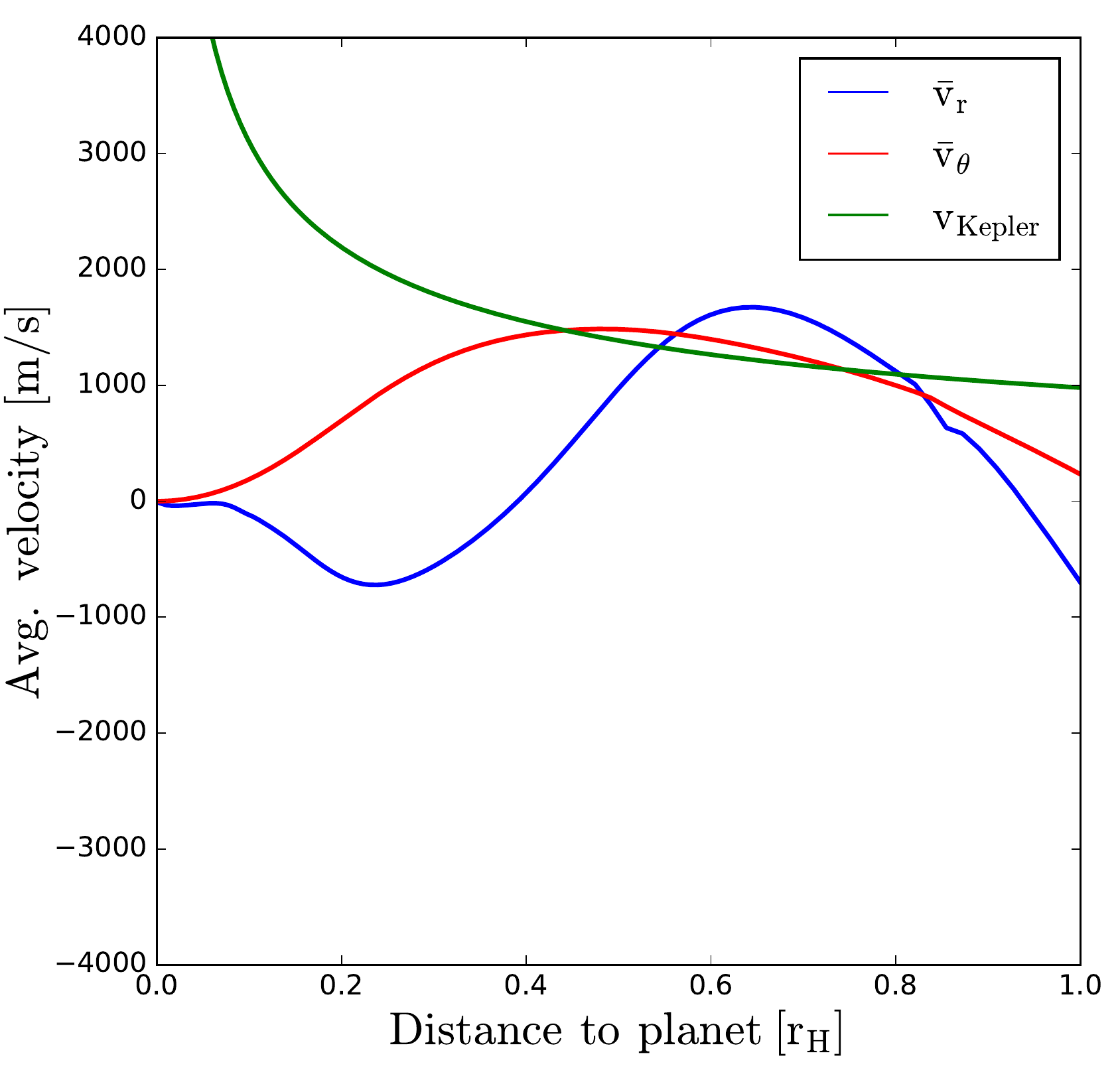}
				\caption{Standard case \\ $\rm \tilde{r}_s=0.1$}
				\label{fig:rotation_sfig1}
		\end{subfigure}%
		\begin{subfigure}{.35\textwidth}
				\centering
				\includegraphics[width=.95\linewidth]{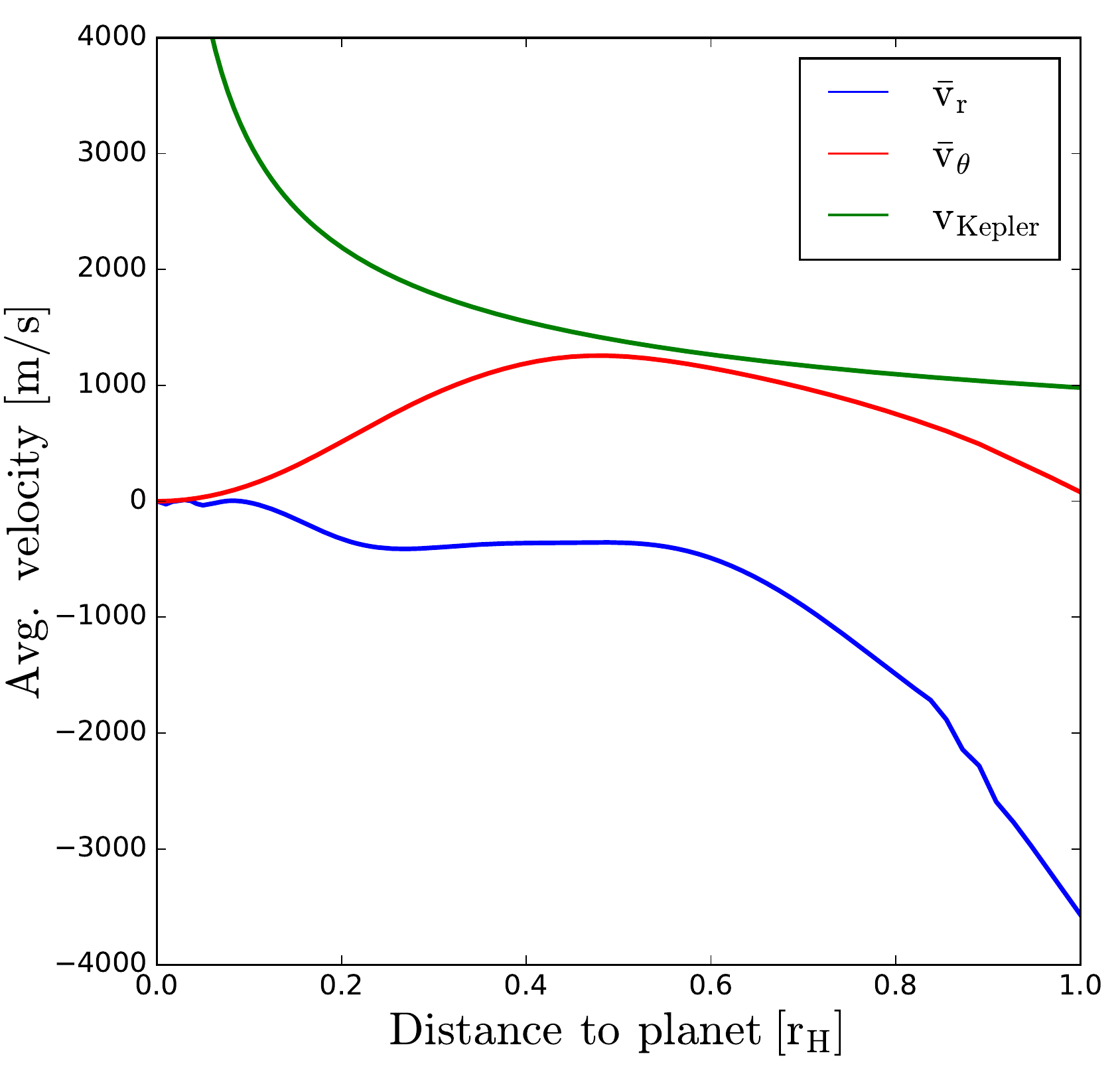}
				\caption{Deeper potential \\ $\rm \tilde{r}_s=0.05$}
				\label{fig:rotation_sfig2}
		\end{subfigure}%
		\begin{subfigure}{.35\textwidth}
				\centering
				\includegraphics[width=.95\linewidth]{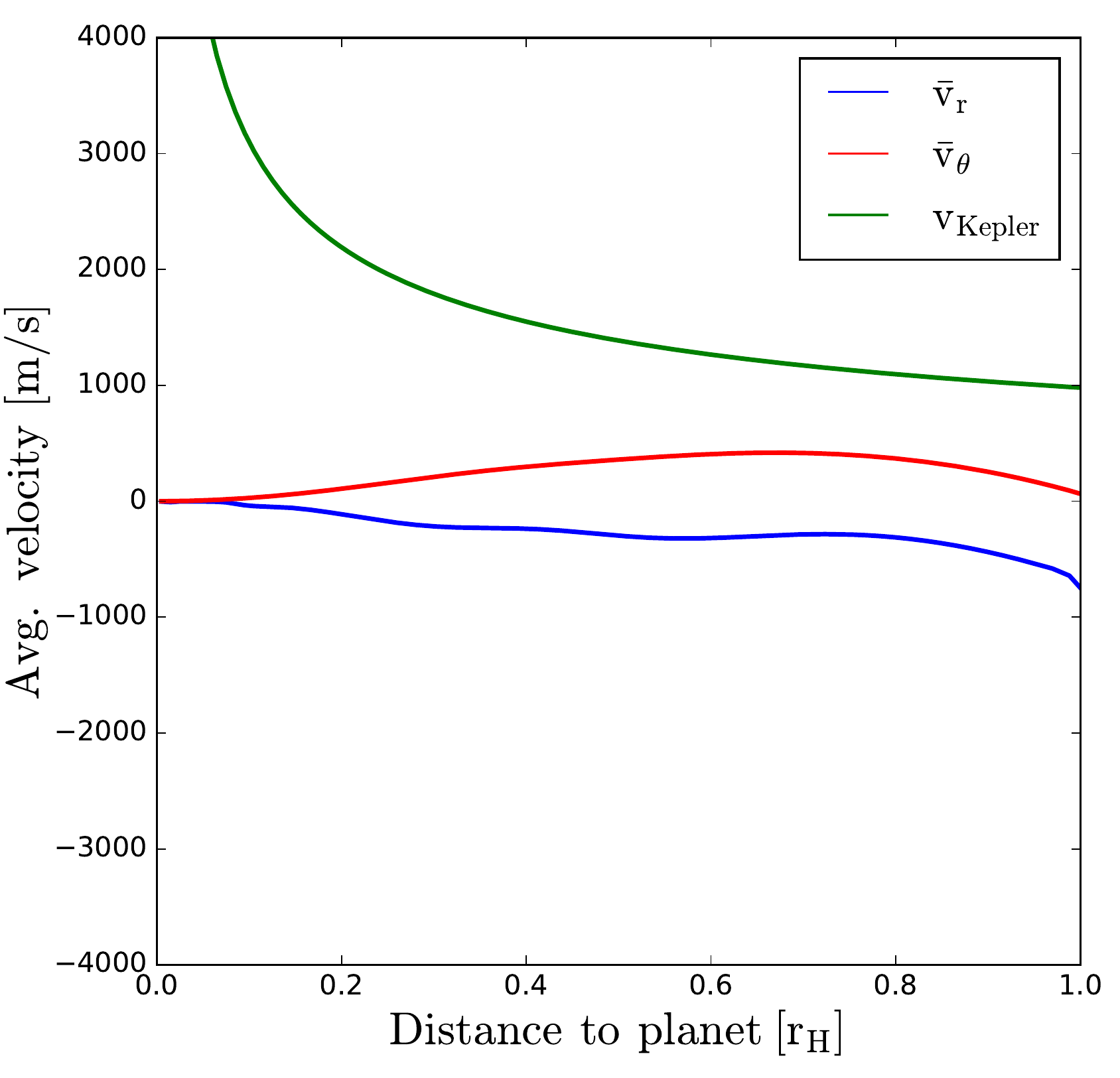}
				\caption{High opacity \\ $\rm \tilde{r}_s=0.1, \; \kappa = 1.0 \cmg$}
				\label{fig:rotation_sfig3}
		\end{subfigure}%
		
\caption{ Rotation profiles in the envelopes for $\kappa=0.01 \cmg$: 
			Our envelopes barely reach Keplerian rotation. There are radial components in the velocity flow that originate from infalling material. The	inner envelope is strongly pressure supported. 
The pressure support is dependent on opacity, as can be expected from higher internal temperatures in the opaque cases. 			
	  }
\label{fig:rotation}

\end{figure*}

\subsection{Rotational properties of the envelope and their evolution}

All our envelopes rotate with sub-Keplerian speed within 0.5 $r_{\rm H}$, as seen in fig. \ref{fig:rotation}. It is conceivable that our choice of large disc viscosity $\alpha$ does not reflect a realistic choice for protoplanetary discs, as probably Saturn had a moon-forming circumplanetary disc, and thus was able to exchange pressure for centrifugal support.

The lack of rotational support may be a consequence of accretion flow coming mainly from above with little or no angular momentum.
However there is still a weak radial inflow that may allow the envelope to accrete high-angular momentum material. 
As seen in the cylindric flow plots, fig. \ref{fig:radvert_const}, there is a net mass flux coming in through the midplane, which can help to speed the envelope up azimuthally, but only if not expelled through the envelope. We now investigate into the physical nature of this flow.

One has to distinguish the evolution of the angular momentum and angular velocity separately. The angular velocity can slow down purely because angular momentum is conserved as mass is accreted. Constant angular momentum with time in spite of accretion is only rarely observed and reflects the originating conditions for the accreted gas: Gas coming from the midplane possesses high angular momentum, while gas from above the planet possesses nearly none.  
As our planetary envelopes usually lose angular momentum over time, we conclude that the mass that is seen flowing in or out through the midplane is in fact only 'passing by', while a significant part of the net accreted mass could conceivably come from the top. 

This is uncertain, however, as the spiral arm shocks have a contribution in changing the angular momentum: As streak-lines are deflected after the shock the relative angular momentum towards the planet also changes. This issue is not settled in our simulation and a follow-up work will investigate the origin of accreted mass by following the streamlines.

Accreted streamlines can transport dust particles of various sizes. In \cite{belllin1994} and \cite{malygin2014} the dust particles correspond to $\rm \mu m$-sized dust. If the same opacity should be provided by larger dust grains (which have a smaller opacity per unit mass), then the total accreted dust mass would need to be higher. At low opacities, this would necessitate the dust grains to be larger than $\rm \mu m$ in size, but then it becomes easier for the pressure bump at the gap edge to capture those particles. It is currently an open question whether moon-forming particles could cross this gap as a result of simple 3D effects, or other physics.

\begin{figure*}
   \centering
	\begin{subfigure}{.45\textwidth}
   \includegraphics[width=1.0\hsize]{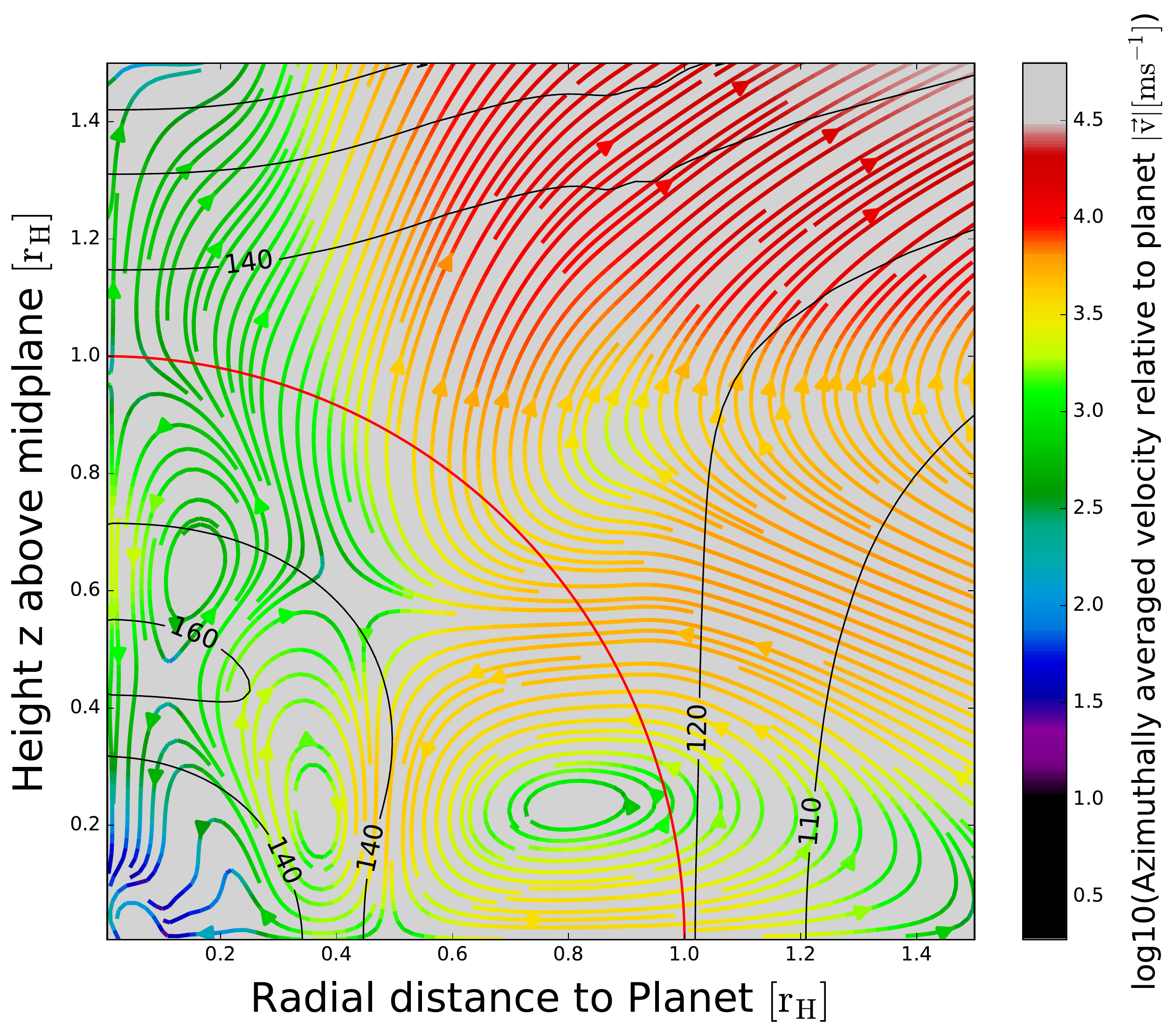}
         \label{fig:buoy2}
				\end{subfigure}
				\begin{subfigure}{.45\textwidth}
				  \includegraphics[width=1.0\hsize]{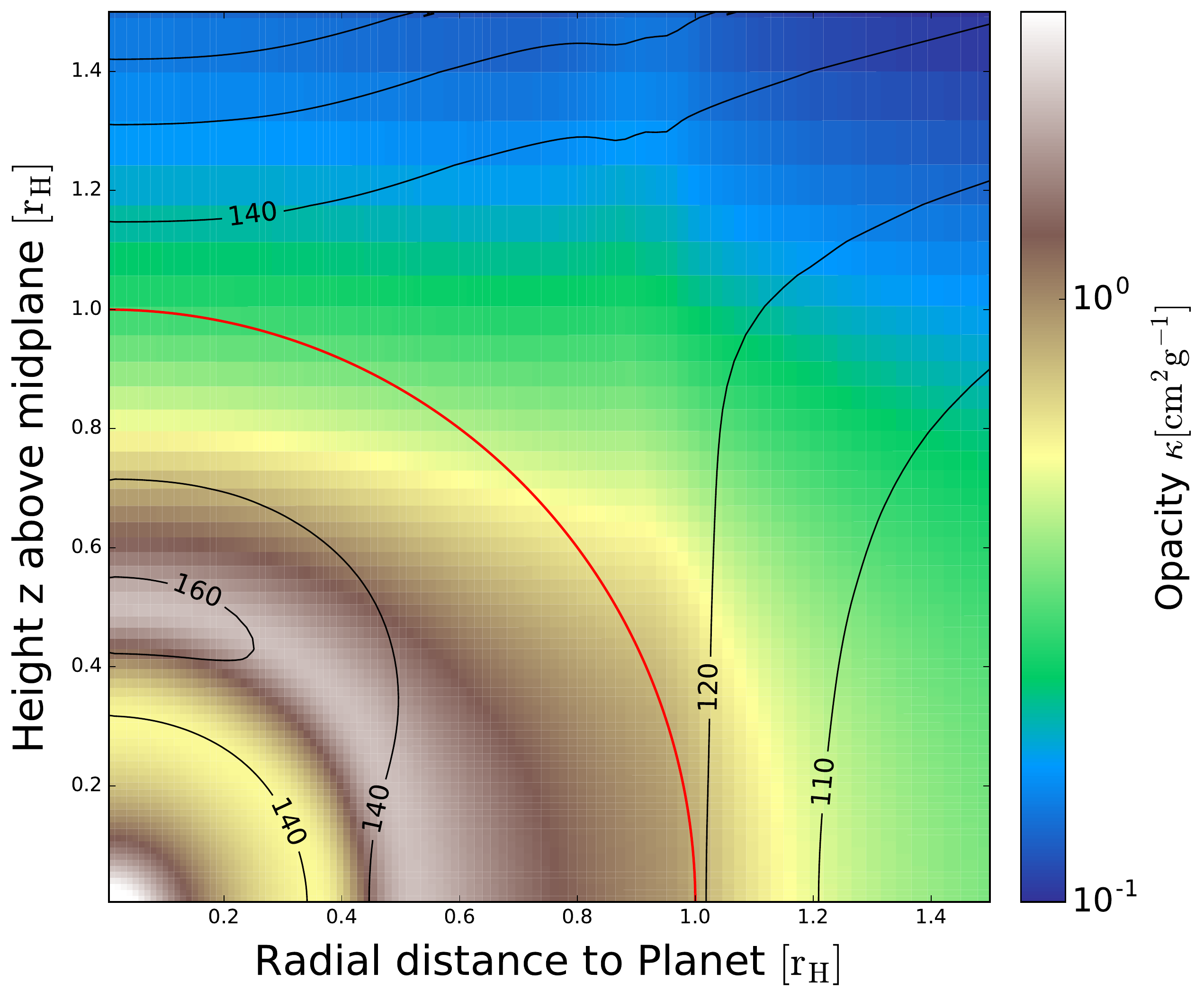}
         \label{fig:opacity}
		\end{subfigure}
		\caption{Potential temperature contours (in black) together with the average flows over 10 orbits (left) and opacity map (right) for Bell \& Lin opacities with $\epsilon=1\%$ and $\tilde{r}_s=0.1$. High optical depth in the vertical direction causes radiative transport to be inefficient, triggering a buoyant instability. This is visible in the irregular vortex structures that the time-average produces, compared to the simpler vortices in fig. \ref{fig:radvert_const} and the potential temperature profile (contours) that is inverted in z-direction between $0.5 \, r_{\rm H} <z<1.0 \, r_{\rm H}$. The	opacity map shows the colocation of potential temperature maxima with the opacity maximma corresponding to a pseudo-iceline according to fig. \ref{fig:opacitiesoverview}. The radial location of the iceline is slightly distorted through the cylindrical averaging process. The net effect of this buoyant instability on the accretion rates in this particular run is weak, as only 10\% of the net accretion rate falls in vertically, which is then blocked out. }
\label{fig:buoy}
\end{figure*}


\section{Discussion}
\label{sec:discussion}

We follow up the main part of our paper with a number of reflections on the results of the various simulation runs.

\subsection{Observability}

Our estimate of the location of the $\tau=1$ surface and the temperature there (see fig. \ref{fig:radvert_1const}), give a first-order approximation of the luminosities and peak wavelength band in which to observe an accreting Saturn-mass planet. For example the run with $\kappa=0.01 \cmg$, $\tilde{r}_s=0.1$ has a luminosity of $3 \times 10^{29} \rm erg/s$ at an effective temperature between 100K-150K, therefore having blackbody-emission maxima between 20$\rm \mu m$ and 30$\rm \mu m$.

Those temperatures are, however, true only for $\tilde{r}_{\rm s}=0.1$, and we have dedicated a significant part of this work to investigate the influence of this numerical parameter, in order to estimate simulation behaviour at realistic smoothing lengths of  $r_{\rm s} \sim r_{\rm Jup}$, corresponding to $\tilde{r}_{\rm s} \sim 0.002$.
Based on the observed scaling behaviour of envelope variables with decreasing smoothing length, the $\tau=1$-surface moves only a small distance, but temperatures there would rise to about $\sim$200K-$250$K.

Those moderate temperatures are partially due to the absence of an accretion shock onto our planet, which can increase luminosities significantly \citep{marleau2017}. This again is due to the high viscosity chosen. The high viscosity imparts through temperature increase a pressure support upon our envelope, which is not bound to vanish when going to realistic $\tilde{r}_s$, quite the contrary. Even deeper potentials will prevent the free-fall of matter from the disc atmosphere through the strengthened pressure support, a trend we have generally observed.

We stress here however, that all this is only true for the Saturn-mass planets. Future work must address the existence of accretion shocks in 3D and observability for a wide range of masses separately.

\subsection{Adiabatic 3D effects}
\label{sec:discussion_adiabatic}

As can be seen in fig. \ref{fig:buoy} there is the possibility that a detached adiabatic region forms in the envelope of a planet. In this case, we found a potential temperature inversion in the simulation run for nominal Bell \& Lin opacities and a potential of $\tilde{r}_s=0.1$. This inversion is directly linked to the existence of an iceline when using this opacity model.
The iceline is at $r=0.5 r_{\rm H}$ and thus at $0.5 \, r_{\rm H} <z<1.0 \, r_{\rm H}$ where the Bell \& Lin opacities are highest, this is a region that is inefficient in radiating energy. The building temperature gradient is sufficient to induce convection, seen in the erratic behaviour of the streamlines in the time-cylindric average. Compare this also to the smooth flows in the constant opacity cases in the the region $0.5 \, r_{\rm H} <z<1.0 \, r_{\rm H}$ in fig. \ref{fig:radvert_const}.

Now checking with fig. \ref{fig:resultssmoothing} we see that this simulation does not differ anomalously in terms of accretion rate from the others in the same opacity function family. The simulation run with constant $\kappa=1.0 \cmg$ and $\tilde{r}_s =0.05$, features a similar adiabatic feature at high z, but a much stronger drop in accretion rate. We speculate that this, however, might be caused by still underresolving this particular simulation.

Different behaviour of the accretion rates would be expected when one would resolve the radiative-convective boundary of a planetary envelope sufficiently: Once resolved, the accretion rate should become nearly constant with deeper and deeper potentials, as convective regions are able to transport high luminosities. Resolving the convective region of the planet would require a much deeper gravitational potential and consequently much higher resolution, than we are able to achieve in this study. This is a topic for future research however.

\subsection{Decrease of accretion rate with time and the problem of the onset of runaway gas accretion}
\label{sec:decrease_accretion}

Our simulation runs experience a slow decrease of accretion rates, as shown in fig. \ref{fig:accretionrates}, even when resolving the smoothing length well. Similar effects have been reported in \cite{pollack1996} (semi-analytical 1D) or in \cite{papa2005} (2D hydrodynamics with newtonian cooling). In the former case the effect was mostly balanced by increasing accretion luminosity through planetesimals, resulting in essentially constant accretion rates with time. The latter work did not include accretion luminosities due to planetesimals, and reports a similar effect as quasi-static contraction of the envelope. They show that gas accretion proceeds from a branch of quasi-static contraction to a branch of runaway accretion as envelope masses increase through ongoing accretion. Those branches are characterized through decreasing mass accretion rate and luminosity on the quasi-static branch (which we would call Kelvin-Helmholtz contraction), and increasing mass accretion rate and luminosity in the runaway branch. 

Later, \cite{ayliffebate2009} presented a long integration run for a $33 \, m_{\oplus}$ core mass where they found a similar but not identical decrease in mass looking similar to ours. They found $\dot{m} \sim t^{-0.6}$, while our data declines as $\dot{m} \sim \exp(-c \dot t)$ with a resolution-dependent constant $c$. Although they employed self-gravity, their envelope was not massive enough to go into runaway gas accretion. Thus although the conditions for future runaway gas accretion were given, the envelope of the $33 \, m_{\oplus}$ core remained on the quasi-static contraction branch of gas accretion. They additionally introduced the notion that the decrease of $\dot{m}$ and $L$ with time that they were observing in their simulations was due to the increased optical depth of the envelope, as more and more material is accreted.

Here we report results that corroborate the finding by \cite{ayliffebate2009}, and that the evolution of the luminosity can explain the decreasing mass accretion rates, see fig. \ref{fig:lumivsmdot}. 
For a well-resolved simulation, the luminosity decreases by the exact same amount as the mass accretion rate, and they can be in fact translated into one another via eq. \ref{eq:gmmdot}. 
The post-prediction of accretion rates from the measured luminosity is off by a factor of $\sim$$1.5-2$, 
but it does reproduce the decline of accretion rates well, and thus suggests that our planetary envelopes lie on the quasi-static contraction branch of gas accretion. Furthermore the the luminosity has been linked to the 
structure of the envelope in sec. \ref{sec:envelope} and can be explained by the increasing mass in the envelope, which moves the $\tau=1$ surface to colder temperatures.

However, our planetary mass of 90 $m_{\oplus}$ would be expected to be in the run-away accretion branch of gas accretion. On this accretion branch the luminosity increases with increasing envelope mass and the accretion accelerates in time. However, as we actually insert a core of mass 90 $m_{\rm \oplus}$, and not a real massive envelope, this core is on the slow contraction branch where the gravitational potential is dominated by the core and it is $m_{\rm env} \ll m_{\rm core}$. We nevertheless expect that the accretion rate that we measure will become independent of the core mass as the envelope mass is increased. Our envelope masses reach at most 20\% of the core mass, but even such a moderate envelope mass should have an accretion rate that approaches the run-away branch with $m_{\rm env} > m_{\rm core}$. Otherwise it would be very puzzling that we find luminosities ($10^{5} \, L_{J} = 10^{-4} \, L_{\odot}$) and envelope accretion rates ($\dot{m} = 10^{-2} \, m_{\oplus}$ for $\kappa=0.01 \cmg$ and $r_s=0.1$) consistent with runaway gas accretion, while being in quasi-hydrostatic contraction.

\cite{ayliffebate2012} employed a very massive protoplanetary disc to force the envelope to accrete rapidly enough to capture the quasi-static contraction as well as the onset of run-away gas accretion. Their simulations behaved identical to \cite{ayliffebate2009} at $m_{\rm env} \ll m_{\rm core}$, thus they demonstrated that the early decreasing mass accretion rates are indeed due to being on the hydrostatic branch of gas accretion. \cite{lambrechts2019} discuss further the two branches of gas accretion and how 3-D simulations on the slow branch can be used to probe the accretion rate on the fast branch as well.

\subsection{Scaling of Saturn-mass accretion rates with constant opacity and disc-limited accretion rates}

According to standard gas accretion theory, accretion rates should scale like $\dot{m} \sim 1/{\kappa}$ \citep{mizuno1980, ikoma2000}. However in 3D radiation hydrodynamics calculations this seems to be only true in the low-mass regime. At higher masses this $\kappa$-dependency gradually weakens, until disappearing at planetary masses much higher than that of Saturn \citep{ayliffebate2009, dangelobodenheimer2013}. The disappearance of the opacity dependence happens when a gas planet reaches the mass accretion limit that can be supplied by the disc.

Our scaling of accretion rates with opacity is $\dot{m} \sim \kappa^{-1/4}$.
This is a similar scaling as \cite{ayliffebate2009} of $\dot{m}$ with opacity, as we have a similar setup and also take gap formation into account.  Therefore we have presumably not reached the limit of what the protoplanetary disc can supply. Also, we have a similar setup to \cite{lambrechts2019}, which state that the total mass flux through the protoplanetary envelopes can be one to two orders of magnitude higher than the accretion rates. Only when the accretion rates would come close to those total fluxes, would we speak of disc-limited accretion.

In fact, the accretion rates in our simulations do not seem to be dictated by the surface densities nor the volume densities at all. Neither the scaling of minimum gap surface densities nor with $\kappa$ or the disc surface densities match with the scaling of $\dot{m}$. Even though we measure strong accretion rates of $\dot{m} \sim 10^{-2} m_{\oplus}\, yr^{-1}$, the disc is able to supply more than that amount of mass. This is also independent of the branch of gas accretion that we find ourselves on, as the disc supply does not care whether it has to supple a certain accretion rate on the slow or fast branch of gas accretion. We have furthermore shown in this paper that the accretion rates are a consequence of envelope properties alone, which makes sense if the disc can supply enough to satisfy any envelope contraction.

There is also the possibility that the sub-linear scaling of accretion rate with inverse opacity is a result of the initial conditions, as a lower opacity implies a more dense midplane and thus the accretion flows can be more massive. This is a hypothesis that will be testable in future work, as we pertain to resolve both lower and higher mass planets in the state of gas accretion.

\subsection{Comparison to isothermal models}
\label{sec:isothermalcomparison}

In a recent work \cite{ida2018} constructed a full model for the gap opening, growth and migration of gas giants in the isothermal framework. While we cannot compare all of our work with an isothermal model, there are certain parts that we can post-predict from our simulation setup and compare results.

The work by \cite{ida2018} does not give a full recipe for the gap shape, but instead provides a formula for the minimum gap depth. Their gap depth is a strong function of $H/r$ and thus the disc temperature. We can use the $H/r$-values from our simulation (see fig. \ref{fig:gapaspectratio}) to compare our actual gap depths $\Sigma(t_{\rm final})/\Sigma_0$ with the ones predicted through the gap depth equation from \cite{ida2018}, which is based on 2D-isothermal simulations presented in \cite{kanagawa2018}. 

For the simulation runs with $\kappa = (1.0,\,0.1,\,0.01) \rm cm^2 /g$ we have unperturbed values of $H/r = (0.038,\, 0.029 ,\, 0.027)$ which would then give $\Sigma/\Sigma_0 = (15, 8, 6)\%$ for the \cite{kanagawa2018}-formula, while we find $\Sigma/\Sigma_0 = (38, 28, 28)\%$ in our simulation runs.
There is work by \cite{bitsch2018} which points out that gaps in 3D isothermal are less deep than in 2D isothermal, which those authors connect to the meridional flows found by \cite{morbidelli2014} presumably repleneshing the gaps with mass. This effect might play a role in our 3D simulations to set our gap depths as well, even if the flow around Saturn-mass planets has different structure (see figs. \ref{fig:radvert_const} and \ref{fig:buoy}, also \cite{lambrechts2019}). Midplane flows towards the planet are not observed at all in isothermal simulations and are due to effects of heating and cooling. The impact on gap depth has not been studied yet, but it is clear that one can not directly compare 3D isothermal with 3D radiative runs.

Using the gas accretion rate of \cite{ida2018} on the runaway accretion branch, we find for our setup an expected $\dot{m} \sim 10^{-1} \, m_{\oplus} \, yr^{-1}$, even considering effect of the gap depth on limiting the accretion rate. Thus there is a clear bifurcation between our work and isothermal models: On one hand the isothermal gaps are deeper, but on the other, the isothermal accretion rates appear to be be higher.

The work of \cite{lambrechts2019} indicates that the mass flow through the envelope is of the order of $\dot{m} \sim 10^{-1} \, m_{\oplus} \, yr^{-1}$ for a Saturn-mass planet, while our actual accretion rates are $\dot{m}_{\rm p} \sim 1$-$3\, \times  10^{-2} \,\rm m_{\oplus} \, yr^{-1}$. This is all the mass that the planet can cool sufficiently while matter flows through the envelope. 
Thus we find, as in \cite{lambrechts2019}, that the flow into the Hill sphere is a poor proxy for the actual gas accretion rate, even for a Saturn-mass planet. Instead, the accretion is mainly regulated by radiative entropy losses, that is not captured in isothermal simulations.

\subsection{Viability of a sink-cell approach}
\label{sec:sinkcell}

In global simulations, it has been a common approach to have a poorly resolved Hill sphere (since otherwise it is impossible to attain results over several hundreds and thousands of orbits) while at the same time implementing a mass sink. 
Usually the planet then has a comparatively flat gravitational potential of $\tilde{r}_s = 0.5-0.8$ and a fraction of the mass inside of $\tilde{r}_s = 0.5$ is simply removed per timestep in order to mimic accretion. A typical recipe \citep[see][]{klahrkley2006} removes mass from the envelope in a more smooth fashion according to some removal function $\dot{m}(r) = f(r) m(r)$. Those approaches are usually generalized under the umbrella term of 'sink-cells'.
As we do find gas accretion with deeper potentials of $\tilde{r}_s = 0.05-0.1$, which decouple the horseshoe region from the planetary envelope, we can aim at justifying or refuting those approaches.

To this end, we define the quantity mass accretion inside fractional Hill-radius $\dot{m}_{\tilde{r}_H}$. This is simply the fraction of the total accretion rate that is accreted by a certain fraction of the Hill radius. In our simulations this is a straightforward quantity to measure, and we see $\dot{m}_{\tilde{r}_H=0.2} = 0.4 \dot{m}$ and $\dot{m}_{\tilde{r}_H=0.1} = 0.2 \dot{m}$ for all constant opacities. This means that 40\% of the total accreted mass 'disappears' into $r=0.2 r_{\rm H}$, and 20\% of the total accreted mass 'disappears' inside $r=0.1 r_{\rm H}$. Those fractions remain remarkably constant for all constant opacity runs.

In the case of Bell \& Lin opacities, we find $\dot{m}_{\tilde{r}_H=0.2} = 0.65 \dot{m}$ for the dusty cases of $\epsilon=1.0, 0.1\%$ and $\dot{m}_{\tilde{r}_H=0.2} = 0.80 \dot{m}$. This means that for a more realistic opacity law one finds more mass being accreted into the gravitational smoothing region, as opposed to outside of it.

This means, if we were to apply a sink-cell with radius $r=0.1 r_{\rm H}$ a mass equivalent of $\dot{m}_{\tilde{r}_H=0.1} = 0.2 \dot{m}$ is needed to be taken out from this region in order for the envelope structure to remain consistent with the remaining simulation domain.
In an effort to partially resolve the physical processes of the Hill-sphere while using a small sink-cell region one can produce inconsistent results, if one had instead applied the full accretion rate to the innermost regions. The full accretion rate is the one consistent with the luminosity according to eq. \ref{eq:gmmdot}. We thus caution against using sink-cells with small radii and full accretion rates.

Drawing conclusions from this for the formation of sub-discs and the mass reservoir available for satellite formation of gas giants, unfortunately our potential depths are insufficient to resolve the inner Hill sphere where moons should form. As 1-D simulations \citep[see the work by][and others]{fujii2017} show, CPDs often only begin at $\tilde{r}_{\rm H}=0.1$. This is the innermost radius where we can draw comparative conclusions for all simulations for. Thus we cannot distinguish between mass delivery towards the circumplanetary disc and mass delivery towards the planet. 

\section{Summary}

In this paper, we have taken a detailed look at the numerical and physical effects that accompany gas accretion into the envelope of a planetary core. We have employed 3-D radiative and hydrodynamical simulations to follow the detailed flow patterns from the protoplanetary disc into the Hill sphere and the gaseous envelope.

The main aim of this paper has been to relax as many approximations as possible, particularly by not using a sink-cell approach, while having a global simulation and resolving the planetary Hill sphere sufficiently. The Hill sphere of the planet has an inner, central boundary condition that is a simple, smoothed gravitational potential characterized by the smoothing length $r_s$.
Our main findings from the simulation framework that we have performed are:
 
\begin{itemize}
\item Numerical resolution of the gravitational smoothing length plays a crucial role in facilitating the accretion process. Insufficient resolution leads to a quick cessation or even negative accretion, likely due to artificial entropy generation in the envelope at the lowest resolutions.\\
\item Simulation runs with 10 or more cells per smoothing length resolution capture accretion physics accurately in the sense that the accretion rate is converged.\\
\item For simulation runs with a constant opacity of $\kappa = 0.01 \cmg$, the gas gaps are optically thin and the planetary envelopes are completely enclosed by their $\tau=1$-surface integrated from infinity towards the midplane. The luminosity that we estimate is lost from this $\tau=1$-surface corresponds well to the mass accretion in accordance with eq. \ref{eq:gmmdot}. We therefore use it as a tool to understand the simulation behaviour when deepening the gravitational smoothing length, while remaining numerically converged. \\
\item We then observe that deepening the gravitational potential increases the energy content of the envelope more than it does the luminosity. As a consequence the accretion rate decreases for a deeper gravitational potential. A signature of this increasingly energetic envelope is the flattening out of its potential temperature gradients which lead to less and less systematic inflows into the envelope. \\
\item Extrapolating the smoothing length $r_s$ to a more realistic size of $\sim$ $1 r_{\rm J}$, we obtain an accretion rate of $10^{-3} \,\rm m_{\oplus} \, yr^{-1}$ for a low opacity $\kappa = 0.01 \cmg$.\\
\item Using constant opacities we find that for high opacities, planetary envelopes remain spherically symmetric and connect smoothly to the disc. At low opacities, the planetary envelope possesses a more complex structure determined by radiating away pressure support and replacing it with centrifugal rotation. The envelope then possesses also a completely enclosed optically thick surface, thus disconnecting it from the disc in terms of radiative energy diffusion.\\
\item The scaling of the accretion rate with constant opacity is $\dot{m} \sim \kappa^{-1/4}$. Considering the agreement of our results with the results from \cite{ayliffebate2009}, this is likely to be an intermediate result between $\dot{m} \sim \kappa^{-1}$ for low-mass planets and $\dot{m} \sim \rm const.$ for Jupiter-mass planets. An investigation of accretion rates versus planetary mass was beyond the scope of this paper and is directed towards future work.\\
\item More complex opacity laws, namely the \cite{belllin1994} and \cite{malygin2014}, differ from the constant opacity case mainly in their shape of the $\tau=1$-surface, the temperatures at that surface, while the luminosities are relatively unchanged, as are the flows inside the envelopes. Thus for the accretion rates the influence of using complex opacity laws seems small in our work, but three important caveats need to be made: 1) In the spirit of constructing self-consistent theories one would rather use the non-constant opacities. 2) In reality line cooling might play an important role in radiating additional thermal energy from the envelope \citep{gustafsson2008} 3) Our simulations do not reach temperatures above $2000$ K as reaching the higher temperatures near the core would require prohibitively high resolution.  
The more realistic non-constant opacity laws would only show major differences with the constant opacities at such high temperatures. It is nevertheless not clear that the opacity so deep in the envelope has any influence on the total radiative entropy loss and mass accretion rate of the envelope.\\ 
\item The three-dimensional gas dynamics through the envelope can be understood by combining pressure and temperature into potential temperature, which is a measure of the entropy that links the local conditions in the envelope to a disc temperature at the same entropy. The potential temperature structure appears to be sensitive to the used opacity function, particularly is the entropy reduction more rapid at low opacity. 
The hydrodynamical flows in the Hill sphere result in heating both by shear and compression, but those values are negligible compared to the compressional work of the deep planetary envelope. In the future, 3-D simulations reaching down to the planetary core would be necessary to analyse this heating budget deep in the convective envelope.\\
\item The role of viscosity and viscous work in our simulations is complex. In underresolved simulations it can outcompete the compressional work by a large factor and drive the envelope into a steady-state with zero accretion but non-zero luminosity. In resolved simulations it becomes clear that most of the viscous work amounts to unity compared to the compressional work and is generated by a shear flow inside the gravitational smoothing region. While it is possible to envision that this shear flow is necessary to dissipate the momentum of accreted gas inside the smoothing region, as it has nowhere else to go, it is not certain whether the amount of heat generated by this process is realistic. \\
\item Our envelopes generally accrete less than the total mass flux through the Hill-sphere. Increasing the viscosity would therefore not help in accreting even more mass through feeding the planetary gap. Instead, increasing the viscosity would help increase the accretion rate via heating the envelope and increasing luminosity, which allows further accretion. \\
\item Planetary envelopes for our simulations generally lack rotational support, as pressure support dominates those envelopes interior of approximately half the Hill sphere. However, we adopted a relatively high disc viscosity $\alpha$$\sim$$0.01$. Future simulations with a lower $\alpha$ would be needed to probe circumplanetary disc formation at viscosities that reflect protoplanetary disc values better.
\end{itemize}

Our investigations pose a challenge for the simplified view of core-accretion where reaching the critical core implies immediate gas accretion and formation of a gas-giant planet. This critical core mass is usually constructed as being a function of disc parameters and envelope parameters \citep{mizuno1980}. Our simulations are set in a regime where any critical core mass is however far exceeded, yet we find ourselves in a state of  Kelvin-Helmholtz contraction. 
Thus, reaching the critical core mass is in itself not a guarantee that the core will accrete substantial amounts of gas. This is also true for low-mass cores, as explored in \cite{lambrechts2017}, where gas accretion can be stalled by high opacities.

Our work demonstrates that the pre-existing concept of a luminosity which results from the temperature structure of planetary envelope, explains our found accretion rates well. This poses a connection point for future radiation hydrodynamics calculations in 1D for structure calculations and full 3D radiation hydrodynamics, as our and similar setups can compute those luminosities self-consistently. Furthermore, future work exploring a broader physical parameter space than what was possible in our work should be able to profit from this.

\begin{acknowledgements}
We thank the anonymous referee in their help improving the quality of the paper and making a number of constructive suggestions.
The authors are indebted to illuminating discussions with Jeffrey Fung, Gabriel-Dominique Marleau and Michiel Lambrechts. The new complex opacity tables have been provided by Nick Malygin himself for which we are very grateful. MS and AJ were supported by a project grant from the Swedish Research Council (grant number 2014-5775) and an ERC Starting Grant from the European Research Council (grant number 278675 - PEBBLE2PLANET).  AJ was further supported by the Knut and Alice Wallenberg Foundation (grant number 2012.0150) and the European Research Council (ERC Consolidator Grant 724687- PLANETESYS). BB thanks the European Research Council (ERC Starting Grant 757448-PAMDORA) for their financial support. All the simulations presented in this work were performed on resources provided by the Swedish National Infrastructure for Computing (SNIC) at Lunarc in Lund University, Sweden.
\end{acknowledgements}

\begin{appendix} 

\section{Radiative fluxes in two-temperature FLD}
\label{sec:appendix_fld}

\begin{figure}
   \centering
   \includegraphics[width=1.0\hsize]{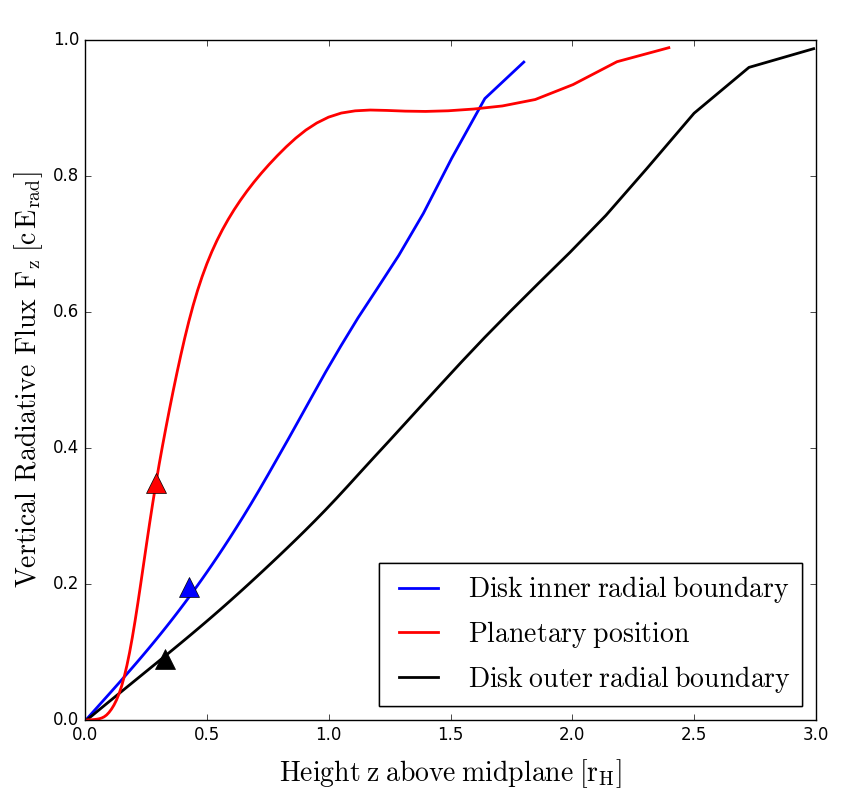}
	\caption{Reduced vertical radiative flux as function of the vertical coordinate. We show that in the far upper regions of the disc atmosphere our code reaches the correct flux values. The computed vertically integrated $\tau=1$-transitions at three different positions in the disc are marked with triangles. The end points of the curves correspond to the $7^{\circ}$-angle of our simulations at those disc positions. Compared to the analytical $|F_{\rm rad}|/c E_{\rm rad} \sim 0.55$  at $\tau=1$ \citep[eq. 82.15 in][]{mihalasmihalas} It is evident that our method locates the optically thin-thick transition further inward than the analytical value. This leads to higher temperatures on the $\tau=1$ surface and thus to an overestimate of L.}
\label{fig:appendix_fred}
\end{figure}

In this appendix we present the behaviour of the radiative fluxes in optically thin regions in the flux-limited diffusion approximation with our two-temperature solver. Radiative fluxes should approach $|F_{\rm rad}| = c E_{\rm rad}$ in the optically thin regions of the disc atmosphere. We briefly comment on how well this limit is approached in our code.

In fig. \ref{fig:appendix_fred} we show how the reduced flux in component $i$ $f_{i,\rm red} \equiv |F_{i}|/c E_{\rm rad}$ evolves with height above the midplane. It is important to note that the correct limit $f_{i,\rm red} = 1$ does not necessarily indicate the correct solution for the fluxes and energies themselves. This needs to be addressed in the future via the computed moments of a direct radiative transport solution.

We probe three different locations in our disc as showcases, the outer and inner rim of our disc simulation boundary, located each $\pm 6 r_{\rm H}$ from the planet, and the position directly above the planet. The curves for both rims show how the open boundary conditions for the radiation in $z$ enforce the correct value of $f_{\rm red}$. Just above the planet, density gradients are stronger than at the rims because of the gap formation and the accreting envelope. Thus above the planet  $f_{\rm red}$ achieves its free-streaming limit much earlier than the boundary, indicating correct cooling behaviour without the action of the boundary conditions.

Our computed optically thin-thick transition is somewhat lower than the analytic value would predict, which impacts our luminosity estimate from Sect. \ref{sec:envelope}, which we shall call here the 'simple' estimate. When the actual optically thin-thick transition lies slightly higher, its temperature will be lower, thus the actually radiated $L_{\rm ex}$ will be less. This is what we see when we compare the simple with the exact luminosity in appendix \ref{sec:appendix_lumi}.

\section{Validation of the luminosity estimate}
\label{sec:appendix_lumi}

\begin{figure}
   \centering
	  \includegraphics[width=1.0\hsize]{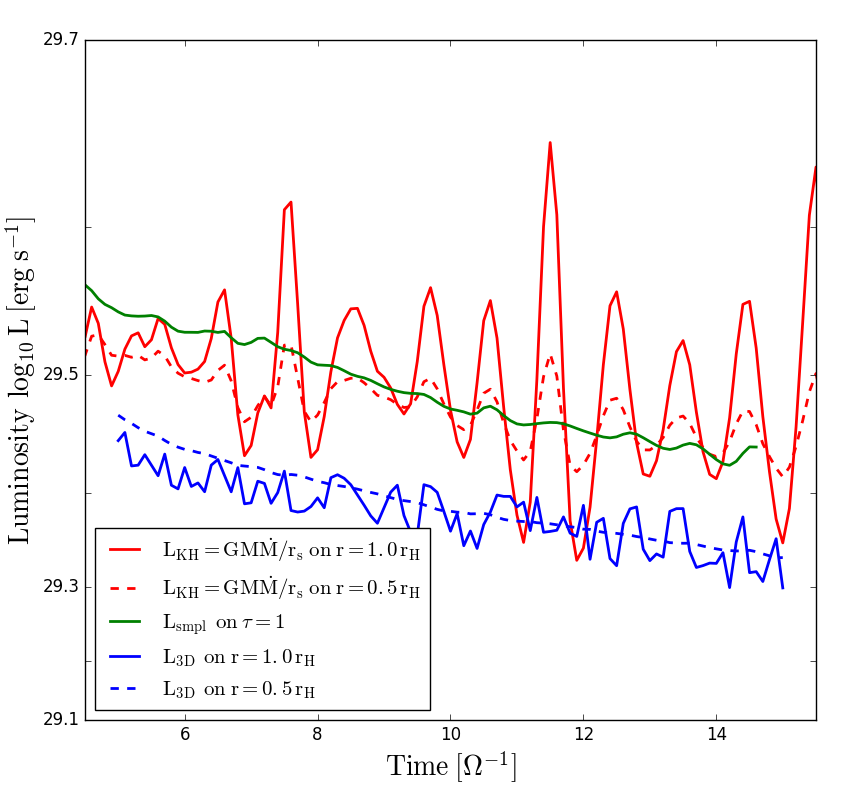}
		\caption{Different luminosities in comparison. The Kelvin-Helmholtz luminosity that would be required to balance the accretion $\dot{M}$ in red, with the accretion rate measured internal to two different shells. Our simple estimate of L from sec. \ref{sec:envelope} is marked in green. The exact measurment of L (shown in blue) differs from the simple one by $20\%$. Thus, even with the misplaced $\tau=1$, the difference in L is quite small and does not impact our main conclusions.}
\label{fig:appendix_lumi}
\end{figure}

This appendix is dedicated to the discussion of the quality of our approximate computation for the luminosity $L$ in Sect. \ref{sec:envelope}. We show a comparison of the simple estimate with the exact luminosities in fig. \ref{fig:appendix_lumi}.

We attempt a validation of our simple method of estimating the envelope luminosity $L_{\rm smpl}$ via the approach from sec. \ref{sec:envelope}, which assumed $L_{\rm smpl} = \sum dA \, \sigma \, T^4_{\tau=1}$. We show that $L_{\rm smpl}$ is correctly estimated down to a systematic uncertainty of $20\%$ and compare it to the luminosity that would be required by eq. \ref{eq:gmmdot}, i.e. the Kelvin-Helmholtz luminosity $L_{\rm KH}$.

To calculate the full 3D luminosity $L_{\rm 3D}$, it is required to transform the data from the spherical, heliocentric grid onto a new spherical, planetocentric grid, in order to compute the integral $L_{\rm 3D} = d_t \int dV E_{\rm rad} = \oint dA F_{\rm r, planet}$. Here $dV$ is the Hill-volume or a fraction thereof, $dA$ are the intersection surface elements of the planetocentric spherical grid with the heliocentric grid cuboids and $F_{\rm r, planet}$ is the planetocentric radial flux that must be computed from the heliocentric flux components.

In this method, the orientation of the surface at the optically thin-thick transition is however non-unique in our implementation of the FLD approximation, because we use the same radiative diffusion coefficient in all three spatial directions. Thus, we can merely measure the luminosity on spherical shells of varying radii around the planet. The planetocentric surface elements $dA$ are then computed semi-analytically from vector geometry and the grid data. Finally, the $F_{\rm r, planet}$ are resulting from the scalar product $\vec n \cdot \vec F$ where $\vec n$ is the planetocentric normal vector on a sphere of arbitrary radius around the planet, and $\vec F$ is the heliocentric flux data coming taken from the simulation.

A remaining difference of about $(10^{29.5}-10^{29.4})/10^{29.5} \approx 20\%$ between $L_{\rm smpl}$ and $L_{\rm 3D}$ is evident. This difference seems to be independent of the chosen shell for the exact flux, because nearly all the radiative flux is generated in the very hot region interior to $r<r_s$. $L_{\rm smpl}$ can only sensibly be computed on the $\tau=1$-surface, so no other comparison value exists.

This comparison was made for the simulation run with parameter values $k=0.01\cmg$ and $\tilde{r}_{\rm s} = 0.1$. We picked this parameter value for the comparison instead of the fiducial $k=0.01\cmg$ and $\tilde{r}_{\rm s} = 0.2$, as the latter was mainly a means of showcasing different numerical resolutions and finding the convergence criterion, while we intend to work with simulations of $\tilde{r}_{\rm s} = 0.1$ in the near future and it is thus reasonable to gauge those correctly.

We conclude that the match of $L_{\rm smpl}$ overestimates the radiative losses, and the fact that it is $L_{\rm 3D} < L_{\rm KH}$ indicates that some energy is lost possibly due to advection of low-density gas from the envelope (thus not negatively impacting the mass accretion rates, see also fig. \ref{fig:overview}). The close fit of $L_{\rm smpl}$ with $L_{\rm KH}$ also occurs for the simulation run with $\kappa =0.01 \cmg$ and $\tilde{r}_s=0.05$, but not with $\tilde{r}_s=0.2$ as seen in fig. \ref{fig:lumivsmdot}. This indicates a larger contribution of radiative transport for the overall cooling vs. advection for hotter envelopes.

The 20$\%$ discrepancy between $L_{\rm smpl}$ and $L_{\rm 3D}$ does not influence our main conclusions of the paper, which were concerning the explanation and analysis of why accretion rates decrease with time and why accretion rates decrease with decreasing $r_{\rm s}$.

\section{Entropy and potential temperature from hydrodynamic data}
\label{sec:appendix_entropy}

In this section we aim to derive the potential temperature and show its relation to other hydrodynamic and thermodynamic quantities.

We start with the first law of thermodynamics, expressed in terms of the Helmholtz potential $H$, for which it is $dH = c_p dT$. We choose this particular potential because it is simply a short-cut, as we don't want to go through a whole class of statistical physics:
\begin{align}
dH = T dS + V dP
\end{align}
and after demanding $dS=0$ for an adiabatic process, we get
\begin{align}
c_p dT = V dP
\end{align}
Now with the ideal gas law $PV = NRT$ and a fixed number of particles $N$ we can change this into
\begin{align}
c_p dT = NRT dP/P
\end{align}
from which we get directly
\begin{align}
\frac{c_p}{NR} d ln T = d ln P
\end{align}
and thus, after integration from the arbitrary state $(T,P)$ to the reference state $(\vartheta,P_0)$ under the adiabatic assumption we find a the temperature $\vartheta$:
\begin{align}
\vartheta = T \frac{P_0}{P}^{\frac{NR}{c_p}}
\end{align}
which is simply the temperature a gas blob would have if we'd move it adiabatically from $(T,P)$ to the pressure $P_0$. We can now take this as a new variable and compute it on the whole simulation domain. It is important to note that we know $NR = cp - cv$, and the adiabatic index $\gamma = cp/cv$. So for $\gamma = 1.4$ we can rewrite $\frac{NR}{c_p} = \frac{c_p - c_v}{c_p} = 1-1/\gamma = 2/7 \approx 0.29 $.\\
In order to do this in post-processing, we rewrite the preceding definition $\vartheta=\vartheta(P)$ into $\vartheta=\vartheta(\rho, T)$ by usage of the ideal gas law. The result with $P = \rho kT / \mu$ is, after a bit of algebra
\begin{align}
\vartheta = T \left(\frac{\rho}{\rho_0}\right)^{-2/5}
\end{align}
The entropy can then be computed via $S = k_B ln \vartheta$.

\end{appendix} 


%
%

\bibliographystyle{aa}


\end{document}